\documentclass[[format=acmsmall,screen=true,review=false,natbib=true]{acmart} 

\usepackage{multibib} 
\AtBeginDocument{%
  \providecommand\BibTeX{{%
    \normalfont B\kern-0.5em{\scshape i\kern-0.25em b}\kern-0.8em\TeX}}}

\setcopyright{acmcopyright} 
\copyrightyear{2021}
\acmYear{2021}
\acmDOI{}

\acmJournal{CSUR}%
\acmVolume{0}
\acmNumber{0}
\acmArticle{0}
\acmMonth{0}

 \usepackage[utf8]{inputenc} 
 \usepackage[T1]{fontenc}    
 \usepackage{lscape}
 \usepackage{pdflscape}
 \usepackage{afterpage}
 \usepackage{multirow}
\usepackage{tabularx}
 \usepackage{threeparttable}
 \usepackage[british]{babel}
 \usepackage{qcircuit}
 \usepackage{makecell}
 \usepackage[colorinlistoftodos]{todonotes}
\usepackage{caption}
\usepackage{subcaption}
\graphicspath{{images/}}
\setcellgapes{3pt}

\usepackage{pifont}
\newcommand{\cmark}{\ding{51}}%
\newcommand{\xmark}{\ding{55}}%

\newcommand{\ie}{{i.e.\ }}
\newcommand{\eg}{{e.g.\ }}

\newcommand{\quotes}[1]{``#1''}
\renewcommand{\b}{\bfseries}
\renewcommand{\H}{\mathcal{H}}
\newcommand{\bb}[1]{\mathbf{#1}}
\newcommand{\ket}[1]{|{#1}\rangle} 
\newcommand{\bra}[1]{\langle{#1}|}
\newcommand{\Ham}{\hat{H}}

\newcommand{\paul}[1]{\hat{\sigma}_{#1}}
\newcommand{\paull}[2]{\hat{\sigma}_{#1}^{#2}}
\newcommand{\inprod}[2]{\langle{#1}|{#2}\rangle}
\newcommand{\outprod}[2]{|{#1}\rangle\langle{#2}|}
\newcommand{\ii}{\mathrm{i}}
\newcommand{\R}[1]{\mathbb{R}^{#1}}

\newcommand{\h}[1]{{#1}}

\newtheorem*{adth*}{Adiabatic Theorem}
\newtheorem{post}{Postulate}

\newcites{A}{Appendix References} 

\begin{document}

\title{A Leap among Quantum Computing and Quantum Neural Networks: A Survey \\
}

\author{Fabio Valerio Massoli}
\authornote{Corresponding author}
\email{fabio.massoli@isti.cnr.it}
\orcid{0000-0001-6447-1301}
\author{Lucia Vadicamo}
\orcid{ 0000-0003-0171-4315}
\email{lucia.vadicamo@isti.cnr.it}
\author{Giuseppe Amato}
\orcid{0000-0001-7182-7038}
\email{giuseppe.amato@isti.cnr.it}
\author{Fabrizio Falchi}
\orcid{ 0000-0001-7182-7038}
\email{fabrizio.falchi@isti.cnr.it}
\affiliation{%
  \institution{Istituto di Scienza e Tecnologie dell'Informazione ``Alessandro Faedo'', CNR}
  \streetaddress{Via G. Moruzzi 1}
  \city{Pisa}
  \country{Italy}
  \postcode{56124}
}

\renewcommand{\shortauthors}{Massoli, et al.}


\begin{abstract} 
In recent years, Quantum Computing witnessed massive improvements in terms of \h{available resources} and algorithms development. The ability to harness quantum phenomena to solve computational problems is a long-standing dream that has drawn the scientific community's interest since the late 80s. In such a context, we propose our contribution. First, we introduce basic concepts related to quantum computations, and then we explain the core functionalities of technologies that implement the Gate Model and Adiabatic Quantum Computing paradigms. Finally, we gather, compare and analyze the current state-of-the-art concerning Quantum Perceptrons and Quantum Neural Networks implementations.
\end{abstract}


\maketitle


 \section{Introduction} \label{introduction}

Artificial Intelligence (AI) has intrigued and puzzled many generations of scientists and largely fueled \h{novelists} imaginations. The modern definition of AI -- as the ensemble of computer systems empowered with the ability to learn from data through statistical techniques -- \h{dates} back to 1959. Machine Learning (ML), a subclass of AI, is a discipline that aims \h{to study algorithms that can learn from data} to perform tasks without following explicit instructions. Often, these algorithms are based on a computational model that belongs to differentiable programming techniques, called Neural Networks (NNs). The success of such algorithms resides in their ability to learn to achieve a specific goal~\citep{lecun1989backpropagation,hinton2006fast}, i.e., they learn to discover hidden patterns and relations among data to fulfill the task at hand~\citep{hastie2009elements,lecun2015deep}.
Mathematically, NNs are made of a sequence of transformations, called layers, composed of \h{linear} operators and elementwise nonlinearities. Then, the goal of learning is to modify the transformations' parameters to fulfill a task successfully. Whenever a model accounts for more than a couple of such layers, it is called a Deep Learning (DL) model or a Deep Neural Network (DNN). Thanks to their representation power and the development of new technologies and training algorithms, DL models obtained astonishing results in the last two decades, achieving superhuman performance on certain tasks~\citep{silver2018general}. 
However, higher performances require more complex models and larger datasets to train them, thus constantly increasing the hunger for resources and power to learn to solve a given problem. 

In this regard, quantum computers might offer new solutions that exploit quantum phenomena such as interference, superposition, and entanglement. Such a characteristic is expected to speed up the computational time and to reduce the requirements for extensive resources, yielding the concepts of \textit{quantum advantage} and \textit{quantum supremacy}~\citep{nielsen2002quantum,preskill2012quantum}. 
The mentioned quantum phenomena are described within the framework of quantum mechanics~\citep{heisenberg1925quantum,schrodinger1926quantisierung,dirac1981principles}, a \quotes{young} physics theory formalized at the beginning of the 20th century. Such a theory unveils the intrinsic statistical characteristic of nature, a behavior that unfortunately is hidden from us, at least in the macroscopic world. 

The quest for a quantum computer started with the ideas of Benoff~\citep{benioff1980computer} and Feynmann~\citep{feynman1982simulating,lloyd1996universal}, in the 1980s, pointing out that quantum computers should be more efficient in solving specific problems. \h{For} example, quantum devices might help in studying very complex, e.g., entangled, systems by emulating or simulating~\citep{georgescu2014quantum,efthymiou2020qibo}
their behavior \h{in} chips that are quantum by their nature~\citep{reiher2017elucidating,wecker2015solving}.

From a computer science point of view, a quantum computer represents a computational model based on the principles and postulates of quantum mechanics. Such techniques aim at embracing the power of quantum theory into a device that can be conveniently \quotes{programmed} to fulfill a given task. Moreover, the result of the computation itself might represent a quantum object encoding different answers, each \h{solving} a specific problem~\citep{deutsch1992rapid}. \h{It is only recently} that researchers \h{have succesfully realized} a quantum processor capable of performing controllable computations based on quantum phenomena. 
\h{Several industrial applications already benefited from such a technology, such as:}
chemistry~\citep{kandala2017hardware}, optimization problems~\citep{moll2018quantum}, finance~\citep{egger2020quantum}, quantum sensing~\citep{degen2017quantum}, and quantum imaging~\citep{chen2012optical}, among others.
\h{Besides the mentioned applications, quantum computers might offer advantages in terms of energy management compared to classical ones}. \h{Indeed, quantum computers are expected to be more energy-efficient than supercomputers considering}~\citep{ajagekar2019quantum}. For that reason, hybrid approaches might offer exciting solutions to lower the energy consumption for a given computation by moving the high-energy portion of the computation on the Quantum Processing Unit (QPU) while leaving the low-energy \h{one} to the cloud~\citep{toosi2014interconnected}. A fundamental result of quantum information theory is the observation that, although quantum phenomena allow solving some classical problems more efficiently than classical computations~\citep{bernstein1997quantum}, quantum computers cannot compute any function which is not turing-computable~\citep{deutsch1985quantum}. 

Concerning the context of Quantum Machine Learning (QML)~\citep{biamonte2017quantum}, a relevant contribution comes from the ability of quantum devices to speed up linear algebra-related tasks. For example, \h{it has been shown that using the Harrow-Hassidim-Lloyd (HHL) quantum algorithm~\citep{harrow2009quantum,zheng2017solving} to sample solutions for a system of linear equations offers an exponential speedup over its classical counterpart. 
} Furthermore, Shor's algorithm~\citep{shor1994algorithms,shor1999polynomial}
 for integer factorization and Grover's algorithm~\citep{grover1997quantum} for searching
unstructured databases are additional examples of procedures that benefit from a quantum formulation. Although general QML is of great interest, it is not the central topic of our work as it has already been covered extensively in the literature. Therefore, we focus on the QML sub-area concerning recent Quantum Neural Networks (QNNs) approaches. \h{Despite the fact that} their name recalls the neural network structure mentioned earlier, they are characterized by a \h{completely different design}. 

Concerning quantum computations, the most commonly adopted paradigms are the Gate Model (GM)~\citep{nielsen2002quantum} and \h{the} Adiabatic Quantum Computation (AQC)~\citep{albash2018adiabatic}. In \h{Inspite of being equivalent} up to a polynomial overhead, they represent two profoundly different computation approaches. The first one is based on the concept of \quotes{gate}\h{: a unitary transformation} applied to one or more quantum bits \h{(\textit{qubits})}, i.e., the basic units of quantum information (see~\autoref{sec:background}). Instead, in the AQC, one typically encodes the desired objective function into a quantum system and then lets it evolve. However, in both paradigms, the \h{QPU state} at the end of the evolution embodies the answer to the given problem. Thus, we can highlight the main differences between the two approaches as follows. The GM allows \h{users} to control the transformation to apply on the qubits directly and is discrete in time, while AQC does not allow to control single qubits directly and does not discretize the time property of the system.

Unfortunately, quantum technologies are still at their dawn, having minimal computational capacity. \h{Furthermore, significant technological challenges arise from the requirement for quantum systems to be isolated from the environment in order to avoid decoherence, which causes lack of information stored in the quantum device}. Therefore, researchers typically rely on quantum simulators to test their ideas while waiting for the next generation of quantum \h{computers}. 
Whether quantum supremacy is real or \h{not} is still an open question. For example, we do not expect quantum computers to solve \h{efficiently} worst-case NP-hard problems like combinatorial \h{ones}. Instead, we do expect that we will be able to find a better approximation to the solution or find it faster than a classical computer~\citep{khot2016hardness,preskill2018quantum}.

Stemming from those considerations, we conceived this survey to offer both the neophyte and the more experienced reader insights into several fundamental topics in the quantum computation field. Moreover, noticing that the literature lacks a detailed discussion about the latest achievements concerning Quantum Perceptrons (QPs) and QNNs, we gather, analyze and discuss state-of-the-art approaches related to those topics.
\h{Notably, we can summarize our work as follows:}

\begin{itemize}
    \item we review the current state-of-the-art regarding QPs and QNNs by discerning among theoretical formulations, simulations, and implementations on real quantum devices;
    \item we report the main achievements by different research groups concerning the topic of quantum supremacy;
    \item \h{we provide a gentle introduction to several basic notions about quantum mechanics, quantum information, and quantum computational models;}
    \item we collect and organize the most relevant papers to this survey on a GitHub\footnote{\url{https://github.com/fvmassoli/survey-quantum-computation}} page allowing the interested readers to easily and quickly browse through the literature.
\end{itemize}

\h{Moreover, to ease the understanding of the paper's content,
in~\autoref{appendix} we summarize basic notions about the Dirac notation, postulates of quantum mechanics, the physical realization of qubits, the Bloch Sphere representation, the Variational Principle, and the Adiabatic Theorem.
We suggest the reader who is unfamiliar with these topics to peruse the~\autoref{appendix} before reading any further. 
}

Concerning the remaining part of the paper, it is organized as follows. \h{In~\autoref{related_works}, we report on other surveys on the topic at hand. In~\autoref{sec:background}, we introduce the fundamental concept of a qubit, and we give the reader a brief overview of the currently most widespread quantum computational models. Then, in \autoref{q_supremacy}, we tackle the concept of quantum speedup.} 
Subsequently, in~\autoref{quantum_neural_networks}, we move to the core topic of this survey, i.e., QNN approaches. Finally, the conclusions are drawn in~\autoref{conclusions}. 
\h{In~\autoref{tab:notation}, we report a summary of the notations used throughout the manuscript, while a summary of the used abbreviations is available in~\autoref{tab:acronym} in the~\autoref{appendix:table}.}

\begin{table}[tbp]
\caption{Summary of notation used throughout this paper.}		\label{tab:notation}
\footnotesize
	\begin{tabular}{p{0.25\columnwidth} p{0.6\columnwidth}} 
		\toprule
		\textbf{Symbol} & \textbf{Definition} \\
        \midrule
        $\text{Re}(z)$, $\text{Im}(z)$ & Real and imaginary parts of the complex number $z$ \\
        ${z}^* $ &  Complex conjugate of the complex number $z$.\\ 
        ${A}^* $ &  Complex conjugate of the matrix $A$.\\
        ${A}^T $ &  Transpose of the matrix $A$.\\ 
        ${A}^\dagger$ &  Complex conjugate transpose of the matrix $A$, i.e.  ${A}^\dagger=(A^*)^T$ .\\
        ${A} \otimes B$ & Kronecker product of the matrix $A$ with the matrix $B$ .\\ 
        $\hat{A}$ & Linear operator whose matrix representation is $A$\\
        $I$, $\hat{I}$ & Identity matrix and Identity operator\\
        $\Ham$ & Hamiltonian \\
        $\xi_i$ & Eigenvalues of the Hamiltonian \\
        $\dot{f}$ & Newton's notation for the derivative of a function $f(t)$ with respect to $t$\\
        $\ket{\psi} $ & Vector of a Hilbert space, also called \textit{ket}.\\ 
        $\bra{\psi}$ & Vector dual to $\ket{\psi}$, also called \textit{bra}. \\
        $\inprod{\psi}{\phi}$ &  Inner product between $\ket{\psi}$ and $\ket{\phi}$ whose results in scalar.\\ 
        $\outprod{\psi}{\phi}$ &  Outer product between $\bra{\psi}$ and $\ket{\phi}$ whose results in a matrix.\\
         $\|\, \ket{\phi}\,\|=\sqrt{\inprod{\phi}{\phi}}$ & Norm of $\ket{\phi}$. \\ 
        $ \delta_{ij} $ & Kronecker delta that is defined to be equal to $1$ whenever $i=j$, and $0$ otherwise  \\ 
        $ \ket{\psi} \otimes \ket{\phi}$, $\ket{\psi}\ket{\phi}$, $\ket{\psi,\phi}$, $\ket{\psi 
        \phi}$ & Tensor product between $\ket{\psi}$ and $\ket{\phi}$ and its abbreviated notations \\ 
        $\ket{\psi}^{\otimes n}$ & Tensor product between $n$ identical states $\ket{\psi}$, \ie $\ket{\psi}^{\otimes n}=\ket{\psi} \otimes \dots \otimes \ket{\psi}$ \\
        $\bra{\psi} \hat{A} \ket{\phi}$ &  Inner product between $\ket{\psi}$ and $\hat{A}\ket{\phi}$ where $\hat{A}$ is a linear operator\\
        $\langle \hat{A} \rangle$ & Average value of an operator w.r.t. to a given state $\ket{\psi}: \bra{\psi}\hat{A}\ket{\psi}$ \\ $\{\ket{0}, \ket{1}\}$ or $\{\ket{\uparrow}, \ket{\downarrow}\}$ & Computational basis\\
        $\ket{+}$, $\ket{-}$ &  Hadamard basis states, defined by $\ket{+}=\frac{1}{\sqrt{2}}(\ket{0}+\ket{1})$, and
        $\ket{-}=\frac{1}{\sqrt{2}}(\ket{0}-\ket{1})$\\ 
        $\rho$ & Density matrix describing a given system's state: $\rho = \sum_i \lambda_i \outprod{\psi_i}{\psi_i}$\\ 
             $\paul{x}$, $\paul{y}$, $\paul{z}$ & Pauli operators\\
        $H$, ${\Qcircuit @C=1em @R=.7em {& \gate{H} & \qw}}$ & Hadamard opertator/gate \\
        $\mathrm{Tr}$ & trace operator\\
        \bottomrule        
	\end{tabular} 
\end{table}

 \section{Other surveys} \label{related_works} 

In the literature, many review papers have been devoted to the realm of quantum computation and information. 
Several surveys address quantum ML and its applications. However, very little has been said about QNNs, and reviews that include this topic often cover it marginally or lack a detailed discussion on the most recent achievements. For \h{this} reason, we \h{devote} our work mainly to \h{reviewing} Variational Learning approaches which encompass Quantum Approximate Optimization Algorithms, Variational Quantum Eigensolver, Quantum Perceptrons, and Quantum Neural Networks. 

Gyongyosi and Imre \cite{gyongyosi2019survey} reviewed the technological status of quantum computation and information systems by discussing several papers that can be useful for researchers interested in delving deep into specific aspects of quantum computations. Savchuk and  Fesenko \cite{savchuk2019quantum} reported a brief analysis of some of the fundamental principles beneath quantum computing. McGeoch \cite{mcgeoch2020theory} presented a comparison between expected results based on a purely theoretical calculation about complexity theory and what is currently achievable and achieved on the modern physical realization of quantum processors based on quantum annealing. They also gave interesting insights on basic notions concerning the quantum annealing process, which is at the heart of the D-Wave quantum platforms. Preskill \cite{preskill2012quantum,preskill2018quantum} debated the concept of \quotes{quantum supremacy} and \h{the} potential of quantum computing.

In the last two decades, there has been a strong interest and commitment in the scientific community to develop quantum algorithms to solve ML problems, giving life to the field of QML. \h{Hitherto,} there is not a universally accepted definition of QML yet. However, A{\"\i}meur et al. \cite{aimeur2006machine} and Adcock et al. \cite{adcock2015advances} provided useful categorizations based on the employed learning paradigms. For example, in \citep{adcock2015advances}, the term QML is used to refer to two learning categories. \h{The first one concerns learning processes performed using classical computations aided by sub-routines relying on quantum computations. The second one concerns approaches that exploit quantum computations only} and for which there are no sub-routines that can be performed equally well on a classical computer. \h{Notwithstanding}, this categorization excludes from the QML class all the learning approaches performed on a classical machine but that use data \h{coming} from quantum devices. 

A viable approach to QML stems from the quantum algorithms that speedup linear algebra-related tasks (see, e.g., \citep{wittek2014quantum,biamonte2017quantum,schuld2015introduction,havlivcek2019supervised}). 
In this context, several quantum \h{extensions} of classical algorithms have already been proposed, for example, for clustering~\cite{kerenidis2018q,otterbach2017unsupervised}, 
Support Vector Machines~\cite{rebentrost2014quantum}, semidefinite programs~\cite{brandao2019QuantumSDP}, gradient descent for linear systems~\cite{kerenidis2020quantum}, PCA~\cite{lloyd2014quantum}, \h{and} 
variational generation for adversarial learning~\cite{romero2021variational}, just to cite some.
Biamonte et al. \cite{biamonte2017quantum}, and Schuld et al. \cite{schuld2015introduction} gave \h{an} 
overview of the achievements and challenges of quantum-enhanced machine learning algorithms, considering aspects of both gate and adiabatic computational models.
Basic algorithms for \h{QML}, such as Grover's algorithm \citep{grover1997quantum}, quantum state "similarity" estimation, and the HLL algorithm \citep{harrow2009quantum}, are 
reviewed in \citep{zhang2020recent}. The authors also illustrate how these algorithms (based on the quantum gate model) have been used in literature to speed up standard ML algorithms, like Support vector machine, $k$-means clustering, {PCA}, and Linear discriminant analysis. 
\h{Furthermore, in \citep{ablayev2020quantum1,ablayev2020quantum2} the authors review quantum solutions for binary classification and k-Nearest Neighbour problems}. Ciliberto et al. \cite{ciliberto2018quantum} covered quantum machine learning approaches with emphasis on theoretical aspects, such as computational complexity analysis, discussion on limitations of QML, and challenging aspects like learning under noise. Theoretical aspects of ML using quantum computers are also discussed in \cite{arunachalam2017guest}. 
Many other survey papers overview quantum algorithms for ML and their applications, see for example \cite{manju2014applications,aimeur2013quantum,bharti2020machine,dunjko2018machine,dunjko2020non,schuld2019quantum,ramezani2020machine} and references therein.

Benedetti et al. \cite{benedetti2019parameterized} reviewed parametrized quantum circuits, which are arguably the quantum analogous of NN models.
The Variational Quantum Algorithms, which employ parameterized quantum circuits, are also discussed in \cite{cerezo2020variational}. 
Kamruzzaman et al. \cite{kamruzzaman2019quantum} reviewed the status of (theoretical) Quantum Neural Networks and their limitations. Allcock and Zhang \cite{allcock2019quantum} focused on quantum generalizations and applications of some popular neural-network concepts, like  Boltzmann machine, generative adversarial networks, and autoencoders. \h{Recently, Magini et al. \cite{mangini2021quantum} gave a concise overview of the main directions that have been taken to develop quantum artificial neural networks. }


\section{Qubit and Quantum Computational Models}\label{sec:background}
In this section, we \h{first introduce the concept of \textit{qubit} and then give} a brief overview of the two main paradigms of quantum computation: the \textit{Gate Model} (GM), or Universal Quantum Computing, and the \textit{Adiabatic Quantum Computation} (AQC). A comprehensive analysis of the mentioned approaches is out of scope for our work. However, in what follows, we try to empower the reader with a basic grasp of those two types of computations.

\subsection{Qubit} \label{sec:qubit}
The \textit{qubit}, \h{short} for \quotes{quantum bit}, represents the fundamental unit of information for quantum devices. \h{It is the quantum correspondent of the bit. However, apart from the last part of its name, the qubit does not share much with its classical cousin.} 
Similar to the bit, which assumes two values only (0 or 1), the qubit can be \quotes{observed} (i.e., measured) in two possible states, typically indicated as $\ket{0}$ and $\ket{1}$.
However, differently from the bit, a qubit can also be in a so-called \textit{superposition} of states before the measurement. 
Intuitively, that means that the qubit can be in either the state $\ket{0}$ or $\ket{1}$ with a certain probability.
Notwithstanding, when measured, the qubit state \quotes{collapses} into one of them.
The qubit is a rather abstract concept since such an object does not exist in the real world. Instead, we relate it with artificial atoms, i.e., physical systems able to emulate the behavior of an atom. Specifically, we are typically interested in emulating the behavior of two-state systems that satisfy a certain number of hard constraints such as small dissipation and isolation from the environment. 

From a mathematical point of view, a \textit{qubit state} $\ket{\psi}$ is a unit vector in $\mathbb{C}^2$. 
The symbol $\ket{\cdot}$ we just used is inherited from Dirac~\citep{dirac1939new} formalism\footnote{\h{please refer to \autoref{app:dirac} in the \autoref{appendix} for an extensive introduction to this formalism.}}
employed in the quantum mechanics. 
In a nutshell, the object $\ket{\psi}$, called \textit{ket}, represents a vector in a Hilbert space. 
Similarly, $\bra{\psi}$, called \textit{bra}, is defined as the adjoint (or dual) of such a ket. 
\h{We remind the reader unfamiliar with the Dirac formalism that the symbol within a ket or bra, e.g., the \quotes{$\psi$} used above, can be an arbitrary label (e.g., a letter or a number).}
Note that often binary labels are used, especially to indicate the vectors of a space basis. For example, the basis $\left\{\begin{bmatrix} 1 \\0 \end{bmatrix},\begin{bmatrix} 0 \\1 \end{bmatrix} \right\}$ of $\mathbb{C}^2$, referred to as the \textit{computational basis}, is typically indicated as $\{\ket{0}, \ket{1}\}$, or sometimes also as $\{\ket{\uparrow}, \ket{\downarrow}\}$.
Using this formalism, a qubit state can be represented \h{as} $\ket{\psi}= \alpha\, \ket{0} + \beta\, \ket{1}$,
where $\alpha$ and $\beta$, called \textit{amplitudes}, are complex numbers such that $|\alpha|^2+|\beta|^2=1$\footnote{
\h{since a qubit has more states available than simply two levels, often it is useful to visualize it as a point on the so-called \textit{Bloch sphere}, which is a unit sphere in a three-dimensional Euclidean space with the north and south pole corresponding to the computational basis states $\ket{0}$ and $\ket{1}$, respectively (see \autoref{sec:bloch} in the Appendix ).}}.
Therefore, a qubit state is a coherent {superposition} of the computational basis states $\ket{0}$ and $\ket{1}$. 
However, this does not mean that a qubit has a value between $\ket{0}$ and $\ket{1}$ but rather that it is not possible to say \h{whether} the qubit is definitively in the state $\ket{0}$, or definitively in the state $\ket{1}$. In fact, when we measure a qubit we observe either $\ket{0}$ with probability $|\alpha|^2$, or $\ket{1}$ with probability $|\beta|^2$. After the measurement, the qubit state collapses to whatever value ($\ket{0}$ or $\ket{1}$) was observed, irreversibly losing memory of the former $\alpha$ and $\beta$ amplitudes. Please note that different kinds of measurements \h{exist;}  
the one we referred to above is called measurement concerning the computational basis, which is among the most widely used.\footnote{ \h{other measurements are discussed in the appendix (\autoref{app:measurements}).}} 

We now turn our attention from the theoretical definition to a more physical one. Any physical system whose state space can be described by $\mathbb{C}^2$ can serve as a qubit's physical realization.\footnote{\h{some physical realizations of qubits are discussed in the appendix (\autoref{sec:physical_realization}). }} These systems are referred to as quantum \textit{two-level} systems, as their state can be described by a vector in a 2-dimensional Hilbert space. 
The state of an isolated quantum mechanical system composed by $n$ two-level system, called \textit{quantum $n$-register}, is described by a vector in a $2^n$-dimensional Hilbert space ($\mathcal{H}^{2^n}$). This means that a state of an $n$-register can represent the superposition of $2^n$ basis states, which is the cornerstone of quantum parallelism. Formally, an n-register state can be described by the linear combination of the basis vector $\{\ket{0}, \dots \ket{2^n-1}\}$:
\begin{equation}
    \ket{\psi}=\sum_{i=0}^{2^n-1} \alpha_i \ket{i}, \qquad \text{with} \quad \sum_{i=0}^{2^n-1} |\alpha_i|^2=1. 
\end{equation}
If we measure the state $\ket{\psi}$ with respect to the computational basis, we get one of the basis state labels $i$ with probability $p(i)=|\alpha_i|^2$ and the corresponding post-measurement state is $\ket{i}$. 
Usually, binary labeling is used to denote the $2^n$ basis states. For example, the state of a two-qubit system can be expressed as 
$
    \ket{\psi}= \alpha_{00}\ket{00}+\alpha_{01}\ket{01}+\alpha_{10}\ket{10}+\alpha_{11}\ket{11}$ {with} $ \sum_{i,j} |\alpha_{ij}|^2=1.
$

The composition of two or more quantum \h{systems} is represented by the tensor product. For example if two systems A and B have the states $\ket{\psi_A}= \alpha \ket{0} + \beta\ket{1}$  and $\ket{\psi_B}= \gamma \ket{0} + \delta\ket{1}$, respectively, then the system C composed by A and B has the state
$
    \ket{\psi_C}= \ket{\psi_A}\otimes \ket{\psi_B}= \alpha\gamma\ket{00}+\alpha\delta\ket{01}+\beta\gamma\ket{10}+\beta\delta\ket{11}
$.
Finally, note that a state $\ket{\psi} \in\mathcal{H}^{2^n}$ of an $n$-register is called \textit{entangled} if it that cannot be decomposed as the tensor product of $n$ single qubit states.

\subsection{Gate Model}\label{gate_model}

At the heart of the GM, there is the concept of \textit{circuit} model, i.e., a sequence of building blocks that realize elementary operations. Such building blocks are called \textit{gates}. \h{Thus, a gate encodes a well-controlled operation acting on a single qubit or a subset of qubits~\citep{nielsen2002quantum,preskill2018quantum} in a given system.} When acting on more than one qubit, these gates can give \h{rise} to the phenomenon of entanglement \h{establishing} a strong correlation among \h{qubits}.

As a direct consequence of the postulates of quantum mechanics\footnote{\h{postulates of quantum mechanics are summarized in the Appendix (\autoref{app:postulates_quantum_mechanics}) }}, quantum gates are represented by unitary operators $\hat{U}$, i.e., $\hat{U}^\dagger \hat{U} = \hat{I}$~\cite{barenco1995elementary,nielsen2002quantum}. 
This property automatically translates into saying that all the quantum gates must be reversible, unlike classical logic gates. However, there are exceptions to such a rule. \h{For example, measurements are transformations on qubits that are allowed not to be reversible. }

\h{Being unitary operators, gates} can be represented in different ways. Although it might be easier to formalize them as matrices, typically the Dirac notation, or the outer product among state vector, is leveraged:
\begin{equation}\label{eq:gates_dirac}
    U = \begin{bmatrix}
            u_{00} & \cdots &u_{0({2^n-1})}\\
            \vdots & \ddots & \vdots\\
            u_{ ({2^n-1})0} & \cdots & u_{({2^n-1})({2^n-1})}
        \end{bmatrix} = 
     \sum_{i=0}^{2^n-1}\sum_{j=0}^{2^n-1} u_{ij} \ket{i}\bra{j}
\end{equation}
where $\{\ket{i}\}_{i=0,\dots, 2^n-1}$ is the currently used basis. Hence, in the simplest case of a single-qubit gate, it is represented by a $2 \times 2$ unitary matrix whose coordinates are completely specified by its action on the basis states. 
More formally, single-qubit gates describe transformations that belong to the Lie group of $2 \times 2$ unitary matrices with determinant 1, called special unitary group of degree 2 ($SU(2)$)~\citep{barenco1995elementary}. 

A particularly useful set of one-qubit gates are the \textit{Pauli gates}\footnote{\h{Please refer to the Appendix (\autoref{tab:gates_examples} and \autoref{tab:gates_examples2} in \autoref{sec-gates_tables}) for the definition of the Pauli gates and other single- and multi-qubit gates mentioned in this manuscript.}}  that, together with the identity operator, span the vector space formed by all one-qubit unitary operators. In other words, any one-qubit gate can be expressed as a linear combination of the Pauli gates. It is worth mentioning that the Pauli gates $\paul{x}$, $\paul{y}$, and $\paul{z}$ correspond to rotations around the $x$-, $y$- and $z$-axes of the Bloch sphere, respectively. As an example, the $\paul{x}$ matrix represents an operation called the \quotes{bit flip}, or NOT gate, which maps the $\ket{0}$ to $\ket{1}$ and vice versa, in the computational basis:
\begin{equation}\label{eq:bitflip}
    \paul{x}\ket{0} = \begin{bmatrix}
            0 & 1 \\
            1 & 0
        \end{bmatrix} \ket{0}= 
    \left(\ket{0}\bra{1} + \ket{1}\bra{0}\right)\ket{0} = 
    \ket{0}\bra{1}0\rangle + \ket{1}\bra{0}0\rangle = 
    \ket{1}
\end{equation}
Interestingly, we can notice that the operation in~\autoref{eq:bitflip} resembles the NOT gate in classical computations. 

Apart from the single-qubit gates, there are unitaries that involve two or more qubits. Perhaps, the most famous ones are the Controlled-NOT (CNOT) and Toffoli gates. 
\h{Note} that many multi-qubit gates are designed to perform \quotes{controlled} operations, meaning that an operation is executed on a qubit, named the \textit{target} qubit, if another qubit (called \textit{control} qubit) is in a specific state. For example, the CNOT gate applies the NOT gate operation to the target qubit only when the control one is in the state $\ket{1}$. It is formalized as: 
$
        CNOT(\ket{\psi} \otimes \ket{\gamma}) = \ket{\psi} \otimes \ket{\psi\ \texttt{XOR}\ \gamma}
$ 
where $\ket{\psi}$ and $\ket{\gamma}$ are the control and target qubit, respectively. 

An \h{exciting} result coming from the GM paradigm is that not all gates are \quotes{fundamental} for computations. Indeed, as it happens in the classical case, there is a set of quantum gates, named \textit{universal}, that can be used to approximate to arbitrary accuracy any quantum circuit~\citep{nielsen2002quantum}. The Hadamard (H), Controlled-NOT (CNOT), phase (S), and $\pi/8$ (T) gates constitute a universal quantum computation set of gates. Although the S-gate can be constructed from the T-gate, it is typically \h{considered an element of the universal set due to its extensive} usage in fault-tolerant circuit construction. 

\begin{figure}[tbh]
    \centering
    \includegraphics[width=0.6\linewidth]{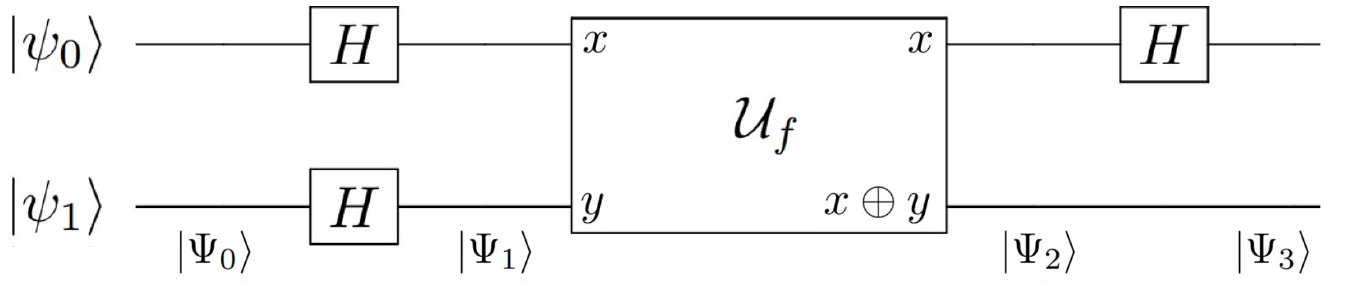} 
    \caption{Quantum circuit implementing Deutsch's algorithm~\citep{deutsch1992rapid,nielsen2002quantum}.}
    \label{fig:gate_circuit}
\end{figure}

As mentioned above, a circuit is made of a sequence of gates acting on a given set of qubits. To characterize a circuit, two metrics are typically reported, namely, the \textit{width} and the \textit{depth}. The first one refers to the number of qubits involved in the calculations, while the second one represents the longest path in the circuit, resembling the largest number of operations applied to a given qubit from the beginning to the end of the computations. In~\autoref{fig:gate_circuit}, we report an example of a circuit in which we applied single- and two-qubit gates. Specifically, the circuit represents one of the most famous algorithms in the scientific community called the Deutsch's algorithm~\citep{deutsch1992rapid}, which determines if a given Boolean function $f:\{0,1\}^n\to \{0,1\}$ is balanced or constant.

As we can see from~\autoref{fig:gate_circuit}, a quantum circuit is depicted as a set of qubits, represented by lines, to which the gates (squares) are applied, and where the order from left to right represents the flow of time. However, it is worth noting that there is not an actual \quotes{flow} of the qubits in a quantum device. Indeed, the state of qubits at each given time represents the state of the QPU, while gates are the transformations applied to them to change the device's status.

\subsection{Adiabatic Quantum Computation}\label{subsec:adiabatic_model}

In the previous section, we introduced the GM in which a qubits' state is evolved by applying a series of gates. In such a design, the time evolution is discretized since the different operations are performed at subsequent time instants. Quite differently, the Adiabatic approach to quantum computation~\citep{farhi2000quantum} leverages a time-continuous evolution of the qubits' state according to the Schr{\"o}dinger equation~\cite{schrodinger1926quantisierung}. Moreover, it does not require nor allow to apply \h{any} transformations, such as gates, directly on qubits. \h{Instead,} everything is encoded into a quantum operator, called \textit{Hamiltonian} ($\Ham$), that describes the forces to apply to a given system of qubits to move it into the desired state over time. However, it is possible to show that these two approaches are polynomially equivalent~\citep{farhi2000quantum,aharonov2008adiabatic} by using the technique introduced in~\citep{lloyd1996universal}. Such a goal is reached by discretizing the evolution time interval and applying the Trotter formula~\cite{farhi2000quantum} to each time segment. 

\h{For a deep dive into the AQC, we refer the reader to the comprehensive and fascinating review by~\cite{albash2018adiabatic}.} In what follows, we \h{assume} 
that the reader knows what a Hamiltonian is or at least its definition in classical mechanics. For our purposes, we need to consider that $\Ham$ can be interpreted as the energy operator, in the sense that its eigenvectors are the energy eigenstates, and its eigenvalues are the energies of the corresponding eigenvectors. Hence, as it will become obvious from the following discussion, in the AQC, any problem is recast as an energy minimization one. The key ingredient of the AQC is that 
$\Ham$ is not constant, rather it varies with time: $\Ham\rightarrow \Ham(t)$. The AQC derives its name from the Adiabatic Theorem~\citep{messiah1962quantum} which asserts how to follow the evolution of a system when $\Ham(t)$ \h{varies slowly enough} over time.\footnote{\h{for the formal definition and proof of the Adiabatic Theorem, we refer the reader to~\autoref{appendix_adiabatic_theorem} in the Appendix.}}$^{,}$ \footnote{\h{the Adiabatic Theorem states that if a system is initially in the $k^{th}$ state of well-defined energy, it will stay in this state when the Hamiltonian is changed sufficiently slowly~\citep{binney2013physics}. However, we focus our attention on the ground state because, in adiabatic quantum computing, a problem is typically encoded as a Hamiltonian whose ground state is the problem solution.  }}~\citep{born1928beweis,farhi2001quantum,aharonov2004adiabatic}. 

\begin{figure}[tbph]
    \centering
    \includegraphics[width=0.6\linewidth]{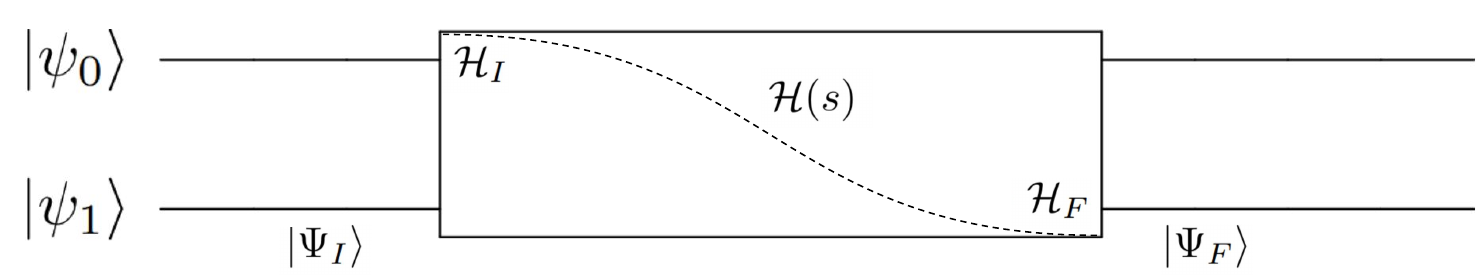}
\caption{Schematic adiabatic evolution for the Deutsch's algorithm~\citep{deutsch1992rapid}.}
\label{fig:adiabatic_circuit}
\end{figure}

We now briefly report \h{what a ``slowly enough varying'' Hamiltonian means}. As previously mentioned, AQC exploits the continuous time-evolution of \h{a system of qubits} within a time interval $[0, T]$, where $T$ represents the end of the adiabatic evolution. 
Commonly, the varying Hamiltonian is expressed as a one-parameter function, given by $\Ham(s)$ where $s=t/T \in [0, 1]$. Note that the eigenvalues of $\Ham(s)$, indicated as $\xi_i(s)$, represent different energy levels of the system ordered with increasing values of $i\geq 0$ (e.g. the ground state energy is $\xi_{0}(s)$). 
To ease the understanding of the \h{following concepts,} we indicate the eigenstate for $\Ham(s)$ related to the energy eigenvalue $\xi_i(s)$ as $\ket{\phi{_i}(s)}$, where we stressed the time dependence through the parameter $s$. Therefore\h{,} the instantaneous eigenstates for $\Ham(s)$ are the states satisfying $\Ham(s)\ket{\phi_i(s)} = \xi_i(s)\ket{\phi_i(s)}$. For example, the ground state of $\Ham(s)$, $\ket{\phi_0(s)}$, is characterized by $\Ham(s)\ket{\phi_0(s)}=\xi_{0}(s)\ket{\phi_0(s)}$.
Given such definitions and restricting our interest to the ground state of $\Ham(s)$, the adiabatic theorem ensures that if it is guaranteed that the difference among the ground and the first excited state energies, $\Delta=\underset{0\leq s\leq1}{\mathrm{min}}\big(\xi_1(s) - \xi_0(s)\big)$, remains large enough throughout the entire adiabatic evolution, then the system is likely to lie in the ground state of the instantaneous Hamiltonian. Specifically, such a requirement directly \h{affects} the length of the adiabatic evolution since $T\propto \Delta^{-2}$. 
As mentioned above, $\Ham(s)$ is interpolated between an initial ($\Ham_I$) and a final ($\Ham_P$), also called the problem, Hamiltonians. Specifically, the \quotes{problem} operator encodes the solution to a given problem in such a way that its ground state configuration,
$\ket{\phi_0(s=1)}$, expresses the solution to the optimization problem at hand. Concerning the AQC, the strategy is first to prepare the qubits in the ground state of $\Ham_I$, and then the operator is evolved towards $\Ham_P$ by following the so-called adiabatic path $\Ham(s) = (1-s) \Ham_I + s \Ham_P$. From a mathematical perspective, $\Ham_I$ and $\Ham_P$ can be defined in terms of the Pauli-matrices. 
Concerning a system made by $n$-qubits, the initial Hamiltonian can be defined as: $\hat{H}_I = \sum_{i=0}^{n-1} \frac{1}{2}(1 - \paul{x}^i)$
where $\paul{x}$ is the Pauli-$x$ operator. The ground state of $\hat{H}_I$ is characterized by all qubits aligned with the $x$-axis, i.e.,
the state $\ket{+}^{\otimes n} = \frac{1}{2^{n/2}}\sum_{z_1}\sum_{z_2} \cdots \sum_{z_n}\ket{z_i}^{\otimes n}$, where $\ket{z_i}\in\{\ket{0}, \ket{1}\}$, which resembles the uniform superposition of \h{the} computational basis states.  
To obtain such a state, one couples a magnetic field, called the transverse or disordering field, in the $x$-direction to each quantum bit. The choice of such a shape for $\hat{H}_I$ relates to specific symmetries that the operator must satisfy to allow for the sought computations. For example, such a choice is mandatory to avoid the so-called \quotes{level crossing}~\citep{farhi2000quantum}. 
Moreover, $\hat{H}_P$ is typically expressed in terms of the $\paul{z}$ operator so that the final configuration of the qubit system is represented by eigenvectors of $\paul{z}$, i.e., each qubit can be found either in the $\ket{0}$ or $\ket{1}$ state.
Such a configuration is nothing more than a classical sequence of bits. Thus, by starting from an initial quantum superposition, the adiabatic process evolves the state into a classical configuration which can then be read to return the solution to the problem at hand. One of the first \quotes{proof of concenpt} to show the effectiveness of such a computation model was first realized in~\cite{farhi2001quantum} for solving instances of the satisfiability problem. Apart from such a first example, many optimization problems can be recast as an energy minimization problem, such as the Grover search algorithm~\citep{grover1997quantum,farhi2000quantum,farhi9612026analog}. We report in~\autoref{fig:adiabatic_circuit} a schematic representation of an adiabatic path representing the Deutsch's algorithm~\citep{deutsch1992rapid}. The reader might compare such an image with~\autoref{fig:gate_circuit} to notice the difference between the two designs.

 \section{The Quest for Quantum Supremacy} \label{q_supremacy}

The first claim about an attainable advantage with a hypothetic quantum computer upon a classical one is attributed to Feynmann~\citep{feynman1982simulating}. However, at that time, such an idea concerned quantum simulations of physical systems rather than universal computations. Indeed, considering the simulation of a dynamical system characterized by $N=2^d$ degree of freedoms, one can immediately see that a classical simulation living in a $2^d$-dimensional Hilbert space requires resources that grow exponentially with \h{the size of the problem}, while only polynomially if using a quantum device~\citep{lloyd1996universal}.

However, soon after, researchers started to think about gaining an advantage over generic classical computations by leveraging quantum effects. 
Thus, the concepts of \quotes{quantum advantage} (or \quotes{quantum speedup}) and \quotes{quantum supremacy}~\citep{preskill2012quantum,harrow2017quantum} appeared in the literature. \h{Despite the fact that} both terms refer to the same principle, loosely speaking, \h{one can distinguish them considering what follows}. The first one typically refers to the ability of a quantum computer to perform a given computation faster than a classical one. \h{Differently, the second one assesses the ability of a quantum computer to} find solutions to a problem not resolvable by a classical computer (or at least not in a reasonable amount of time). Thus, even though both terms express the same principle, sometimes they are used with slightly different purposes.

One of the most important phenomena for quantum computations is the entanglement, which allows propagating the action on a qubit to others and compressing the resource requirements to describe a given state. Highly entangled states cannot be simulated efficiently by classical systems, \h{therefore proving the advantage} of quantum devices over classical ones.
Interestingly, assuming the existence of computational tasks beyond any classical computer's capability that can be solved with a universal quantum computer, then in that case, it could be possible to refute the \quotes{extended Church-Turing thesis}.
As we mentioned previously, concerning an algorithmic point of view, the size of the Hilbert space grows exponentially for a classical algorithm while only polynomially in the quantum case. Moreover, quantum phenomena such as entanglement, superposition and tunnelling, help to navigate such a vast space and should allow for a quantum speedup~\citep{lloyd1996universal,abrams1999quantum,harrow2009quantum}. Concerning the GM, Deutsch~\citep{deutsch1985quantum} was among the first ones to show that a quantum circuit can be built of reversible quantum logic gates to compute any classical function $f$ defined on $n$ bits. In 1992, the first example of quantum advantage in computation was formalized in~\cite{deutsch1992rapid} in which the authors showed that for a \h{specific} 
class of problems, a quantum device required exponentially less time to solve them \h{than} any deterministic classical approach.

When talking about quantum advantage or speedup, two algorithms are typically cited: Grover's search~\citep{grover1997quantum}
and Shor's factoring~\citep{shor1994algorithms} algorithms.
However, there is a fundamental difference in \h{how} they obtain the advantage. Indeed, while for the first one the \quotes{modest} quadratic speedup over the classical formulation is mathematically demonstrated, concerning the second one this is not the case. Specifically, all that can be said about Shor's algorithm is that it reaches an exponential speedup over the most efficient classical analogous available today. However, nothing forbids that a new classical formulation might reduce or eliminate such a gap.

Generally speaking, there are several definitions of quantum speedup. \quotes{Provable quantum speedup} stems from the existence of a mathematical proof that there is not any classical algorithm that can perform better than the quantum one~\citep{grover1997quantum,bennett1997strengths}.
A different concept is expressed by the \quotes{strong quantum speedup}~\citep{papageorgiou2013measures} which considers the performance of the best classical algorithm. Shor's algorithm~\citep{shor1994algorithms} is an instance of such a definition. Indeed, although classical algorithms require super-polynomial cost in the number of digits, the proof of an exponential lower bound for classical factorization has not been found yet. Thus, one typically adopts the concept of quantum speedup by referring to \h{the comparison} among the best known classical algorithm, which might not \h{be} the best possible one, and its quantum counterpart. 

Another class of problems probed to prove and harness a quantum advantage is the sampling problems' class~\citep{harrow2017quantum}, such as: constant-depth circuits~\citep{terhal2002adaptive}, boson sampling~\citep{aaronson2011computational}, and random quantum circuits containing commuting and non-commuting gates~\citep{shepherd2009temporally,bremner2011classical,boixo2018characterizing}. Such examples are somehow in between the factoring algorithm~\citep{shor1994algorithms} and analog quantum simulators~\citep{cirac2012goals,georgescu2014quantum,cheuk2016observation}. 

In~\citep{ronnow2014defining}, the authors performed experiments on a \quotes{D-Wave Two} quantum annealer. As a benchmark, they tasked the QPU with finding the ground state of a 2D planar graph of an Ising spin glass model, which is known to be an NP-hard problem~\citep{barahona1982computational}. The authors compared the results from simulated annealing~\citep{kirkpatrick1983optimization}, simulated quantum annealing~\citep{santoro2002theory,martovnak2002quantum}, and the actual quantum device from D-Wave~\citep{harris2010experimental,johnson2010scalable,berkley2010scalable,johnson2011quantum}. However, they did not observe any evidence for a genuine quantum speedup \h{from the experiments}. Recently, 
\h{Google} claimed to have reached quantum supremacy~\citep{arute2019quantum}. In their work, the authors proposed an ad-hoc experiment involving a QPU containing 53 qubits and a circuit depth of 20 and claimed that a classical computer would have required thousands of years to simulate the obtained results. However, several research groups immediately replied to such a claim by showing \h{that} it was possible to simulate such a quantum computer in just slightly more than two days~\cite{pednault2019leveraging} or even less~\cite{zhou2020limits}. Such \h{profoundly} different results \h{witness} the difficulty that researchers typically face when trying to estimate the real power of 
Noisy Intermediate-Scale Quantum (NISQ) devices.

We conclude this brief overview over the concept of quantum supremacy and the various attempts to harness it by observing that, despite the tremendous efforts that scientists are devoting to embracing the quantum phenomena, such a goal is still far from being reached.
Moreover, it is clear that any claim about quantum supremacy must undertake a detailed analysis based on the most recent achievements in classical simulation algorithms~\citep{bremner2017achieving,rahimi2016sufficient,kalai2014gaussian}.

 \section{Quantum Learning} \label{quantum_neural_networks}

The research on Quantum Neural Networks is a quest started more than twenty years ago~\citep{altaisky2001quantum,faber2002quantum,schuld2014quest,schuld2015simulating}.
Studies from Behrman et al.~\cite{behrman1996quantum} and Toth et al.~\cite{toth1996quantum} are examples of seminal works that introduced the concept of QNNs. Behrman et al.~\cite{behrman1996quantum}\h{,} described a mathematical model based on quantum dot molecules and showed by simulations that such a model was able to realize any classical logic gate. Instead, Toth et al.~\cite{toth1996quantum}\h{,} proposed a more biologically inspired architecture, where quantum dots were coupled to form a cellular structure in which near-neighbor connectivity allowed the information to flow. The realization of a quantum \h{analogue} of an Artificial Neural Network (ANN)\h{,} as an algorithm able to combine features from both the classical and quantum worlds\h{,} is a long-standing dream for the scientific community~\citep{schuld2014quest}. 
In the classical realm, a feedforward ANN is a universal approximator of continuous functions~\citep{hornik1989multilayer}. \h{Thus,} a QNN should at least satisfy such a requirement. 

Classically, an ANN \h{is made by} 
a sequence of layers. \h{Each of them} applies a specific mathematical transformation \h{to} its input and produces an output taken as input by the subsequent layer in the network. \h{Moreover}, the various layers are \h{typically} interleaved by non-linear functions \h{to enhance the ANN representation power.} Indeed, if only linear operations are considered, one could reduce the entire network to a single affine mapping. ANNs owe their name to their structure loosely inspired by the \h{human} brain. Moreover, still based on \h{a} biological \h{analogy}, the basic building block of ANNs is the artificial neuron, called \emph{perceptron}, a computational model \h{proposed initially} by Rosenblatt~\citep{rosenblatt1957perceptron}. These algorithms typically deal with large amounts of data aiming at finding patterns to describe them by surfing the parameters' space. Thus, they are perfect candidates to leverage the advantage of quantum phenomena that allows \h{navigating} more efficiently high \h{dimensional spaces.} 

\h{It is worth mentioning that classical deep ANNs have been exploited to solve quantum problems in disciplines such as chemistry~\citep{pfau2020ab,schutt2019unifying,li2018density} and physics~\citep{nomura2021helping,chen2018equivalence,cai2018approximating} among others. In such contexts, a DL model can be trained, for example, to predict the quantum mechanical wavefunction for a given many-body system~\citep{carleo2017solving}, to predict the interatomic potential energy surfaces~\citep{smith2017ani}, to identify distinct phases of matter and the transitions between them~\citep{carrasquilla2017machine}, to mention some.
In spite of this topic being fascinating, a detailed discussion about these techniques is out of scope for our work. Thus, in what follows, we will focus on approaches based on quantum models only.}

\h{Concerning} the quantum realm, in 1994, Lewenstein~\cite{lewenstein1994quantum} was among the first ones to propose a quantum-mechanical perceptron as an ideal basic element that processes input quantum states through unitary transformations. Since then, several studies have been conducted \h{to formulate} QNNs as made up of basic cells such as QPs or as \h{entirely} new computing architectures.
Indeed, the design of QNNs can be much different from that of classical ANNs meaning that the first ones do not require to be made of several layers of QPs. For example, they can be realized as variational circuits (see~\autoref{sec:qnn}) without relying on direct implementation of a basic cell resembling the quantum analogue of a perceptron.

\h{
In the following sections, we first introduce the concept of {quantum embedding} (\autoref{sec:qemb}) and then we describe variational algoritmhs  (\autoref{sec:qo}). Subsequently, we report several designs for QPs (\autoref{sec:qp}) and QNNs (\autoref{sec:qnn}). Finally, we discuss trainability issues, e.g., barren plateaus (\autoref{sec:barren}).
We focus primarily on the GM computational paradigm since it is currently the most widely adopted approach in such a context, and it offers a closer analogy to classical approaches, especially from a computer scientist's point of view. We also report some interesting formulations based on AQC. We leave a more detailed analysis of the results obtained with the AQC paradigm for future work.
}

\subsection{\h{Encodings of data}}\label{sec:qemb}

\h{A quantum machine learning algorithm\footnote{Following the categorization used in \cite{aimeur2006machine,schuld2018supervised} we use the term quantum machine learning for approaches that use a quantum device to process classical or quantum data.}, such as a QNN,} accepts quantum states as input by its nature. Hence, classical data must be translated into quantum states and transferred from a classical memory to the quantum device, a process called \quotes{state preparation}. 

\h{
The \textit{basis} and the \textit{quantum amplitudes encodings}~\citep{schuld2018supervised} are two fundamental examples of quantum data embeddings.
The basis encoding (also called bit encoding) associates each data input with a computational basis state of a qubits system. This is equivalent to saying that each data item $\bb x \in\mathbb{R}^\h{d}$ is first transformed to a binary string $\bb b$ of length $n$ and that each binary string is uniquely associated with a computational basis state of an $n$-qubit system, $\ket{\bb b}$. 
For example, given a vector $\bb x=[0.3, -0.4]$ we may use a bit to encode the sign and $T$ bits to encode each number in the interval $[0, 1)$ according to a binary fraction representation $\sum_{k=1}^T b_k \dfrac{1}{2^k}$ of precision $T$. In the case of $T=4$ we obtain $\bb 0.3 \to 0\, 0100 $ and $-0.4 \to  1\, 0110$, therefore the vector $\bb x$ is mapped to the binary string $\bb b=0010010110$ that is represented by the quantum state  $\ket{0010010110}$ of 10 qubits.
}
Amplitudes encoding\h{, instead,} associates \h{real vector components} with \h{quantum amplitudes} with the only caveat that \h{the original real vectors} must be normalized.
Formally, given a 
vector $x\in\mathbb{R}^\h{d}$, where $\h{d=2^n}$, we have the following mapping:
\begin{equation}\label{eq:ampl_enc}
    \bb x = \begin{bmatrix}
           \,\, x_0 \,\, \\ 
           \,\, x_1 \,\, \\
           \,\, \vdots \,\, \\
           \,\, x_{\h{d}-1} \,\,
            \end{bmatrix} \ \ \ \ \rightarrow \ \ \ \ \ket{x} = \dfrac{1}{\| \bb x\|} \sum_{i=0}^{\h{d}-1} x_i \ket{i}
\end{equation}
where $\ket{i}$ represents the $i$-th computational \h{basis state for an {$n$}-qubit system.} 
\h{For example, if $\bb x=[0.3, -0.4]$ then $\ket{x}=\left(0.3\,\ket{0}-0.4\,\ket{1}\right)/0.5$, upon normalization.}
\h{Note that the amplitude encoding requires exponentially fewer qubits than the base encoding to embed a vector into a quantum state.}
\h{An entire dataset can be represented in the computational basis by considering the amplitude encoding of the concatenation of all the input data.}
Hence, the main advantage of this encoding is that it only requires $\mathrm{log}\h{_2}M\h{d}$ qubits to encode a dataset of size $M$ and dimensionality \h{$d$}~\cite{schuld2018supervised}.

\subsection{Variational Algorithms}\label{sec:qo}

Currently, the leading approach to the optimization of quantum circuits~\citep{benedetti2019parameterized} exploits Variational Quantum Algorithms (VQA)~\citep{peruzzo2014variational,mcclean2016theory,guerreschi2017practical,cerezo2020variational}, hybrid quantum-classical algorithms that leverage both quantum as well as classical components. \h{Their name is} inspired by the well known result from physics called the \textit{Variational Principle}, which provides an upper bound of the ground state's energy of a quantum system\footnote{\h{we refer the interested reader to~\autoref{app:variational_principle} in the Appendix where we report the formal definition and proof of the Variational Principle. }}. Generally speaking, the Variational Principle gives a recipe to construct a {trial state} (i.e., a quantum state parameterized by learnable parameters) for optimization purposes. 

\begin{figure}[tbph]
    \centering
    \includegraphics[width=\linewidth]{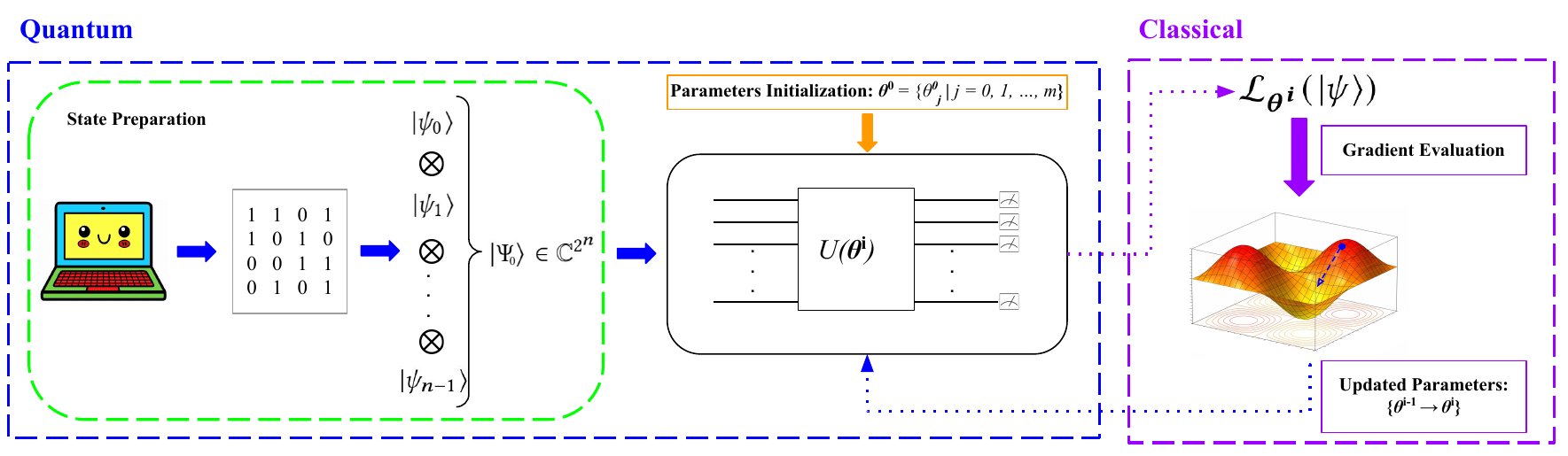}
\caption{Schematic view of a VQA. \h{The quantum device implements a parametrized quantum circuit $U(\boldsymbol{\theta})$ that given an initial state $\ket{\Psi_0}$ prepares the trial state $\ket{\Psi(\boldsymbol{\theta})}=U(\boldsymbol{\theta}) \ket{\Psi_0}$. Observable measurements on the trial state are used to return estimates of expectation values (e.g., the state of a certain qubit), which 
are used to evaluate an objective function \h{$\mathcal{L}_{\boldsymbol{\theta}}$}. A classical optimization algorithm (e.g., gradient descent) is employed to update the circuit parameters. }}
\label{fig:vqa}
\end{figure}

From a computer science point of view, one can formulate a VQA as follows. A hybrid quantum-classical architecture comprising a set of quantum operations, the \textit{ansatz}, that are applied to the initial qubits state \h{to} produce a final state, called \textit{trial state}. The ansatz is characterized by a set of gates controlled by some \h{classical} parameters \h{$\boldsymbol{\theta}$} 
whose \h{values are} modified with the help of a classical procedure, e.g., gradient descent. 
Therefore, the design of such algorithms typically relies on three main components. 
The first one is the ansatz, \h{also referred to as "parametrized quantum circuit" or "variational circuit"}, which sometimes includes the embedding of classical data into a quantum state. Subsequently, one must specify an objective function \h{$\mathcal{L}_{\boldsymbol{\theta}}$}, whose value is used to drive the optimization procedure, and finally a classical optimization procedure \h{is} used to modify the ansatz's parameters \h{$\boldsymbol{\theta}$} that define the transformations applied on qubits. Hence, it is clear \h{that VQAs represent} the practical embodiment of the idea \h{of training} quantum computers as we train \h{classical} neural networks. We report in~\autoref{fig:vqa} a schematic view of a VQA.

The first application of such \h{class of methods} 
is the \h{\textit{Variational Quantum Eigensolver} (VQE)}~\citep{peruzzo2014variational,kandala2017hardware,wang2019accelerated}\h{, initially proposed to find the lowest energy state (the ground state) of a quantum system. Given the Hamiltonian $\hat{H}$ of a system, the Variational Principle tells us that the ground state $\ket{\psi^*}$ corresponds to the normalized state $\ket{\psi}$ that minimizes the \h{quantity} $\langle \psi |\hat{H}|\psi\rangle$ expressing the energy of the system in the state $\ket{\psi}$. That is $\ket{\psi^*}= \underset{\ket{\psi}}{\text{argmin}}\langle \psi |\hat{H}|\psi\rangle$, and $E_{min}=\langle \psi^* |\hat{H}|\psi^*\rangle$. Therefore, the ground state can be approximated by optimizing the parameters $\boldsymbol{\theta}$ of an ansatz $U(\boldsymbol{\theta})$ in order to minimize the expectation value  $\mathcal{L}_{\boldsymbol{\theta}}=\langle \psi(\boldsymbol{\theta}) |\hat{H}|\psi(\boldsymbol{\theta})\rangle$, where $\ket{\psi(\boldsymbol{\theta})}=U(\boldsymbol{\theta}) \ket{\Psi_0}$ for a given initial state  $\ket{\Psi_0}$. In other words, given $\boldsymbol{\theta}^*=\underset{\boldsymbol{\theta}}{\text{argmin}}\, \mathcal{L}_{\boldsymbol{\theta}}$, the state $\ket{\psi(\boldsymbol{\theta}^*)}$ approximates  $\ket{\psi^*}$ and $\mathcal{L}_{\boldsymbol{\theta}^*}$ provides an upper bound for $E_{min}$. VQE can solve general optimization problems as long as the problem is formulated using an Hamiltonian such that its ground state corresponds to the solution of the original problem. For example, to find the binary solution $x$ that minimizes a cost function $C(x)$ one may first find an Hamiltonian $H_C$ that encodes the cost function $C(x)$, i.e. $ H_C\ket{x}=C(x)\ket{x}$, and then use the VQE to approximate the ground state of this Hamiltonian. 
}

The \h{\textit{Quantum Approximate Optimization Algorithm}} (QAOA)~\citep{farhi2014quantum} is a famous 
\h{variational algorithm} that defines an ansatz to generate approximate solutions to a given \h{problem}. It was \h{initially} proposed to solve combinatorial optimization \h{problems}, and then it was generalized as a standard ansatz~\citep{hadfield2019quantum,cerezo2020variational}. QAOA finds the trial state by repeatedly applying unitary evolutions according to a two-terms Hamiltonian~\citep{farhi2014quantum}, namely, $\Ham_C$ and $\Ham_B$. The first one encodes the classical cost function \h{$C(x)$} to be 
\h{minimized}, and it is typically diagonal in the computational basis. \h{In contrast,} the second one is the mixing Hamiltonian which coherently moves the system through different configurations in the Hilbert space seeking the ground state of $\Ham_C$. Typically, $\Ham_B$ is a global transverse field, i.e., $\Ham_B = -\sum_i\paull{x}{i}$. 
Thus, given the problem Hamiltonian $\Ham_C$, one can use QAOA to find its ground state by applying the evolution operator $\hat{U}$:
\begin{equation}\label{eq:qaoa_ansatz}
    \ket{\psi(\boldsymbol{\beta}, \boldsymbol{\gamma})} =
    \hat{U}(\boldsymbol{\beta}, \boldsymbol{\gamma}) \ket{\psi_0} =
    \prod_{i=0}^p  e^{-i\beta_i \Ham_B}e^{-i\gamma_i \Ham_C} \ket{\psi_0}=
    e^{-i\beta_p\Ham_B}e^{-i\gamma_p\Ham_C} \cdots e^{-i\beta_1\Ham_B}e^{-i\gamma_1\Ham_C}\h{\ket{\psi_0}}
\end{equation}
where the set of angles $\{\boldsymbol{\beta}, \boldsymbol{\gamma}\}$ are the parameters to be optimized. Starting from 
\h{an} initial state \h{$\ket{\psi_0}$}, e.g., the uniform superposition of the basis states \h{$H^{\otimes n} \ket{0}$}, one can evolve it by using~\autoref{eq:qaoa_ansatz} and obtain $\ket{\psi(\boldsymbol{\beta}, \boldsymbol{\gamma})}$. The parameters are then adjusted such as to minimize the objective function expressed as:
\begin{equation}\label{eq:qaoa_obj}
    \h{\mathcal{L}}(\boldsymbol{\beta}, \boldsymbol{\gamma}) = 
    \bra{\psi(\boldsymbol{\beta}, \boldsymbol{\gamma})}\Ham_C\ket{\psi(\boldsymbol{\beta}, \boldsymbol{\gamma})} =
    \bra{\psi_0}\hat{U}^\dagger(\boldsymbol{\beta}, \boldsymbol{\gamma})\Ham_C \hat{U}(\boldsymbol{\beta}, \boldsymbol{\gamma})\ket{\psi_0}
\end{equation}
\h{The parametrized evolution operator $\hat{U} (\boldsymbol{\beta}, \boldsymbol{\gamma})$ can be viewed as a layerized variational circuit: 
\begin{equation} 
{\Qcircuit @C=1em @R=.7em {\ket{\psi_0} & &\gate{U_C(\gamma_1)}  & \gate{U_B(\beta_1)} & \qw &\cdots &  &\gate{U_C(\gamma_p)}  & \gate{U_B(\beta_p)} & }}
\end{equation}
with $p$ repetitions of alternating cost layers, $U_C(\gamma_i)=e^{-i\gamma_i \Ham_C}$, and mixing layers, $U_B(\beta_i)=e^{-i\beta_i \Ham_B}$. Note that alternating the cost layers with the mixer layers allows enlarging the space of quantum states that can be generated with the circuit (since a circuit consisting of only cost layers is equivalent to a circuit with only one parameterized cost layer). The layers of QAOA correspond to the trotterized segments of the time evolution operator associated to the adiabatic evolution $\hat{H}(s)=(1-s)H_B + s H_C$, $s\in [0,1]$. The initial QAOA state $\ket{\psi_0}$ is set as the ground state of the initial Hamiltonian $H_B$ (e.g., $\ket{\psi_0}=H^{\otimes n} \ket{0}$ if $\Ham_B = -\sum_i\paull{x}{i}$) so that the adiabatic process evolves to the ground state of the problem Hamiltonian $H_C$.
} 

A fundamental ingredient for QAOA is the parameter $p$ in~\autoref{eq:qaoa_ansatz} since it regulates the precision of the approximate solution. Growing $p$ means reducing the \quotes{distance} from the optimal solution. \h{However, for large $p$, decoherence starts} to play a significant role in the computation since the ansatz will contain more gates. We recommend the reader the original paper~\citep{farhi2014quantum} for a \h{complete} discussion about the impact of the $p$ parameter and QAOA in general. 

In the \h{following} two sections, we move to the topics of QPs and QNNs. \h{Unlike} classical learning approaches, quantum \h{ansatzes} are typically specialized to build perceptrons or represent an entire network. From such an observation, we first delve deep into the QPs and then to full QNNs proposals. 

\subsection{Quantum Perceptron} \label{sec:qp}
\h{From the classical perspective, a perceptron is a computational model characterized by a non-linear response to its input and is parametrized as:
\begin{equation}\label{eq:cls_perceptron}
    s_j = 
    f \left( \bb{w}_j^T\bb{x} +  b_j 
    \right)
\end{equation}
where $\bb x\in \R{d}$ is the input, $\bb w_j \in \R{d}$ and $b_j\in \R{}$ are learnable parameters (weights and bias), and $f$ is a non-linear activation function. Depending on the value of $s_j$, the perceptron is said to be either active or at rest. 
One of the simplest forms of an artificial neural network is the Multilayer Perceptron (MLP), i.e., a network composed of multiple layers of perceptrons (called inner-product or fully-connected layers). Each layer is composed of many perceptrons whose action is a linear projection of the input followed by a non-linear element-wise activation function, i.e., 
$\bb{ s}= f \left( \bb{W}^T\bb{x} +  \bb{b} \right) \in \R{m}$, where $\bb{W} \in \R{d\times m}$ and $\bb{b} \in \R{m}$ are the weight matrix and the bias vector, respectively. The fully-connected layer is one of the basic building blocks for Deep Neural Networks, so the perceptron can be considered a "fundamental unit" of classical NNs. Therefore, the research in the field of QNN initially focused on defining quantum algorithms capable of reproducing the functionalities of classical perceptrons, e.g., for classification tasks.
}

The implementation of the Quantum Perceptron is still \h{an active research topic}. 
We know that classically there is one reference implementation of such objects (\autoref{eq:cls_perceptron}). However, due to the intrinsic differences among the \h{available} quantum hardware platforms, different formulations report advantages and disadvantages depending on the context in which they are used, thus allowing for distinct implementations equally valid in the respective fields of application. To ease the reader's understanding of the current state-of-the-art QPs, we report a summary in~\autoref{tab:q_perceptron_summary}. It should be noted that some approaches have been implemented on real QPUs (e.g., IBM Q 5), while other methods have only been tested using numerical simulations so far. 

\renewcommand{\arraystretch}{1.2}
\begin{table}[tbph]

    \centering
      \caption{Summary comparison among various quantum perceptron proposals. The symbol \quotes{nr} stands for \quotes{not reported} meaning that the authors did not explicit reported that specific information. In the column \quotes{\h{Impl.}} we reported whether the method was implemented on an actual QPU or tested using numerical simulations only (reporting in brackets the library used when the information was available).}
    \label{tab:q_perceptron_summary}
    \small
     \begin{threeparttable}
    \begin{tabular}{
    >{\raggedright\arraybackslash}p{0.175\columnwidth}
    >{\centering\arraybackslash}p{0.6\columnwidth}
  >{\centering\arraybackslash}p{0.085\columnwidth}
  >{\centering\arraybackslash}p{0.035\columnwidth}
    }
    \toprule
    \b Work & \b Formulation & \b Impl.  &  \b  Code  \\ \midrule
        \multicolumn{4}{c}{\textit{Using the Gate Model}}\\         \midrule
        \h{Schuld et al. \cite{schuld2015simulating}} & \h{Quantum circuit using inverse quantum Fourier transform to reproduce a step activation function. The 
        input is encoded into an $n$-qubit state} & \h{Simulation}  & \h{\xmark} \\ 
        
        Cao et al. \cite{cao2017quantum} 
          & 
        Circuit based on RUS (using a tangent-based non-linear function). The \h{$d$-dimensional} input data \h{is} encoded into an \h{$d$}-qubit state & {Simulation} 
          & \xmark \\ 
          
       Hu \cite{hu2018towards} 
          & 
        Circuit based on RUS (using a sigmoid-based non-linear function). 
          The \h{$d$-dimensional} input data \h{is} encoded into an \h{$d$}-qubit state. & IBM Q 5 
          & \xmark \\ 
          
        Du et al. \cite{du2018implementable}
        & Variational quantum circuit based on a parametrized Grover oracle learned with a classical optimization approach. & {Simulation {\footnotesize (pyQuil)}} & \xmark \\ 
       
       Wiebe et al. \cite{wiebe2018quantum} 
       & Quantum search problem, based on the version space interpretation of the perceptron, based on the Grover's algorithm. & nr & \xmark \\ 
       
        Liu et al. \cite{liu2019unitary}
          & Given a supervised dataset, construct the weights' matrix as the tensor product of input and target and then apply SVD to it. & nr & \xmark \\ 
          
        Wiersema et al. \cite{wiersema2019implementing}
          & Qubit represented as a density matrix that describes the classifier as a Boltzmann-like finite-temperature distribution. & {Simulation  {\footnotesize(Tensorflow)}} & \cmark $\,^*$\\ 
          
     Tacchino et al. \cite{tacchino2019artificial}   
          & Quantum variational circuit in which the $\h{d}$-dimensional input datum is encoded into the $\h{d}$ coefficients of an \h{$n$}-qubit state \h{($d=2^n)$}. 
          & {IBM Q 5 \quotes{Tenerife}} & \cmark $\,^\circ$ \\
         
         \midrule
         \multicolumn{4}{c}{ \textit{Using the Adiabatic Model}}\\         \midrule
    
      Soloviev et al. \cite{soloviev2018adiabatic}  
       &   Perceptrons and synapses are implemented as JJ-based superconducting circuits. & nr & \xmark \\
         
       Torrontegui et al. \cite{torrontegui2019unitary} 
          & Two-level system (qubit) with a non-linear excitation response to the input field. & {Simulation {\footnotesize(Tensorflow)}} & 
       \cmark $\,^\bullet$\\
        
    \bottomrule
 \end{tabular}
    \begin{tablenotes}\footnotesize
    \item[] \textbf{Code:} \xmark $\,$ the code is not publicly available, \cmark$\,$ the code is available either on GitHub ($*$), upon request to the authors of the original work ($\circ$), or in the supplementary material of the original work ($\bullet$).
\end{tablenotes}
    \end{threeparttable}
\end{table}

Due to the linear, unitary, and non-dissipative nature of the quantum transformations, one of the hardest challenges to design a QP resides in \h{designing and implementing} non-linear activation functions for quantum computations. 
 \h{Schuld et al. \cite{schuld2015simulating} proposed one of the first designs for a quantum perceptron model imitating the step-activation function of a classical binary perceptron}. Their main idea was \h{to encode} a normalized version $\psi\in[0,1)$ of the input signal $\bb w^T \bb x$ into the phase of a quantum state of $n$ qubits and then use a phase estimation algorithm with a precision of $\tau$, implemented using \h{an} inverse quantum Fourier transform. They proved that measurements on the first qubit of the \h{output quantum state could have been used to estimate if $\psi$ was larger} than $1/2$, hence reproducing the behavior of a step activation function.


In 2017, Cao et al.~\cite{cao2017quantum} proposed an approach based on the \textit{Repeat-Until-Success} (RUS)~\citep{paetznick2014repeat,bocharov2015efficient} technique to synthesize quantum gates.
As we mentioned previously, the classical perceptron \h{evaluates the non-linear transformation ${s}= f(\bb{w}^T \bb{x} + b)$ that, once thresholded, defines the activation status of the perceptron.}
In such a context, \h{$\theta=\bb{w}^T \bb{x} + b$ is the input signal to the neuron and} $f$ represents the non-linear operation whose output $\h{s}$ lies in $[l, u]$, where typically $l=-1$ or $0$ and $u=1$. 
\h{In \cite{cao2017quantum}}, such an operation \h{is} mapped to the quantum realm by defining the perceptron as a qubit whose state is defined as: $\hat{R}_y(\h{s}\frac{\pi}{2}+\frac{\pi}{2})\ket{0} = \mathrm{cos}(\h{s}\frac{\pi}{4}+\frac{\pi}{4})\ket{0} + \mathrm{sin}(\h{s}\frac{\pi}{4}+\frac{\pi}{4})\ket{1}$, where $\h{s}\in [-1, 1]$ and $\hat{R}_y$ represents a rotation around the $y$-axis as described by $\paul{y}$. 
Note that as far as the extreme values of $\h{s}$ are concerned, they recover a classical behavior being the system in the state $\ket{0}$ or $\ket{1}$. Instead, in all other cases it behaves quantum-mechanically. \h{The quantity $s=f(\theta)=f(\bb{w}^T \bb{x} + b)$ is computed using a quantum circuit where the input data is encoded into a state vector $\ket{x}=\ket{x_1\cdots x_d}$} acting as the control register over \h{an} ancilla qubit. \h{Initially, a rotation $R_y(2w_i)$ conditioned on $\ket{x_i}$ followed by a rotation $R_y(2b)$ is applied on the ancilla qubit, that is equivalent to say that  $R_y(2\theta)$ is applied on the ancilla qubit conditioned on the state $\ket{x}$. Then\h{,} the non-linearity is realized by using the rotation $R_y(2f(\theta))$ which is approximated by a class of RUS~\citep{paetznick2014repeat} circuits each implementing a rotation with an angle equal to $\text{arctan}(\text{tan}^2(\theta))$. Repeating the RUS circuit $k$ times gives a rotation with an angle $\text{arctan}(\text{tan}^{2^k}(\theta))$ that is a sigmoid-like, tangent-based, non-linear function.   } 
Stemming from the same approach, Hu~\cite{hu2018towards} implemented a different RUS circuit to represent a sigmoid-based activation function, $f(x)=\mathrm{arcsin}(\sqrt{\mathrm{sigmoid}(x)})$, thus avoiding the drawbacks of using periodic functions, such as the tangent. 
Wiebe et al.~\cite{wiebe2018quantum} \h{adopted} 
a very interesting and completely different perspective. Indeed, the authors reported that the only way to unveil a quantum advantage is to formalize approaches that do not try to emulate classical algorithms in the quantum world. Instead, a new point of view is needed. Thus, they based their approach on the \emph{Version Space} interpretation of the perceptron and Grover's search algorithm~\citep{grover1997quantum}. 
Specifically, they exploited a general method referred to as amplitude amplification~\citep{brassard2002quantum} for which Grover's algorithm represents a particular case. Using the \emph{Version Space}, the authors posit the problem of training a perceptron as a search problem (i.e., search for the optimal parameters rather than \quotes{learn} them), hence fully exploiting the quantum advantage offered by the amplitude amplification technique. 
Wiersema et al.~\cite{wiersema2019implementing} exploited the formalism of density matrices to cast the perceptron learning problem in terms of a quantum cross-entropy:
\begin{equation}\label{eq:qce}
   \mathcal{L}_q = -\sum_\mathbf{x}q(\mathbf{x}) \mathrm{Tr}\{\eta_\mathbf{x}\ln\rho_\mathbf{x}\}
\end{equation}
where $q(\mathbf{x})$ is the empirical distribution of observing the datum $\mathbf{x} \h{\,\in \{1,-1\}^d}$, $\eta_\mathbf{x} \equiv \ket{\Psi}\bra{\Psi}$, with $\ket{\Psi}=\sqrt{q(y|\mathbf{x})}\ket{0} + \sqrt{q(-y|\mathbf{x})}\ket{1}$, is the density matrix associated with the distribution $q(y|\mathbf{x})$ of observing the label $y\in\{1, -1\}$ given $\bb{x}$, while $\rho_{\mathbf{x}} \equiv \rho(\bb{x}, \bb{w}; \bb{y})$ is the density matrix associated to the model \h{conditional} distribution $p(y|\mathbf{x};\mathbf{w})$. Note that such a formulation resembles the classical cross-entropy objective: $\mathcal{L}_{cl} = -\sum_\mathbf{x}q(\mathbf{x})\sum_yq(y|\mathbf{x})\ln p(y|\mathbf{x};\mathbf{w})$.
The matrix $\rho_{\mathbf{x}}$, which embodies the perceptron's description, represents a finite-temperature description of a qubit formulated as: 
\begin{equation}
    \rho_{\mathbf{x}} = \exp{(-\beta\sum_{\h{k\in\{x,y,z\}}} h_{\h{k}} \paul{k})}/Z,
\end{equation}
where \h{ the inverse temperature $\beta$ is set equal to $-1$, $\paul{k}$ are the Pauli matrices, the term $h_{{k}} \in \mathbb{R}$ is evaluated as ${\mathbf{w}_k^T}\mathbf{x}$ (i.e., for each Pauli matrix there is one set of weights), and $Z$ is such that $\mathrm{Tr}\{ \rho_{\mathbf{x}}\}=1$}. Thus, there is a direct dependence from the classical input data of the perceptron state. When the dataset $\{\mathbf{x} \}$ is linearly separable, the convergence point is similar to the classical case. Instead, in the presence of label noise, there is a contribution also from Pauli operators other than $\paul{z}$ that allow the \h{quantum} perceptron to perform better than its classical counterpart. 

Torrontegui et al.~\cite{torrontegui2019unitary} based their approach on the similarity between the classical perceptron and a single qubit interpreted as a two-levels system. Indeed, they implemented \h{a perceptron\ with sigmoid non-linear activation} as a qubit that undergoes a unitary transformation\h{, parametrized by an external input field $\hat{\theta}_j,$} given by
\begin{equation}
    \hat{U}_j(\hat{\h{\theta}}_j; f) = \exp{\left( \h{\ii}\,  \arcsin \left(f(\hat{\h{\theta}}_j)^{1/2}\right)\h{(\paul{y})_{j}} \right)}\ \ \ \ \mathrm{with}\ \  \ \ \hat{\h{\theta}}_j = \sum_{k<j} w_{jk}\h{(\paul{z})_{k}} - \h{b}_j
\end{equation}
where \h{$j$ is the label of $j$-th perceptron of a layer, $\hat{\theta}_j$ is the quantum field generated by neurons in earlier layers,} 
$\paul{y}$ and $\paul{z}$ are the Pauli matrices, and $f(\cdot)$ encodes the non-linear activation function.
The authors formalized the implementation of such a model using an adiabatic process. Specifically, they proposed implementing the \h{system's evolution} as a single adiabatic path for an Ising model of interacting spins. \h{The} authors proved that such a perceptron acts as a universal approximator with \h{such a definition.} Moreover, they report how it is possible to treat such objects as the basic building block for large neural networks. 

In Liu et al.~\cite{liu2019unitary}, the authors proposed a quantum algorithm based on unitary weights, suitable to implement a one-iteration learning approach. The weights' operator is constructed as the tensor product $\hat{w} = \sum_j\ket{y_j}\otimes\bra{x_j}$, where $\bra{x_j}$ represents the conjugate transpose matrix of a training input and $\ket{y_j}$ is the desired output. The Singular Value Decomposition procedure is then applied to $\hat{w}$ to decompose it as the product \h{$\hat{F} \Sigma \hat{w}_{\text{new}}$, where $\hat{F}$ and $\hat{w}_{\text{new}}$ are unitary matrices, and $\Sigma$ is the diagonal matrix of singular values. Since $\Sigma$ may not be unitary it is replaced with a diagonal unitary matrix and the output of the quantum perceptron on an input $\ket{x}$ is defined as $\hat{F} \Sigma_{\text{new}} \hat{w}_{\text{new}}\ket{x}$.} 
The authors showed how to approximate several quantum gates with such an algorithm \h{in their work.} A similar approach is also exploited by Khan et al.~\cite{khan2020derivative}.
Du et al.~\cite{du2018implementable} proposed a variational QP implementing a parametrized oracle for a Grover-like search algorithm. The parametrized quantum circuit \h{learns} features from the input dataset, and a classical optimization algorithm guides the parameters update.  
Differently from the previously introduced approaches, Soloviev et al.~\cite{soloviev2018adiabatic} reported a discussion about how to exploit superconducting circuits equipped with Josephson junctions \h{(JJ)}~\citep{josephson1962possible,josephson1974discovery} to physically realize perceptrons and synapses to build up a neural network. Specifically, the authors exploited the operational principles of adiabatic superconducting cells to implement multilayer perceptrons. The current-phase relation is used to implement the non-linear response \h{in such a design.} \h{Instead, the system's input is supplied through magnetic fluxes to which the circuits, operating at a few kelvins, couple. }

Finally, in 2019, Tacchino et al.~\cite{tacchino2019artificial} proposed a binary perceptron model \h{(where both input and weight vectors are limited to binary values)} that can be implemented on NISQ~\citep{preskill2018quantum} devices. To reduce the resource demand of the algorithm, the authors formalized a new procedure to generate multipartite entangled states based on the so-called hypergraph states~\citep{rossi2013quantum,ghio2017multipartite}.
\h{First, the input vector $\bb x \in \{-1,1\}^d$ and the weight vector $\bb w \in \{-1,1\}^d$ are encoded on the quantum states $\ket{\psi_\h{\bb x}}$ and $\ket{\psi_\h{\bb w}}$, respectively, using $n=\log_2 d$ qubits (see Eq. \autoref{eq:ampl_enc}). Then, given} the preparation state as \h{$\ket{0}^{\otimes n}$}, the authors evolved the state employing two unitaries, $\hat{U}_\h{\bb x}$ and $\hat{U}_{\bb w}$\h{, such that $\hat{U}_\h{\bb x} \ket{0}^{\otimes n}=\ket{\psi_{\bb x}}$ and $\hat{U}_\h{\bb w} \ket{\psi_{\bb w}}=\ket{1}^{\otimes n}$. The authors proved the final N-qubit state $\ket{\psi_{\bb x, \bb w}}=\hat{U}_\h{\bb w}\ket{\psi_{\bb x}}$, obtained by applying $\hat{U}_\h{\bb w}$ after $\hat{U}_\h{\bb x}$, contains the scalar product among  the input and the weight vectors, up to a normalization factor, in its last amplitude coefficient, i.e., $\inprod{1 \cdots 1}{\psi_{\bb x, \bb w}}=\inprod{\psi_{\bb w}}{\psi_\h{\bb x}}=\frac{1}{n}\bb w^T\bb x$. }
Finally, the non-linearity was implemented as a measurement whose output represented the probability for the system of being activated by the given input pattern. The authors successfully proved that for any given weight vector, $\bb w$, the perceptron was able to single out from the 16 possible input patterns only $\bb i=\bb w$ (with output probability of 1, i.e., the perfect activation of the neuron), while all other inputs gave outputs smaller than 0.25.

\subsection{Quantum Neural Networks}\label{sec:qnn}

Before delving deep into the most recent achievements in the field of QNNs, we briefly mention a few seminal works that, although 
published several years ago, played a relevant role \h{in advancing} the discipline at hand. One of the first proposals concerning a hybrid algorithm for QNNs, \h{dates} back to 2000 with the work of Narayanan and Menneer~\cite{narayanan2000quantum}. Using simulations, the authors compared the performance of a classical artificial network in which, from one experiment to the \h{following,} different components were substituted with their quantum analogous. Interestingly, the authors found that a fully quantum network did not report any tangible advantage compared to a \quotes{partially} quantum one. Ventura and Martinez~\cite{ventura2000quantum} proposed a quantum version of associative memories to harness the exponential increase in storage capacity, originated from superposition effects that quantum mechanics allows, concerning classical algorithms. To their aim, the authors exploited a generalization of Grover's algorithm~\cite{grover1997quantum} able to fulfill the search and completion tasks.  

\h{\textit{Parametrized Quantum Circuits} (PQCs), which include variational quantum algorithms, are} the most commonly adopted choice to design quantum neural networks. In such a context, the qubits' states are manipulated by unitaries characterized by a set of parameters that are classically learned at training time. Finally, the value of a given observable is accessed through a measurement that outputs a discrete quantity. Even if we focus on those algorithms in what follows, it is worth mentioning that another class of quantum algorithms exploits continuous variables, namely Continous-Variable (CV)~\citep{killoran2019continuous,bondesan2021hintons} models. These models do not use qubits to store quantum information. Instead, they exploit continuous degrees of freedom such as the amplitude of the electromagnetic field, thus making them suited for photonic hardware. 
\h{In what follows, we review the state-of-the-art QNN models. To ease the comparison among the analyzed approaches, we report in~\autoref{tab:q_nn_summary} a summary of the most relevant properties of the models. }

    	

 \afterpage{%
		\clearpage
\begin{landscape}
\centering 
\renewcommand{\arraystretch}{1.3}
 \begin{table}[!hp]
 \footnotesize
  \caption{\h{Comparison among the analyzed quantum neural networks designs. The symbol \quotes{\texttt{nr}} stands for \quotes{not reported} meaning that the authors did not explicit reported that specific information or that the specific attribute cannot be applied to the referred work.}}
    \label{tab:q_nn_summary}
    \begin{threeparttable}
    \begin{tabular}{
    >{\raggedright\arraybackslash}p{0.12\columnwidth} 
    >{\centering\arraybackslash}p{0.13\columnwidth} 
    c
     >{\centering\arraybackslash}p{0.08\columnwidth}
    >{\centering\arraybackslash}p{0.07\columnwidth}
    >{\raggedright\arraybackslash}p{0.02\columnwidth}
     >{\raggedright\arraybackslash}p{0.02\columnwidth}
    >{\centering\arraybackslash}p{0.04\columnwidth}
    >{\centering\arraybackslash}p{0.035\columnwidth}
     >{\centering\arraybackslash}p{0.08\columnwidth}
    }
    \toprule
    \multirow{ 2}{*}{\b Work}
    & \multirow{2}{*}{\b Aim}  & \multirow{ 2}{*}{\b Loss Function} & \multirow{2}{*}{\b \makecell{Variational \\ Parameters \\ ($n$-qubits)}} & \multirow{ 2}{*}{\b Optimization} &  \multicolumn{2}{c}{\b Impl.} & \multirow{2}{0.043\columnwidth}{\b {Appl. Field}} & \multirow{2}{0.035\columnwidth}{\b {Input Type}} &\multirow{2}{*}{\b  \makecell{Test\\ Datasets}}\\ 
    \cline{6-7}
    &&&&& \textbf{Sim.} & \textbf{QPU} &  & 
    &\\
    \midrule
         
        Perez et al. \cite{perez2020data}
        & {PQC as a Universal  Quantum  Classifier} & 
        $\sum_{i=1}^{D}\left(1-\left| \bra{\psi_i^{\text{true}}} \hat{\mathcal{U}}{(x_i; \Theta)}\ket{\psi_0} \right|^2 \right)$ 
        & \texttt{nr} & L-BFGS-B & \cmark & \xmark & CS & C 
        
        & \\ 
        
        Schuld et al. \cite{schuld2018circuit}
         & {PQC for Binary Classification} & 
       $\frac{1}{2}\sum_{i=1}^{D}\left|y_{\text{out}}(x_i; \Theta) - y_i\right|^2$
        & $\mathcal{O}((\mathrm{log}\ n)^k)$ & Gradient Descent & \cmark & \xmark & {CS} & C & {\scriptsize MNIST${\,}^*$, CANCER, SONAR, WINE, SEMEION}
        \\ 

        Macaluso et al. \cite{macaluso2020variational}
        & {Quantum MLP for Binary Classification} & 
        $\sum_{i=1}^{D}\left|y_{\text{out}}(x_i; \Theta)- y_i\right|^2$ 
        & 
        \texttt{nr} & Nesterov &  \cmark $\,^{\mathsection \,\ddagger}$ & \cmark $\,^{\bullet}$ & {CS} & C
        &\\

        Beer et al. \cite{beer2020training}
        & Quantum Dissipative Multilayer Perceptron 
        & $\frac{1}{D}\sum_{i=1}^{D}\bra{\psi_i^{\mathrm{true}}}\rho_i^{\mathrm{out}}\ket{\psi_i^{\mathrm{true}}}$ & $\mathcal{O\O}(2^{2N})$ & {Gradient Descent} & \cmark & \xmark & {QI} & Q 
        &
        \\ 
        
         Verdon et al. \cite{verdon2019quantum}
         & {PQC for Learning Hamiltonian Dynamics\tnote{$+$}} & 
        $1 - \frac{1}{B}\sum_{j=1}^{B}|\bra{\psi_j^{\text{true}}}U^j(\Theta)\ket{\psi_0}|^2{\,\,\,}^{\star}$  
         & \texttt{nr} & Adam & \cmark & \xmark & Phys. & Q &
         \\
        \midrule

        {Li et al. \cite{li2020quantum}} 
        & {Quantum CNN for image recognition}
        & {$-\sum_{k}^C (\bb{y}_i)_k \log\left(g(x_{i};\Theta)_k\right) {\,\,\,}^\dagger$}%
        & nr
        & Adam
        & \cmark
        & \xmark
        & {CS}
        & {C}
        & {\scriptsize MNIST, GTSRB${\,}^*$}
        \\  

        {Henderson et al. \cite{henderson2020quanvolutional}}
        & {Hybrid classical-quantum CNN for image recognition}  
        & \texttt{nr} 
        & \texttt{nr} & \texttt{nr} & \cmark  $\,^{\dagger \dagger}$ & \xmark & {CS} & C 
        & {\scriptsize MNIST }
        \\ 
        
        Cong et al. \cite{cong2019quantum}
        & {CNN-inspired QNN for quantum data analysis} & 
        $\frac{1}{D}\sum_{i=1}^{D} 
        \left|y_{\text{out}}(\ket{x_i}; \Theta) - y_i\right|^2$
        & $\mathcal{O}(\mathrm{log}\ n)$ & {Gradient Descent} & \cmark & \xmark & Phys. & Q &
        \\ 
        
        Grant et al. \cite{grant2018hierarchical}
        & {Hierarchical Quantum  Classifiers} & 
        $\frac{1}{D}\sum_{i=1}^{D}\left|y_{\text{out}}(\ket{x_i}; \Theta)-y_i\right|^2$
        & \texttt{nr} & Adam & \xmark & \cmark $\,^*$ & {CS, Phys.} & {C, Q} & {\scriptsize MNIST, IRIS }
        \\
        \midrule
        
        Jaderberg et al.~\cite{jaderberg2021quantum}
        & {Quantum Self-Supervised Learning} & 
        $\frac{1}{D}\sum_{i=1}^{D} \mathrm{log}\frac{-\mathrm{exp}(\bb z_i^1 \cdot \bb z_i^2 / \tau)}{\mathrm{exp}(\bb z_i^1 \cdot \bb z_i^2 / \tau) + \sum_{\alpha}\mathrm{exp}(\bb z_i^1 \cdot \bb z_i^\alpha / \tau)}$
        & nr & 
        Adam
        &  \cmark & 
        \cmark ${\,\,}^\circ$
        & {CS} & C 
        & {\scriptsize CIFAR10}$^*$\\

        Mari et al. \cite{mari2020transfer}
        & {Transfer Learning} & $-\frac{1}{D}\sum_{i=1}^{D} p_i[y_i-\mathrm{log}(\sum_k^C e^{y_k})]$ 
        & \texttt{nr} & Adam & \cmark $\,^\mathsection$ &  \cmark $\,^{*\, \diamond}$  & {CS} & C 
        &ImageNet$^*$, {\scriptsize CIFAR10}$^*$\\

        \midrule

          Wan et al. \cite{wan2017quantum}
         & {Quantum Feedforward Networks} & 
         $\sum_{i=1}^{3}\sum_{j\in\mathrm{\{q_{out}\}}}\eta_{ij}\big(\langle\paull{i}{(j)} \rangle_{\mathrm{out}} - \langle\paull{i}{(j)} \rangle_{\mathrm{true}} \big)^2
         {\,\,\,}^{\ddagger}$
         & \texttt{nr} & {Gradient Descent} & \cmark & \xmark & \makecell{QI} & Q
         &\\

        Chen et al. \cite{chen2021universal}
        & {Quantum State Discrimination} & $J(P_{\mathrm{succ}}(\ket{\psi_0}), P_{\mathrm{err}}(\ket{\psi_0}), P_{\mathrm{inc}}(\ket{\psi_0})){\,\,\,}^{\bullet}$ & $\mathcal{O}(n^k)$ & Adam &  \cmark & \xmark & {Inf. Theory} & Q 
        &\\

         Romero et al. \cite{romero2017quantum}
        & {QAE for quantum data compression} & $\sum_{i=0}^{S-1} p_i \cdot F\Big(\mathrm{Tr}_A\big[U_\theta\ket{x_i}\bra{x_i}_{AB}U_\theta^\dagger\big], \ket{\phi}_B\Big) {\,\,\,}^{\mathsection}$  & $\mathcal{O}(n^2)$ & {L-BFGS-B, Basin-Hopping} &\cmark $\,^\dagger$ & \xmark & Chem. & Q 
        &\\

    \bottomrule
  \end{tabular}
  \begin{tablenotes} \scriptsize 
  
  \item[]\h{\textbf{Loss Function:}
  $D$: training dataset size, 
  $B$: batch size, 
  $C$: number of the output classes,
  $\Theta$: QNN parameters,
  $\ket{\psi_{0}}$: initial state of the quantum system,
  $x_i$: $i$-th classical training data,
  $\ket{x_i}$: $i$-th quantum training data,
  $y_i$: the ground truth label of the $i$-th training data, 
  $\ket{\psi_i^{\text{true}}}$ is the ground truth label state of the $i$-th training data,
  $y_{\text{out}}(z; \Theta)$: the expected QNN output value for the input $z$ ($z$ may be a quantum input $\ket{x_i}$ or a classical input $x_i$ depending on the considered case),   
  $\star$ average fidelity evaluated over a batch \h{of size $B$}, 
  $\dagger$ $\bb{y}_{i}=[(y_i)_1,\dots,(y_i)_C]$ is the one-hot encoding of the ground truth label of the $i$-th training data and $g(x_i; \Theta)\in \R{C}$ is the probability distribution of the QNN output for the $i$-th classical training data, 
  $\ddagger$ $j$ runs on the output qubits only, 
  $\bullet$ $P_{\mathrm{succ}}(\ket{\psi_0}),\ P_{\mathrm{err}}(\ket{\psi_0}),\ \mathrm{and}\ P_{\mathrm{inc}}(\ket{\psi_0})$ are the probabilities of correct/erroneous/inconclusive measurement outcome,$\mathsection$ sum runs over an ensemble $S$ of input states. }
  
  \item[]\h{\textbf{Implementation - Sim.:} Numerical Simulation. $\dagger$ QuTiP library, $\dagger \dagger$  QxBranch library, $\mathsection$ PennyLane library, $\ddagger$ QASM library. 
   $\quad$
   \textbf{Implementation - QPU.:} $\bullet$ IBM Q 5 \quotes{Vigo}, $\circ$ IBM Q Paris, $*$ IBM Q 4, $\diamond$  Rigetti Q 4.}
 
  
  \item[]\h{\textbf{Application Field:} Quantum Information (QI), Chemistry ({Chem.}), Computer Science (CS), Physics (Phys.), Information Theory ({Inf. Theory}). $\quad$
  \textbf{Input Type:}  Quantum (Q). Classical (C)}
  \item[] \h{\textbf{Test Datasets:} we reported only real-world and publicly available datasets (synthetic datasets are omitted in the table), $*$  a subset of the dataset is used.}

\item[$+$] We only considered the Quantum Graph Recurrent Neural Networks

\end{tablenotes}
\end{threeparttable}
 \end{table}
 	\end{landscape}
  	\clearpage
}

\subsubsection{\h{The expressivity of PQC models}} 
\begin{figure}[!hp]
   \centering
     \begin{subfigure}[b]{0.35\textwidth}
         \centering
         \includegraphics[width=\textwidth]{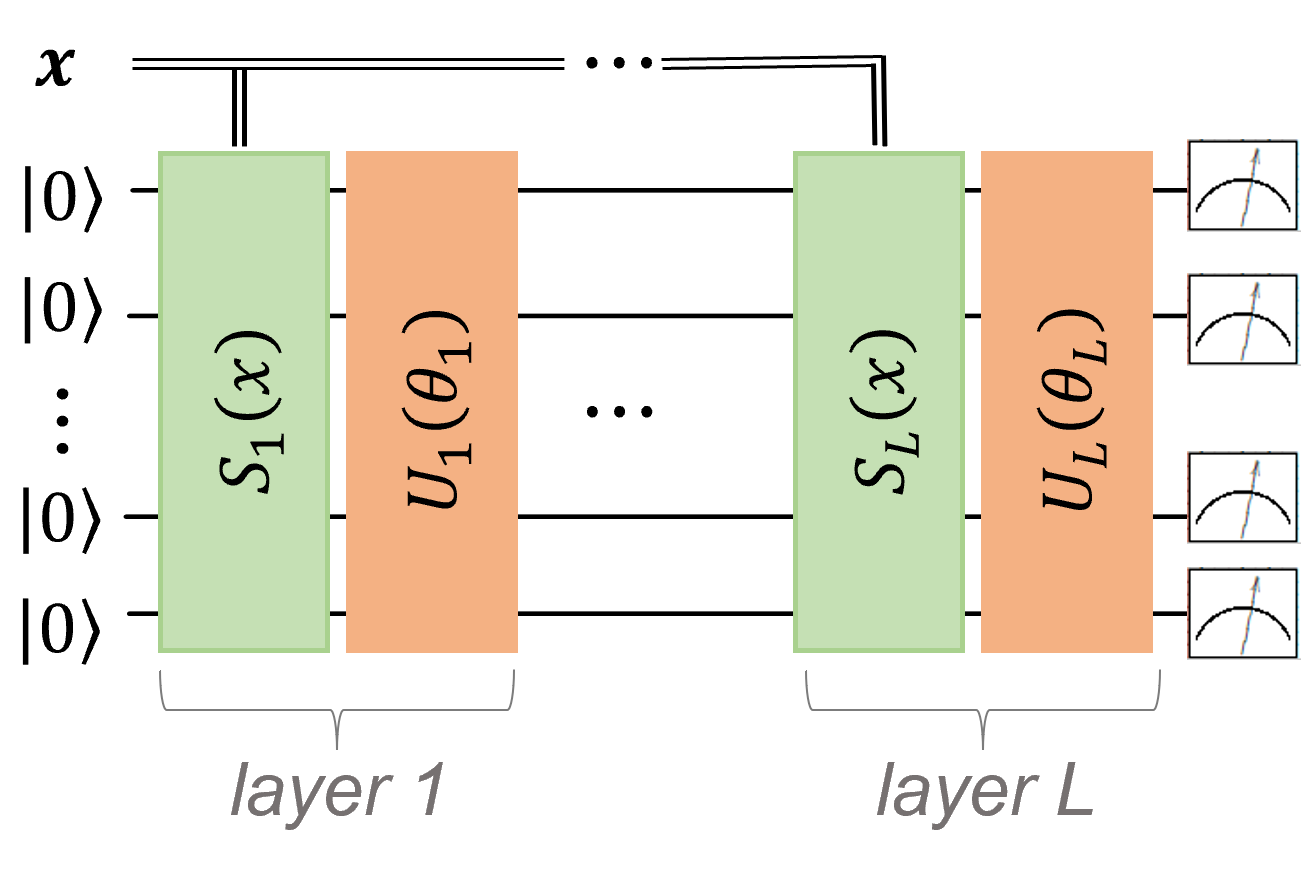}
         \caption{PQC as a universal function approximator \cite{schuld2021effect,perez2020data}}
         \label{fig:PCQ_UFA}
     \end{subfigure}
    \hfill
    \begin{subfigure}[b]{0.6\textwidth}
         \centering
         \includegraphics[width=\textwidth]{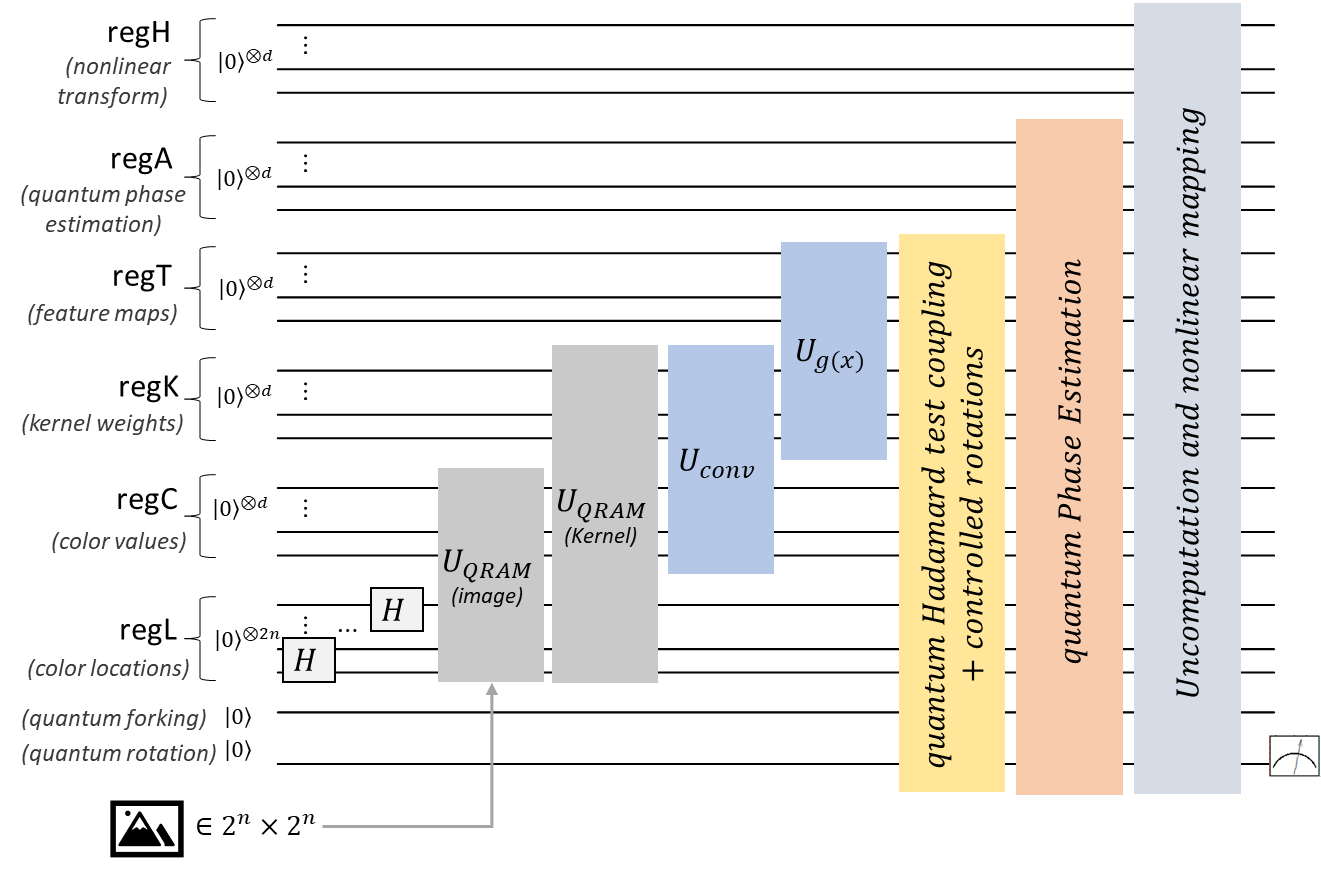}
         \caption{QDCNN (direct quantum implementation of a CNN) \cite{li2020quantum}}
         \label{fig:QDCNN}
     \end{subfigure}\\ 
     \par\bigskip
     \begin{subfigure}[b]{0.4\textwidth}
         \centering
         \includegraphics[width=\textwidth]{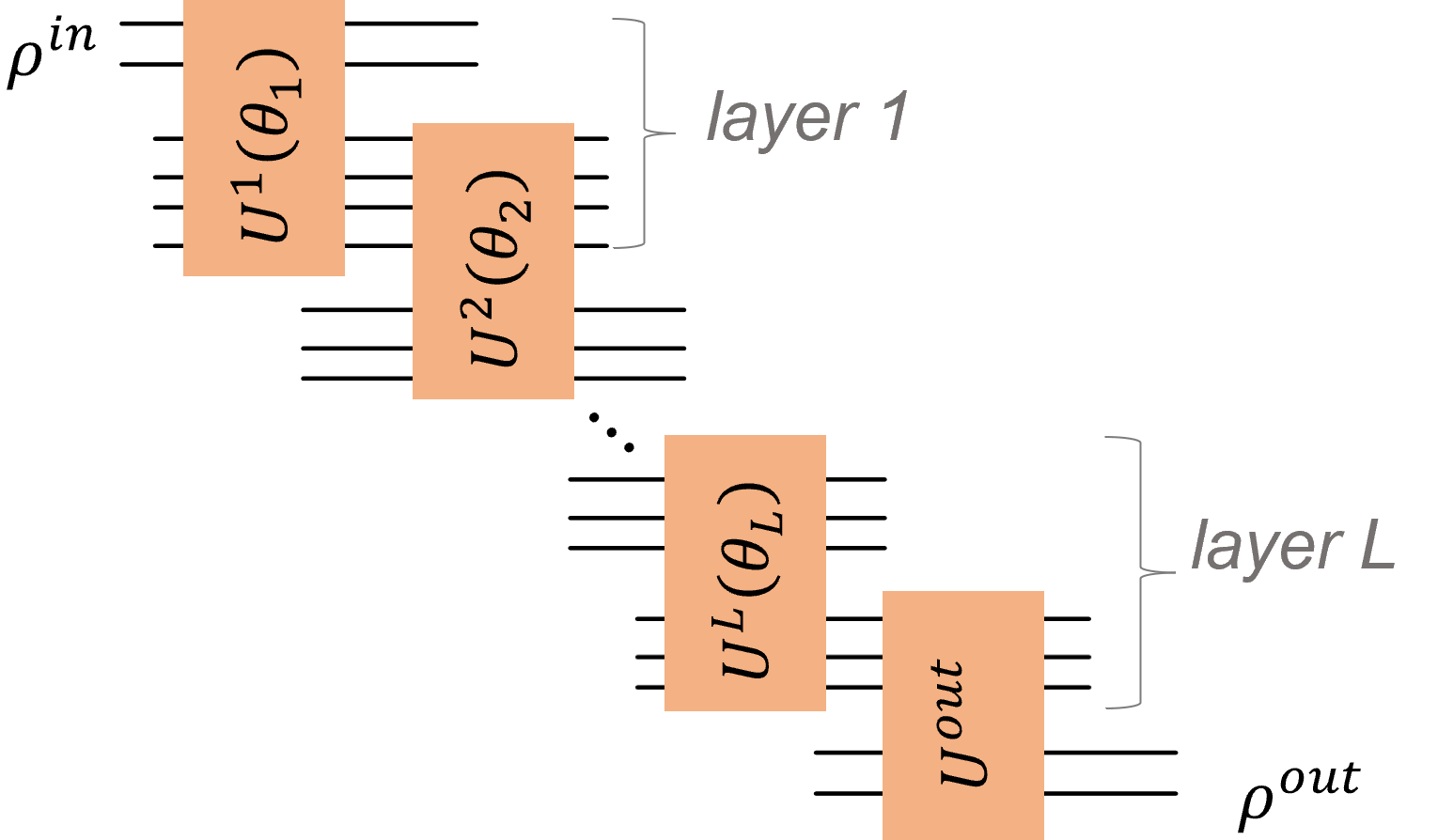}
         \caption{Quantum Dissipative Multilayer Perceptron \cite{beer2020training}}
         \label{fig:beer}
     \end{subfigure}
         \hfill
           \begin{subfigure}[b]{0.55\textwidth}
         \centering
         {\includegraphics[width=\textwidth]{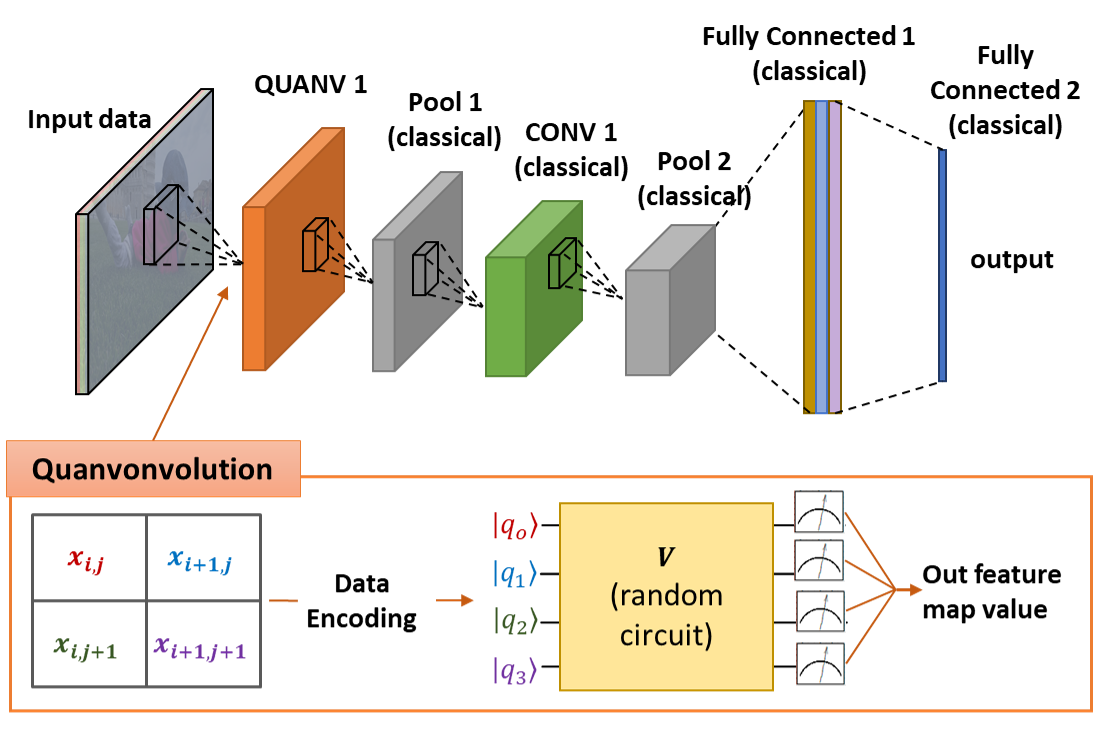}}
         \caption{Example of a Quanvolutional Neural Network (QVNN) \cite{henderson2020quanvolutional}}
         \label{fig:QVNN}
     \end{subfigure}\\ 
     \par\bigskip
      \begin{subfigure}[b]{0.4\textwidth}
         \centering
         \includegraphics[width=\textwidth]{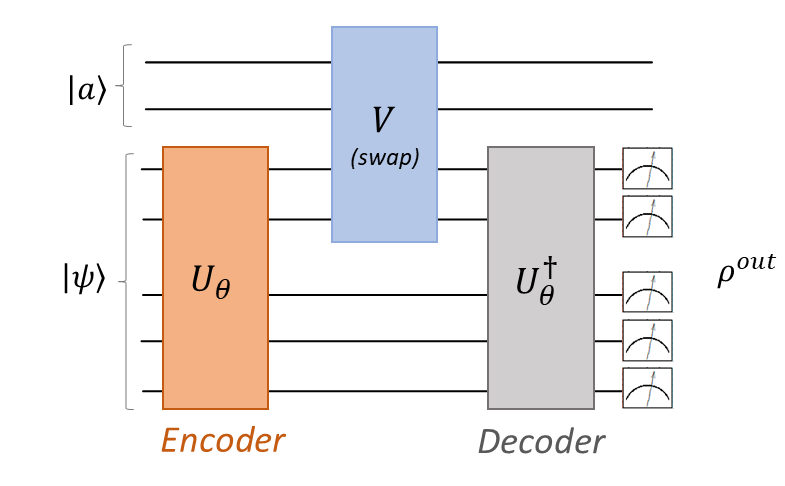}
         \caption{Quantum AutoEncoder (QAE)\cite{romero2017quantum}}
         \label{fig:romero}
     \end{subfigure}
     \hfill
     \begin{subfigure}[b]{0.55\textwidth}
         \centering
         {\includegraphics[width=\textwidth]{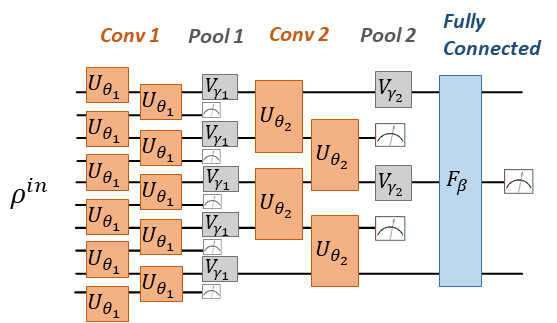}}
         \caption{Quantum Convolutional Neural Network (QCNN) \cite{cong2019quantum}}
         \label{fig:QCNN}
     \end{subfigure}
    \caption{\h{Schematic circuit representations of some representative QNNs. }}
    \label{fig:my_label}
\end{figure}

\h{
Classical feedforward networks have a layered structure where each layer processes the previous layer's output using functions depending on some trainable parameter. Since a layered structure with learnable parameters also characterizes PQC-based algorithms, they have been widely used as quantum versions of neural networks. Typically, each layer of a PQC accounts for three types of blocks: data-encoding circuit blocks $\hat{S}_i(\bb x)$ (depending on the input variable), parametrized circuit blocks $\hat{U}_i(\boldsymbol \theta_i)$ (depending on a set of trainable parameters), and non-parametrized circuit blocks $\hat{V}_i$ (e.g., entangling operators). 
Given these blocks, a PQC with $L$ layers is usually described by a unitary $\hat{\mathcal{U}}(\bb x; \Theta)=\prod_i^L\hat{V}_i\hat{U}_i(\boldsymbol \theta_i)\hat{S}_i(\bb x)$. The prediction from such a model is then evaluated as the expectation value of an observable $O$, over $m$ runs of the circuit, with respect to the final state of the quantum circuit, e.g.,  $y_{\text{out}}(\bb x; \Theta)=\langle \bb \psi_0|\hat{\mathcal{U}}(\bb x; \Theta)^\dag O \hat{\mathcal{U}}(\bb x; \Theta)|\bb \psi_0\rangle$ where $\ket{\psi_{0}}$ is some initial state of the quantum system. 
\h{Different choices for the implementation of the data encoding, parametric and non-parametric circuit blocks, and their mutual interaction correspond to different QNN architectures.} Hereafter, to ease the notation, we merge the non-parametrized blocks with the parametrized ones, i.e., we will use $\hat{U}_i(\boldsymbol \theta_i)$ to denote the composition $\hat{V}_i\hat{U}_i(\boldsymbol \theta_i)$ referred to as a \textit{trainable block}.

An important difference between this type of QNNs and classical NNs is that the information does not flow from one layer to the next only. Instead, the input data $\bb x$ is uploaded at each input layer $\hat{S}_i(\bb x)$ (see, e.g., \autoref{fig:PCQ_UFA}). Such a design directly impacts the class of functions the PQC can approximate \cite{schuld2021effect,perez2020data,gil2020input}. 
Indeed, note that a PQC made by only trainable blocks would be equivalent to a circuit made by a single trainable layer (since all the $\hat{U}_i(\boldsymbol \theta_i)$ are linear unitary operators, thus their composition is a linear unitary operator as well).
To draw a parallel with deep learning, data-encoding layers play a role reminiscent of the non-linearity of activation functions in classical DNNs, increasing the expressive power of models as function approximators. Indeed, a classical NN composed only of linear layers is equivalent to a network consisting of only one linear layer. In order to have a "deep" network, i.e. a network composed of multiple layers potentially able to generate progressively more abstract features with a higher representation power, the linear operations must be interleaved with non-linearities.} 

\h{Recently, Schuld et al. \cite{schuld2021effect} proved that if the data-encoding block $S(\bb x)$ of a PQC can be expressed via a Hamiltonian time evolution, i.e., $\h{S}(\bb x)=e^{-\ii H_1x_1} \otimes \dots \otimes e^{-\ii H_dx_d}$, and it is the same in every layer, then the PQC is capable of computing a partial Fourier sum,  $y_{\text{out}}(\bb x; \Theta)=\langle \bb \psi_{0}|\hat{\mathcal{U}}(\bb x; \Theta)^\dag O \hat{\mathcal{U}}(\bb x; \Theta)|\bb \psi_{0}\rangle= \sum_{\omega} c_\omega (\Theta) e^{\ii \omega^T \bb x}$. Thus, it is able to approximate any square integrable function in a given interval. In such a context, the output of the PQC model is a linear combination of complex exponential functions with a set of frequencies $\omega$ and coefficients $c_\omega$ determined by the topology of the network. In particular, trainable blocks determine the expressiveness of Fourier coefficients $c_\omega$, while the spectrum  $\Omega=\{\omega\}\subset \R{d}$ of the available frequencies is determined only by the data-encoding blocks of the circuit. In particular, the available frequencies are a combination of the eigenvalues of the Hamiltonians $\{H_j\}$ defining the data-encoding blocks. Moreover, the size of the spectrum, defined as the number of independent non-zero frequencies, increases linearly with the number of data encoding repetitions: the more times the original data is loaded, the richer the frequency spectrum will be. Finally, note that there are mainly two ways of repeating the data encoding: in series (by uploading several times the inputs while quantum computations proceed, as discussed above) or in parallel (by repeating the encoding on different subsystems at the same layer, so the depth of the circuit will be less but more qubits will be used) \cite{schuld2021effect}. Connections between Fourier series and PQC models with repeated data encoding have also been established by Gil Vidal, and Theis \cite{gil2020input}.}

Similar results \h{on the expressivity of PQCs} were reported by Perez-Salinas et al.~\cite{perez2020data}, 
\h{who} showed that it is possible to use a single-qubit quantum circuit to implement a universal quantum classifier. The Universal Approximation Theorem (UAT)~\citep{hornik1989multilayer} represents a milestone in the context of machine learning. Practically speaking, it ensures \h{that a} neural network comprising a single hidden layer with enough units can approximate any continuous function. Interestingly, \h{Perez-Salinas et al.~\cite{perez2020data}} presented an equivalence between UAT and the data re-uploading strategy.
\h{In their framework}, data are uploaded by single-qubit rotations followed by another set of unitaries representing the evolution of the quantum state. These two operations can be compressed into one by defining quantum layers with tunable parameters as: \h{$\hat{\mathcal{L}}(i) = \hat{U}(\boldsymbol{\theta}_i+\mathbf{w}_i^T\mathbf{x})$}, where $\mathbf{x}$ represents the input data, $\mathbf{w}_i$ are weights that play a similar role as in artificial neural networks, while $\boldsymbol{\theta}_i$ are a set of variational parameters. Although, at first sight, it seems that there is no non-linearity in such a construction, the authors reported that they come directly from the structure of the used gates. 
\h{The output state generated by the circuit is $\hat{\mathcal{U}}(\bb x; \Theta)|\bb \psi_{0}\rangle$, where $\hat{\mathcal{U}}(\bb x; \Theta)=\prod_{i=1}^L{\hat{\mathcal{L}}(i)}$ and $\Theta=\{\boldsymbol{\theta}_i,\mathbf{w}_i \}_i$ is the set of all circuit parameters. }
To train their classifier, the authors employed the L-BFGS-B optimization algorithm and used the average fidelity among the output state and the expected one, \h{$\ket{\psi^{\text{true}}}$}.

\h{In 2018, Schuld et al.~\cite{schuld2018circuit} proposed a low-depth parametrized quantum circuit acting as a binary classifier. Their} approach is particularly relevant since it was among the first ones to apply a quantum ansatz to solve machine learning problems using variational circuits. \h{The proposed circuit uses a data-encoding block $\hat{S}(\bb x)$, based on the amplitude encoding technique, to embed classical input data into the state of a quantum system, which is then evolved by a trainable ansatz $\hat{U}(\boldsymbol\theta)$.  }
The circuit ansatz accounted for a set of single- and single-controlled qubit gates divided into the so-called \quotes{cyclic codes}. Once the initial state is evolved, the first qubit is measured, and the probability of finding it in the state $\ket{1}$ is used to infer the input label. Specifically, such a probability is given by:
\begin{equation}
    p(q_0=1, \bb x; \boldsymbol \theta) = \sum_{k=2^{d-1}+1}^{2^d}\left|\left(\h{\hat{U}(\boldsymbol\theta)\ket{\phi(x)}}\right)_k\right|^2
\end{equation}
where \h{$\ket{\phi(x)}=\hat{S}(\bb x) \ket{\bb 0}$ is the state prepared by the data-encoding block, }
and $\left(\h{\hat{U}(\boldsymbol \theta})\ket{\phi(x)}\right)_k$ is the $k$-th entry of the result after the application of the operator $\h{\hat{U}(\boldsymbol\theta)}$ to $\ket{\phi(x)}$. The sum over the second half of the resulting vector corresponds to the single-qubit measurement resulting in state 1. 
Postprocessing adds the bias $b$ and thresholds to compute a binary prediction: $\h{f}(x;\boldsymbol \theta,b)=p(q_0=1, x; \boldsymbol\theta)+b$. To estimate such a probability, the circuit must be run several times. The overall operation is equivalent to evaluate the expectation value of the $\paul{z}$ operator acting on the first qubit of the evolved state: $\mathbb{E}(\paul{z})=\bra{\phi(x)}\h{\hat{U}(\boldsymbol \theta})^\dagger(\paul{z}\otimes\mathbb{I}\otimes...\otimes\mathbb{I})\h{\hat{U}(\boldsymbol \theta})\ket{\phi(x)}$. To validate their ansatz, the authors' employed numerical simulation on several UCI datasets, namely CANCER, SONAR, WINE, and SEMEION, and on the MNIST dataset. While the first two are binary classification problems, the last three are multiclass problems. In such last cases, the authors employed the \quotes{one-vs-all} approach, thus allowing them to use their binary classifier in these cases too. 

Building on top of~\h{\cite{schuld2018circuit}}, 
Macaluso et al.~\citep{macaluso2020variational} are among the first ones to propose a learning framework for a quantum Single Layer Perceptron (qSLP) that is implemented on an actual quantum device, representing the quantum analogous of a single layer perceptron containing two neurons. Thus, resembling the first attempt at a quantum neural network. Specifically, the circuit computes linear operations in superposition, each characterized by a different set of parameters. The approach is based on three quantum registers\h{:} control, data, and temporary. At the end of the computation, the second one stores the results of the two linear operations in superposition. The final response of the qSLP can be obtained as the expected value of $\paul{z}$ acting on the data register. Notably, a non-linear activation function is not implemented in the quantum circuit \h{in such a work.} Instead, it is part of post-processing operations. Finally, to train the qSLP, the authors used the Sum of Squared Errors (SSE) loss and the Nesterov~\cite{ruder2016overview} accelerated gradient method.

Instead, Beer et al.~\cite{beer2020training} \h{proposed a quantum MLP, also referred to as q\textit{uantum dissipative neural network}. In their model, each quantum perceptron is} formulated as a unitary operator acting on $n$ input and $m$ output qubits. The idea was to stack this kind of perceptrons 
to form \quotes{layers} which emulated the behavior of a classical deep model.
\h{Each layer interacts with the following one before being dissipated (discarded); the interaction is implemented by coupling the qubits in the $i$-th layer to some ancillary qubits in the $(i+1)$-layer (\autoref{fig:beer}).}
The overall QNN can be then described as follows: $\mathcal{\hat{U}}=\prod_l^L\hat{U}^l$, where $\hat{U}^l=\prod_i\hat{U}^l_i$ represents the set of unitaries of layer $l$, i.e., the QPs of that specific layer. The QNN is applied to the input state, $\rho^{\mathrm{in}}$ \h{and produces an output state, $\rho^{\mathrm{out}}$.}
To assess the performance of their approach, the authors used numerical simulations with a QNN concerning input and output spaces of 2 and 3 qubits. To train their model, they employed a gradient descent technique, while as a loss function, they employed the 
\h{fidelity among the output state and the desired (ground-truth) state averaged over the training data.}

Graph Neural Networks~\citep{sperduti1997supervised} were introduced to exploit the geometrical structure in the input data and accounted for the application of neural networks to acyclic graphs. Based on a similar principle, a general \h{PQC} 
was proposed  by Verdon et al.~\citep{verdon2019quantum}, named \textit{Quantum Graph Neural Networks} (QGNN), consisting of a sequence of $Q$ different Hamiltonian evolutions repeated $P$ times:
\begin{equation}
    \hat{U}({\boldsymbol{\eta};\boldsymbol{\theta}}) = \prod_p^P\Big[\prod_q^Q e^{-i\eta_{pq}\Ham_q(\boldsymbol{\theta})} \Big]
\end{equation}
where the $\boldsymbol{\eta}$ and $\boldsymbol{\theta}$ are the sets of variational parameters, and the $\Ham_q(\boldsymbol{\theta})$ are the Hamiltonians encoding the problem to be solved. The authors proposed different variants for the general QGNN ansatz concerning the following tasks: learning Hamiltonian dynamics of quantum systems, learning how to create multipartite entanglement in a quantum network, unsupervised learning for spectral clustering, and supervised learning for graph isomorphism classification. Concerning the first task, they specialized the ansatz\h{,} naming it Quantum Graph Recurrent Neural Network \h{(QGRNN),} %
\h{by tying } the temporal parameters between iterations \h{($\eta_{pq}\to \eta_{q}$)}, thus resembling the classical sequence networks in which the parameters are shared among the various time step of the representation evolution. \h{Given a target Hamiltonian $\hat{H}_\text{target}$ the authors train the model parameters by using the Adam~\citep{kingma2014adam} optimizer to minimize the average infidelity :}
\begin{equation}
    1 - \frac{1}{B}\sum_{j=1}^B|\bra{\psi_{T_j}}\hat{U}^j(\Delta,\theta)\ket{\psi_0}|^2
\end{equation}
\h{where the $\ket{\psi_{T_j}}=e^{-iT_j\Ham_\text{target}}\ket{\psi_0}$ is the state evolved from $\ket{\psi_0}$ for a time $T_j$, $\hat{U}^j(\Delta,\theta)\ket{\psi_0}$ is the state obtained by evolving $\ket{\psi_0}$ according to the QGRNN for $P\sim T_j/\Delta$ iterations with $\Delta$ being a hyperparameter determining the Trotter step size, and the average is evaluated over batch sizes of $B$ different times $T_j$.}

\subsubsection{\h{CNN-inspired Quantum Neural Networks}}
\h{Convolutional Neural Networks (CNNs) are feedforward networks particularly suitable for data that have a grid-like topology such as images or time-series data. This class of models owes its name to the use of convolutional operations in at least one of their layers. A convolutional layer takes a tensor as input and processes it using 
a set of trainable kernels (filters) that slide along the input tensor, computing the inner product between the filter and the input entries, producing as output the so-called feature map. Each element of the output is then transformed by applying a non-linear activation function. Finally, a pooling operation is typically used to replace portions of the feature map with a summary statistic (e.g., max, average, sum) - which makes it more manageable in the successive layers (since the dimension is reduced), the network more resilient to small input changes, and helps to avoid overfitting issues. Convolutional layers can then be followed by fully-connected layers where each perceptron receives information from all the units of the previous layer. This latter typology of a layer is also used to produce the output of the network, e.g. a vector containing the probabilities that the input data belong to certain classes for classification tasks or real values for regression tasks.
}

\h{
Given the success of deep CNNs on the task of pattern recognition~\citep{lecun1995convolutional,krizhevsky2012imagenet}, many researchers proposed quantum algorithms inspired by these models.
}
\h{In \cite{li2020quantum,kerenidis2019quantum}, the authors proposed quantum models, running on an actual quantum hardware, capable of performing all the classical CNN operations. The main limitation of these approaches is that they usually require a QRAM~\cite{giovannetti2008quantum} to interface classical data and quantum states and may suffer from classical data encoding bottleneck. Moreover, the overall algorithm may require subroutines that are not easy to implement with many qubits (e.g., quantum phase estimation).
For example, 
Li et al.~\cite{li2020quantum} proposed the \textit{Quantum Deep Convolutional Neural Network} (QDCNN), based on a PQC, whose architecture consists of several successive quantum convolutional layers and a quantum classifier layer (\autoref{fig:QDCNN}). Their model does not implement a quantum version for the pooling layer, however the dimension of the generated feature maps is reduced by preserving some qubits in the location register and disentangling the other ones. Quantum measurement is performed to output a category prediction, and a hybrid quantum-classical training algorithm is used to optimize the parameters of the circuit. The classical input (e.g. an image) and the convolutional kernels are prepared using a QRAM. 
Similarly,} Kerenidis et al.~\cite{kerenidis2019quantum} proposed to vectorize the convolution operations in a CNN to exploit linear-algebra subroutines to carry out the calculation exploiting quantum phenomena. To their aim, the authors proposed algorithms both for the forward and backward passes while training a QNN. 
\h{D}ata and convolutional kernels are stored into QRAMs registers that are resumed when needed.

\h{Henderson et al.~\cite{henderson2020quanvolutional}} proposed \h{a hybrid classical-quantum approach} to empower a classical CNN with \h{layers} containing filters implemented as quantum circuits\h{, called quanvolutional layers, that can be stacked on top of any layer of a traditional CNN (see \autoref{fig:QVNN})}. 
\h{The resulting} architecture was named \textit{Quanvolutional Neural Network} (QVNN). Such a proposal stemmed from the observation that, on the one side, machine learning algorithms benefit from random non-linear features in terms of accuracy and training time, while on the other, quantum circuits can model complex functional relationships. Each quanvolutional layer \h{is} characterized by a certain number of filters\h{,} designed as a sequence of single- or single-controlled qubit gates\h{, and produces feature maps by locally transforming input data}. Specifically, each circuit \h{is} built by considering each qubit as a node in a graph and assigning what the authors named a \quotes{connection probability} between each pair of qubits. A key difference of such a formulation of QVNN compared to others is the lack of a variational tuning of the circuits' parameters \h{(they use random quantum circuits in the quanvolutional filters)}.
\h{Interestingly, such an architecture does not require a QRAM, thus reducing the overhead due to its usage, and it is compatible and directly integrable with existing classical architectures}. 
QVNNs were tested against the MNIST dataset using the QxBranch Quantum Computer Simulation System, and, unfortunately, the authors did not find any advantage over the classical counterpart. Indeed, the performance of the QVNN was indistinguishable from that of a \h{model using a classical random non-linear transformation instead of the quanvolutional layer}.

\h{In} Cong et al.~\citep{cong2019quantum} a variational ansatz\h{, named \textit{Quantum Convolutional Neural Network} (QCNN),} is proposed \h{ for analysing quantum data} \h{(\autoref{fig:QCNN})}.
As typically happens for variational circuits, hyperparameters such as the number of gates are fixed, while the parameters characterizing them are learned by classical optimization. The similarity between \h{CNN} and QCNN stems from the interpretation given to each \quotes{layer} of the last one. 
However, it is fundamental not to confuse the concept of layer between the classical and quantum architectures. Indeed, concerning the last one, a layer is represented by a unitary operator applied to the data register. With such a difference in mind, one can think of the QCNN structure as a sequence of layers as follows. The first layer applies a single quasilocal unitary, $\hat{U}_\theta$, in a translationally invariant manner followed by a measurement on a subset of the input qubits, which resembles the pooling operation. The outcome of such an operation determines the unitary operation, $V_\gamma$, applied to the nearby qubits. These two operations are applied a few times until the system is small enough. A final unitary, $F_\beta$, is then applied to emulate the action of a fully connected layer. Finally, the outcome of the QCNN is obtained by measuring a specific set of output qubits. Such a model was applied successfully to solve the problem of Quantum Phase Recognition, which aims at recognizing whether a given input quantum state $\rho_i$ belongs to a particular quantum phase of matter. Given a QCNN, the training set comprised training vectors $\{(\ket{\psi_i}, y_i) | i=\h{1, \dots, D}\}$, where the binary labels $y_i$ identified a given input state vector. As objective, the authors exploited the mean squared error between the label $y_i$ and the \h{QCNN output.} 
Thus, the learning process consisted of initializing all the unitaries and then optimizing them until convergence, for example, via gradient descent.

In Grant et al.~\cite{grant2018hierarchical}, the authors exploited the tensor networks' hierarchical structure to represent quantum circuits\footnote{\h{This method can be interpreted as a CNN-inspired model since tensor networks with hierarchical structure exhibit many similarities with neural networks and in some cases have been shown to be equivalent to CCNs with rectified linear activation \cite{grant2018hierarchical}}}. Specifically, they employed tree tensor networks~\citep{shi2006classical} and the multi-scale entanglement renormalization ansatz (MERA)~\citep{vidal2008class} to implement a quantum classifier subsequently tested against the Iris, MNIST, and synthetic quantum datasets. The classifiers were implemented on the ibmqx4 quantum computer and trained using the mean squared error loss evaluated between the circuit's output and the expected one. Specifically, concerning the first quantity, it represented the outcome of a measurement on the target qubit: $\langle \hat{M}_\theta(\ket{\psi})\rangle = \bra{\psi} \hat{U}_\theta^\dagger \hat{M} \hat{U}_\theta\ket{\psi}$. To minimize the objective function, the authors employed the Adam~\citep{kingma2014adam} optimizer. As an example of hierarchical structure, the tree tensor networks-inspired circuit begins by applying a set of two-qubit unitaries among the nearest qubits. After each operation, one qubit is discarded, thus halving the total number of qubits to the next layer. After a sequence of such unitaries, the state of the final remaining qubit is measured.

\subsubsection{\h{Hybrid approaches for different learning paradigms}}
Jaderberg et al.~\cite{jaderberg2021quantum} proposed a VQA to solve a classification task by employing a self-supervised training approach~\cite{de1994learning}. They formulated the ansatz to be part of a hybrid encoder network tested against the classification task on 
\h{five classes} of the CIFAR10 dataset. To train the model, the authors exploited the contrastive learning procedure~\cite{oord2018representation,he2020momentum} applied to the representations generated by their models. As the objective function, they used the normalized temperature-scaled cross-entropy loss~\cite{sohn2016improved}. As mentioned, the QNN was part of a hybrid network whose first part was classical. Specifically, the overall architecture began with a ResNet-18~\cite{he2016deep} \h{followed by a single-layer convolutional network} that compressed the features representation down to an eight-dimensional vector, \h{$\bb v$,}  such that \h{it was encoded in an} entangled state \h{$\ket{\psi_{\bb v}}$} comprising $\h{n}=8$ qubits. 
\h{The quantum data encoding is achieved by applying a single qubit rotation $\hat{R}_x$ (as generated by the $\paul{x}$ operator) to each qubit of the register, i.e., $\ket{\psi_{\bb v}}=\otimes_i^n\hat{R}_x(v_i) \ket{\bb 0}$, where $v_i$ is the $i$-th element of the $n$-dimensional features vector $\bb v$. The state is evolved using an ansatz made by parametrized controlled qubit rotations.}  Finally, to get the output from the QNN, the authors measured the qubits state in the computational basis.
The authors tested their approach \h{on quantum state vector simulations and real quantum devices}. 
In both cases, they observed that the hybrid architecture reported performance similar or higher than the full classical implementation, thus giving an interesting hint about the capability of such hybrid models applied to computer vision-related tasks.

Transfer learning is a widely used technique in deep learning that exploits pre-trained neural models on a different task. Interestingly, such an approach can be \quotes{hybridized} by combining classical and quantum algorithms. Indeed, Mari et al.~\citep{mari2020transfer} considered three different scenarios for transfer learning, namely, classical-to-quantum, quantum-to-classical, and quantum-to-quantum. In what follows, we focus on the classical-to-quantum configuration. In such a case, a ResNet-18~\citep{he2016deep} network trained on the ImageNet dataset~\citep{deng2009imagenet}, was used as a features extractor for the quantum circuit. The 
\h{parametrized quantum circuit} was interpreted as made by three components: $Q=\mathcal{M}\circ\mathcal{Q}\circ\mathcal{E}$. The first component, $\mathcal{E}$, represents the quantum embedding of the classical features vector into the Hilbert space depending on $\mathbf{x}$\h{. The} second one, $\mathcal{Q}$, is a variational circuit of a given depth composing single and two-qubit gates, while the $\mathcal{M}$ term represents the measurement on the circuit's output \h{that maps a quantum state to a classical vector}. The authors trained such a circuit to classify images of bees and ants utilizing the Adam~\citep{kingma2014adam} optimizer and the cross-entropy loss. The authors assessed the performance of their models through numerical simulations by using the PennyLane framework and then implemented the quantum circuit on the ibmqx4 and the Aspen-4-4Q-A quantum processors from IBM and Rigetti, respectively. 
The authors' results demonstrated that there was already the possibility to investigate efficient algorithms to process high-resolution images in the NISQ era~\citep{preskill2018quantum} concerning the classical-to-quantum transfer learning scheme.

\subsubsection{\h{Other approaches}}
Differently from the variational approach, a quantum generalization of the classical feedforward neural networks was proposed by Wan et al.~\cite{wan2017quantum}. The quantum neural network accounted for a reformulation of the perceptron model by making each transformation reversible and unitary. Although different from VQA, such networks have shown, by employing numerical simulation, to be trainable using gradient descent for a given objective function. In such a case, the goal was to minimize the difference between the quantum circuit's output and the expected one. A viable way to physically realize such an architecture was employing quantum photonics. For example, to prove the effectiveness of their approach, the authors proposed a quantum autoencoder to compress two quantum states with high accuracy.

Quantum State Discrimination (QSD)~\citep{barnett2009quantum} requires the design of measurements to optimally identify the $\rho$ that represents the unknown state of a quantum device which is believed to belong to a non-orthogonal set of states $\{\rho_i\}$. Such a capability has a broad range of applications. For example, quantum state discrimination by itself plays a crucial role in quantum information processing protocols and is used in quantum cryptography~\citep{bennett1992quantum}, quantum cloning~\citep{duan1998probabilistic}, quantum state separation, and entanglement concentration~\citep{barnett2009quantum}. Typically, distinguishing among non-orthogonal states is challenging, and specialized measurements are required. An exciting approach for solving such a problem was proposed by Chen et al.~\cite{chen2021universal}, who formalized a framework to learn to simulate the unknown structure of the generalized quantum measurement, or Positive-Operator-Value-Measure (POVM)~\citep{nielsen2002quantum}. The circuit designed by the authors contained only a single-qubit and CNOT gates, and the proper non-linearities were introduced by measuring qubits states. Specifically, they assessed the performance of their circuits by employing numerical simulation on classical machines and by using the Adam~\citep{kingma2014adam} optimizer to learn the circuit's parameters. However, differently from the previously cited variational approaches, in the current case, the authors aimed at maximizing their circuit's generalization performance given a specific range of the parameters rather than looking through all the available space. 

Classical AutoEncoders are tasked with data reduction, i.e., given an $(n+k)$-bit representation, the encoder generates a compressed $n$-bit datum with the same discrimination power of the original input. Romero et al.~\citep{romero2017quantum} proposed a \textit{Quantum AutoEncoder} (QAE) as a solution to the compression problem concerning quantum states represented by $(n+k)$-qubits. Specifically, they were meant to facilitate quantum data compression, especially concerning quantum simulations. Such models should recognize patterns due to quantum phenomena, such as superposition and entanglement, not accessible to classical algorithms. 
\h{The authors} tested their architecture to compress ground states of the Hubbard model and molecular Hamiltonians. The task of a QAE was then to get rid of a certain number of qubits, say $k$ while maintaining the same amount of quantum information in the remaining ones. Thus, the QAE must entangle qubits, and for such an aim, the authors exploited unitary gates to build the variational circuit \h{(\autoref{fig:romero})}. As a cost function, QAE leverages the expected fidelity~\citep{nielsen2002quantum}, $F(\ket{\psi_i}, \rho_i^{out})$, which quantifies the deviation from the initial state $\ket{\psi_i}$ to the output one $\rho_i^{out}$. The authors used a hybrid approach to train the model in which the quantum device took care of the state preparation and the measurement while the optimization procedure was carried out through classical techniques. The authors exploited two different optimizers \h{to train the circuit}, namely, L-BFGS-B and Basin-Hopping. After training, the authors proved that QAE could compress the ground state of the two-sites Hubbard model from 4 qubits down to 1 with an error of below $10^{-4}$. It is relevant to notice that such a compression allows moving from a space of size $2^4$ to one of size 2.

Finally, it is worth mentioning \h{that there exist approaches} based on an optimization procedure different from the variational approaches we have seen so far. For example, Silva et al.~\cite{silva2010superposition,da2012classical} proposed a learning algorithm in which all the input patterns were presented concurrently in superposition to the neural networks. Their model was based on a quantum version of the RAM-based neural networks~\cite{aleksander1966self}, termed quantum-RAM (q-RAM) node~\cite{de2009quantum}, a weightless network in which the learning procedure consisted upon simply writing the proper output value in the corresponding look-up table entries of the q-RAM node. The model was designed as a quantum circuit based on the gate model paradigm. Moreover, the authors commented on the possibility of \h{stacking} mode nodes to form a deeper network. A fundamental step in the training procedure accounted for Grover's algorithm~\cite{grover1997quantum} 
applied to the input until the desired output was returned.

\subsection{\h{Trainability Issues and Barren Plateau}}\label{sec:barren}
\h{
As discussed in the previous section, the most common approach for implementing a QNN is using a PQC characterized by a set of parameters $\boldsymbol \theta$ that are typically learned by optimizing a cost function $C(\boldsymbol \theta)$ using a classical optimization algorithm. One of the main issues with the trainability of a QNN is that the gradient of the cost function with respect to $\boldsymbol \theta$ may exponentially vanish during the training as a consequence of the progressive flattening of the cost function's landscape -  a phenomenon referred to as \textit{barren plateaus}. 
This phenomenon is particularly present when using a random initialization for \quotes{deep} circuits, or more precisely, for circuits characterized by a number of layers that depend polynomially on the number $n$ of qubits 
\cite{mcclean2018barren}. Indeed, McClean et al. \cite{mcclean2018barren} stated that for a wide class of PQCs with a sufficient depth and number of qubits, the probability that the gradient along any reasonable direction is non-zero to some fixed precision is exponentially small as a function of the number of qubits.
In other words, the expected value of $\frac{\partial C(\boldsymbol \theta)}{\partial \theta_i} $ is zero and its variance decreases exponentially with $n$. 
Unfortunately, the region where the gradient is zero does not correspond to local minima of interest. 
Instead, it corresponds to a large plateau of states for which no interesting search direction can be pursued to exiting the barren plateau, or more precisely, an exponential precision is required to determine a minimizing direction in order to navigate the cost landscape.
The use of a shallow circuit does not solve the problem a priori because as long as a \textit{global} cost function is used, i.e. a measure obtained using an observable that involves all the qubits of the circuits, the barren plateau can occur in any layer of the circuit \cite{mcclean2018barren}. This is one of the main differences between the gradient vanishing problem for classical DNNs versus QNNs: in the case of classical DNNs, the gradient can vanish exponentially in the number of layers, while for QNNs, the gradient can vanish exponentially in the number of qubits \cite{mcclean2018barren}. 
  
Cerezo et al. \cite{cerezo2021cost} showed that a promising approach to overcome the problem of exponentially vanishing gradients 
is to use shallow circuits with \textit{local} cost functions, which are constructed by observing only a limited set of qubits at a time. Indeed, they demonstrated that local cost functions lead, at worst, to a polynomially vanishing gradient if the circuit depth is $O(\log n)$, with $n$ being the number of qubits, and therefore the circuit will be trainable with a polynomial scaling with the system size (i.e. a polynomial number
of shots per optimization step are needed to estimate the gradient).
However, note that using local cost functions alone is not sufficient to avoid the problem: deep networks with local observables can still give rise to barren plateaus. 

Interestingly, some QNNs have not exhibited barren plateau in the cost function landscape, such as dissipative quantum neural networks with shallow local perceptrons \cite{beer2020training,sharma2020trainability} and Quantum Convolutional Neural Networks \cite{grant2018hierarchical, cong2019quantum, pesah2021absence}.
Moreover, several strategies have been proposed for mitigating barren plateaus,  including the use of local cost function \cite{cerezo2021cost}, specific parameter initialization/pre-training \cite{grant2019initialization,verdon2019learning}, clever problem-inspired ansatz \cite{hadfield2019quantum,bharti2021iterative}, layerwise learning \cite{skolik2021layerwise}, and parameter correlation \cite{volkoff2021large,holmes2021connecting}, just to name a few. 
 
All the methods mentioned so far concern the problem of noise-free barren plateaus since they do not consider quantum hardware noise. However, a different kind of barren plateau can occur in the presence of noise. Wang et al. \cite{wang2020noise} provided an analytical study on noise-induced barren plateaus for a generic variational ansatz. They proved that in the presence of local noise, the magnitude of the gradient decays exponentially with the depth of the circuit. Therefore, while noise-free barren plateaus are mainly connected to the structure of the ansatz (number of qubits, parameter initialization, locality of the cost function), the noise-induced barren plateaus are mainly affected by the ansatz depth. Moreover, for noise-induced barren plateaus, the magnitude of the cost function's minimum is also flatting, in addition to its gradient. Due to the diverse nature of noise-induced barren plateaus, strategies to mitigate noise-free barren plateaus, e.g. \cite{cerezo2021cost,grant2019initialization,volkoff2021large,skolik2021layerwise}, do not solve the problem, and the strategies to adopt should rather rely on simplifying the complexity of the circuit by reducing its depth or reducing the hardware noise level.
}

\begin{table}[h!]
\caption{\h{Performance comparison between some QNNs on the MNIST dataset. The original MNIST dataset contains 70,000 (60,000 training and 10,000 test) 28-by-28 greyscale pixel images of handwritten digits (10 classes). The test accuracy of the QNNs is reported from the original papers. }}		
\label{tab:mnist}
\small
	\begin{threeparttable}
	\begin{tabular}{
	p{0.165\columnwidth} 
	p{0.1\columnwidth}
	p{0.17\columnwidth}
	p{0.17\columnwidth}
	 >{\centering\arraybackslash}p{0.14\columnwidth}
	 >{\raggedleft\arraybackslash}p{0.1\columnwidth}
	}
		\toprule
		\multirow{2}{*}{\textbf{Reference}} & \multirow{2}{*}{\textbf{Impl.}} & \multirow{2}{*}{\b \makecell{Input dim\\ (after preprocessing)}} & \multirow{2}{*}{\textbf{$\#$ Samples}} & \multirow{2}{*}{\textbf{$\#$ Classes}}  & \multirow{2}{*}{\textbf{Accuracy (\%)}} \\ \\   \midrule
		 Schuld et al. \cite{schuld2018circuit}${\,}^*$ &
		 Simulation	& 
		 256 real values&
		 2,766 (train/test split not reported)	&
		 10&
		 $67.00$\\
		 
		Li et al. \cite{li2020quantum}&
		Simulation&
		28$\times$28 (784 real values)&
		60,000 train/10,000 test&
		10&
		$98.97$\\

		Henderson et al. \cite{henderson2020quanvolutional}${\,}^\circ$&
		Simulation&
	    28$\times$28 (784 real values)&
	    60,000 train/10,000 test&
	    10&
	    $\sim98.00$  \\
	    
		\multirow{4}{*}{Grant et al. \cite{grant2018hierarchical}$\, ^\star$}&	\multirow{4}{*}{QPU (IBM Q4)}&
		\multirow{4}{*}{8 real values}&
		\multirow{4}{*}{60,000 train/10,000 test}	&
		2 (is $> 4$ or  $< 4$ )	
		& $79.10 \pm 0.90$ \\
		&&&& 2 (is even or odd)	& $84.85 \pm 0.20$\\
        &&&& 2  (is $0$ or $1$)	& $99.87 \pm 0.02$ \\
        &&&& 2 (is $2$ or $7$)	& $98.86 \pm 0.07$\\
		        \bottomrule

\end{tabular}
\begin{tablenotes} \scriptsize 
   \item[$*$] multilabel classification problems casted as a set of “one-versus-all” binary classification subtasks;  fivefold cross-validation with one repetition was carried out
   \item[$\circ$] approximate accuracy estimated from the graph of test accuracy versus training iterations available in the original paper
   \item[$\star$] four distinct binary classification tasks were considered in the paper 
  \end{tablenotes}
\end{threeparttable}
\end{table}

\subsection{Final Remarks}
As we have seen, \h{PQC} currently \h{represents} the most adopted approach to QNNs. Perhaps, their capability to allow implementing learning circuits on NISQ devices and the similarities between the ansatz and the architecture of classical neural networks are two of the reasons for the variational \h{algorithms' success}. However, other research lines try to exploit intrinsic properties of quantum systems without the necessity of emulating a classical procedure, such as in the work of~\cite{wiebe2018quantum}. Hence, as we have seen so far, the research for QNNs that one day will be applied to real-world problems represents a very exciting and active field. 

\h{Unlike the literature on classical neural networks, identifying state-of-the-art QNNs for solving a given task is not easy. This issue is mainly due to the lack of benchmark procedures for comparing these algorithms. Many approaches have only been tested on synthetic data generated by authors, and only a few have publicly released their code. Furthermore, the use of different implementations (simulations or different QPUs) makes a direct comparison among the various approaches very difficult. Even when the same benchmark dataset is used, see, for example, the MINST dataset for handwritten recognition in \autoref{tab:mnist}, it is not easy to compare the performance of the various methods because there is a variety in the number of samples used, the classes or the specific tasks considered.}

Due to the \h{limitations of current} quantum platforms, several works \h{have} remained only theoretical, waiting for QPUs to become powerful enough and less noisy to allow \h{the exploitation }of a large number of entangled units. Indeed, the dragon of decoherence is always on the hunt and, unfortunately, qubits easily fall \h{prey}, thus strongly limiting the \h{number} of computations that can be performed. Perhaps, future quantum devices will open the frontiers to a new generation of neural networks based on quantum phenomena.
 \section{Conclusions} \label{conclusions}

Artificial intelligence has played a central role in academic and industrial debates in the last couple of decades, especially concerning machine learning and neural network techniques. Although these algorithms have shown an incredible generalization capability when adequately trained to solve a given problem, how and why these algorithms behave as they do is a topic that still dazzles scientists worldwide. 
Nevertheless, notwithstanding astonishing results, these algorithms will always and irremediably be tied to the classical interpretation of the real world. 

Differently, quantum computations might overcome such a limit and exploit phenomena such as superposition and entanglement that belong to the quantum realm. Although still at their dawn, quantum technologies represent a promising and fascinating alternative to classical computing techniques, perhaps realizing practical quantum supremacy soon.

Stemming from these considerations, we conceived this survey to offer both the neophyte and the more experienced reader insights into several fundamental topics in the quantum computation field such as the qubit, the Gate Model, and the Adiabatic Quantum Computation paradigms, to mention some. Moreover, noticing that the literature lacks a detailed discussion about the latest achievements concerning Quantum Perceptrons and Quantum Neural Networks, we gather, analyze and discuss state-of-the-art approaches related to these topics.
From our work, it is clear that quantum neural networks and algorithms, in general, are still far from proving decisive supremacy over the classical ones. Too often, such a goal has been claimed. Nevertheless, it is still an open quest. \h{It} is not clear if quantum devices will replace classical chips, becoming the core of a new generation of personal computers. However, recent improvements for quantum hardware and algorithms have opened the gates towards a new world, the quantum world, characterized by phenomena that were completely unknown until the last century.

\begin{acks}
The work was partially supported by H2020 project AI4EU under GA 825619, by~H2020 project AI4Media under GA 951911, by WAC@Lucca funded by Fondazione Cassa di Risparmio di Lucca. 
\end{acks}

\bibliographystyle{ACM-Reference-Format}
\bibliography{bibl_sorted_by_year_abbr}

\newpage
\appendix

    \section{APPENDIX: Table of Abbreviations Used }\label{appendix:table}
\begin{table}[h!]
\caption{Summary of abbreviations and acronyms used throughout this paper.}		
\label{tab:acronym}
\small
	\begin{tabular}{
	p{0.2\columnwidth} 
	p{0.43\columnwidth}
	p{0.2\columnwidth}
	}
		\toprule
		\textbf{Acronym} & \textbf{Definition} & \textbf{Section Introduced} \\        \midrule
      AI & Artificial Intelligence & \autoref{introduction}\\
      ANN & Artificial Neural Network & \autoref{quantum_neural_networks} \\
      AQC & Adiabatic Quantum Computation & \autoref{introduction}\\
      AT &  Adiabatic Theorem & \autoref{subsec:adiabatic_model}\\
      CNN & Convolutional Neural Network & \autoref{sec:qnn}\\
        DL & Deep Learning  & \autoref{introduction}\\
      DNN & Deep Neural Network & \autoref{introduction}\\
      NISQ & Noisy Intermediate-Scale Quantum &  \autoref{q_supremacy}\\
      GM & Gate Model & \autoref{introduction}\\
      ML & Machine Learning  & \autoref{introduction}\\ 
      NN & Neural Network & \autoref{introduction}\\
      {PQC} & {Parameterized Quantum Circuits} & {\autoref{sec:qnn}}\\
      QAE & Quantum AutoEncoders & \autoref{sec:qnn}\\
      QAOA & Quantum Approximate Optimization Algorithm & \autoref{sec:qo}\\
     QCNN &  Quantum Convolutional Neural Network & \autoref{sec:qnn}\\
     QGNN & Quantum Graph Neural Networks & \autoref{sec:qnn}\\
      QML & Quantum Machine Learning & \autoref{introduction}\\
      QNN & Quantum Neural Network & \autoref{introduction}\\
      QP &  Quantum Perceptron & \autoref{introduction}\\
      QPU&  Quantum Processing Unit & \autoref{introduction}\\
      QRAM & Quantum Random Access Memory & \autoref{sec:qnn} \\
      QVNN & Quanvolutional Neural Networks & \autoref{sec:qnn}\\
      RUS & Repeat-Until-Success & \autoref{sec:qp}\\
      qSLP & Single Layer Perceptron & \autoref{sec:qnn}\\
     TI & Trapped Ions & \autoref{sec:physical_realization}\\
      UAT & Universal Approximation Theorem & \autoref{sec:qnn}\\
      VQA & Variational Quantum Algorithm & \autoref{sec:qo}\\
      {VQE} & {Variational Quantum Eigensolver} & \autoref{sec:qo}\\
        \bottomrule        
	\end{tabular} 
\end{table}

\newpage
\section{APPENDIX: Background on Quantum Mechanics and Computation}\label{appendix}

\subsection{$\ket{\mathrm{ket}}$, $\bra{\mathrm{bra}}$, and $\H_{\mathrm{ilbert}}$ spaces}\label{app:dirac}
The \textit{Dirac} or \textit{bra-ket notation}~\citepA{dirac1939new} is a convenient and concise formalism introduced by the physicist Paul Dirac in quantum mechanics to aid algebraic manipulations on Hilbert spaces \h{and} their dual space. Before describing this notation, we recall the inner product and Hilbert space \h{concepts}.

An \textit{inner product} on a complex vector space $\mathcal{V}$ is a binary operation 
$(\cdot,\cdot):\mathcal{V} \times \mathcal{V} \to \mathbb{C}$ that satisfies the following properties\footnote{We follow the convention of defining an inner product as linear in the second component and \h{conjugate}-linear in the first component, as typically done in physics. However, please note that in mathematics\h{,} an inner product is often defined as linear in the first coordinate and \h{conjugate}-linear in the second coordinate.}
\begin{align*}
   &(\bb v, \alpha \bb w_1 + \beta \bb w_2)=\alpha(\bb v,  \bb w_1)+\beta (\bb v, \bb w_2)  &\text{\textit{linearity in the second argument  }}\\
   &(\bb v,\bb w)={(\bb w,\bb v)}^* &  \text{ \textit{conjugate symmetry}}\\
   & (\bb v,\bb v)\geq 0 \quad \text{(The equality holds if and only if $\bb v=\bb 0$)} & \text{ \textit{positive definiteness}}
\end{align*}
For example, in $\mathbb{C}^n$ the most used inner product between two column vectors $\bb{v}=[v_1,\dots, v_n]^T$ and $\bb{w}=[w_1,\dots, w_n]^T$ is defined as $\sum_{i=1}^n v_i^*w_i$, which is equivalent to the matrix multiplication between the transpose-conjugate vector of $\bb v$ and the vector $\bb w$.

An inner product naturally induces a norm $||\bb v||=\sqrt{(\bb v,\bb v)}$ and a metric $d(\bb v, \bb w)=|| \bb v- \bb w||$ on the vector space on which it is defined, which is referred to as an \textit{inner product space}. A \textit{Hilbert space} $\H$ is a inner product space which, as a metric space, is complete.
In the notation introduced by Dirac
\begin{itemize}
    \item a particular Hilbert-space vector, specified by a label $v$, is denoted by $\ket{v}$ (instead of using arrows ${\vec {v}}$ or boldface letters $\bb v$);
    \item 
    the dual vector to a Hilbert-space vector specified by the label $v$ is denoted by $\bra{v}$  (instead of using notation like $\bb v^\dagger$);
    \item the inner product between a pairs of vectors, labeled as $v$ and $w$, is compactly denoted by $\inprod{v}{w}$ (rather than $(\ket{v}, \ket{w})$ or other common notations like $(\bb v, \bb w)$ used above)\h{.}
\end{itemize}
Originally, Dirac proposed the words \quotes{\textit{bra}} and \quotes{\textit{ket}} as the names for new symbols $\langle$ and $\rangle$. However, we follow the common convention of using the terminology \textit{ket} for vector $\ket{v}$ and \textit{bra} for dual vector $\bra{v}$. It is worth mentioning that two general rules in connection with the Dirac notation should be noted: "\textit{any quantity in brackets $\langle \, \rangle$ is a number, and any expression containing an unclosed bracket symbol $\langle$ or $\rangle$ is a vector in Hilbert space}" \citepA{dirac1939new}. Therefore $\ket{v}$ is simply a vector in a Hilbert space, where $v$ is the label of the vector, and $\bra{v}$ is the complex conjugate transpose of $\ket{v}$, which is a vector of a different Hilbert space\footnote{Actually, the symbol $\bra{f}$ is more generally used to denote a functional $ f:\H\to \mathbb{C}$ specified by the label $f$. We recall that for a complex vector space $\mathcal{V}$ all the continuous conjugate-linear functional $f:\mathcal{V} \to \mathbb{C}$ form the so\h{-}called \textit{dual space} $\mathcal{V}^*$. By the \textit{Riesz representation theorem}, any element $f$ of the dual space of a Hilbert space can be represented as an inner product to some fixed vector. Therefore when working on Hilbert space $\H$ we can identify $\H$ with its dual $\H^*$ by using the one-to-one antilinear correspondence between vectors and continuous linear functionals $\ket{v} \to f_v$ where  $f_v$ is the functional
\begin{align*}
   f_v:\H &\to \mathbb{C}\\
 \ket{w} &\to \inprod{v}{w}   
\end{align*}
In other words, $\bra{v}$ is the functional (named $f_v$ above) associated to the vector $\ket{v}$. Moreover, using Dirac notation, the inner product between two vectors with labels $v$ and $w$ can be easily obtained by the graphically joining of $\bra{v}$ and $\ket{w}$.
}.
For example, for finite vector spaces, the ket $\ket{v}$ can be identified with a column vector and the bra $\bra{v}$ with a row vector. Therefore, expressions like $\ket{v}+\bra{v}$ are not valid as they have no meaning. Note that the $v$ inside the symbols $\ket{\,\, }$ and $\bra{\,\, }$ is just a label and that any label is valid, although\h{,} in the context of quantum mechanics\h{,} we often use Greek letters, such as $\phi$ and $\psi$, or binary labeling associated to a space basis. For example, if $\H$ has an indexed basis that, using the old notation, is indicated with $\{\bb e_1, \dots, \bb e_{n}\}$, then we indicate it with $\{\ket{e_1}, \dots, \ket{e_{n}}\}$ or more compactly $\{\ket{1}, \dots, \ket{n}\}$. {The \textit{canonical basis} of $\mathbb{C}^n$, also called \textit{computational basis}, is typically denoted $\{\ket{0}, \dots, \ket{n-1}\}$, where
$ \quad
\ket{0}=[ 1, 0 , \cdots , 0 ]^T, \quad \cdots, \quad 
\ket{n-1}=[ 0, \cdots 0, 1 ]^T
$}. To avoid ambiguity between vector basis and the zero vector, as an exception, we do not use the notation $\ket{0}$ for the zero vector that is denoted simply by $\bb 0$. 

We also observe that if we restrict our attention on the case of a finite-dimensional vector space $\mathcal{V}$ 
and we assume that for a given \textit{orthonormal basis}
$\{\ket{e_1}, \dots, \ket{e_{n}}\}$ the vectors $\ket{v}$ has the  coordinates 
 \begin{equation}
\begin{bmatrix} v_1 \\ \vdots \\ v_n \end{bmatrix} \qquad \qquad 
 \end{equation}
then 
\begin{itemize}
    \item $\ket{v}= v_1\ket{e_1} + \dots+ v_n\ket{e_n}$
     \item $\|\, \ket{v}\,\|=\sqrt{\sum_{i=1}^n |v_i|^2}$
     \item $\{\bra{e_1}, \dots, \bra{e_{n}}\}$ is an orthonormal basis for the dual space $\mathcal{V}^*$ 
    \item $\bra{v}=v_1^*\bra{e_1} + \dots+ v_n^*\bra{e_n}$, i.e.\h{,} it has the coordinates $\begin{bmatrix} v_1^*,  \cdots, v_n^* \end{bmatrix}$
    \item $\inprod{e_i}{e_j}=\delta_{ij}$ where $\delta_{ij}$ is the Kronecker delta, and $\inprod{v}{w}=\sum_{i=1}^n v_i^*w_i$
\end{itemize}
 
 Let $\mathcal{V}$ and $\mathcal{W}$ two Hilbert spaces with orthonormal basis  $\{\ket{e_j}\}_{j=1, \dots,n}$ and $\{\ket{f_i}\}_{i=1,\dots,m}$, respectively. Let $\ket{v}=\sum_{j=1}^n v_{j}\ket{e_{j}} \in \mathcal{V}$ and $\ket{w}=\sum_{{i}=1}^m w_{i}\ket{f_{i}}\in \mathcal{W}$. The \textit{outer product} of $\ket{v}$ and $\ket{w}$ is calculated as
   $$
    \outprod{v}{w}=\begin{bmatrix} v_1 \\ \vdots \\ v_n \end{bmatrix}\begin{bmatrix} w_1^*,\,  \cdots,\, w_m^* \end{bmatrix}=
    \begin{bmatrix} v_1w_1^* & \dots & v_1w_m^* \\ 
    \vdots & \ddots & \vdots  \\ 
    v_n w_1^*& \dots & v_n w_m^*  
    \end{bmatrix}
    $$
   and it corresponds to a linear operator that when applied to $\ket{u} \in \mathcal{V}$ acts as $\left(\outprod{v}{w}\right)\ket{u}=\inprod{w}{u}\, \ket{v}$. Thus, the outer product of a vector $\ket{v}$ with itself, that is $\outprod{v}{v}$, corresponds to the operator \h{projecting} any vector to the 1-dimensional subspace spanned by $\ket{v}$, \ie\h{,} it is an \textit{orthogonal projector}. 
  
 If $\hat{A}:\mathcal{V} \to \mathcal{W}$ is a linear operator represented by the $m\times n$ matrix $A=(a_{{i}{j}})$, 
\ie\h{,} $\hat{A}(\ket {e_{j}})=\sum_{i=1}^m a_{{i}{j}}\ket{f_{i}}$, then 
for any ket $\ket{v}=\sum_{{j}=1}^n v_{j}\ket{e_{j}}$ we have $\hat{A}(\ket {v})=\sum_{{i}=1}^m \sum_{{j}=1}^n a_{{i}{j}} v_j \ket{f_{i}}$, which is compactly denoted by $A\ket{v}$. Note that any matrix can be expressed as a linear combination of outer product between the basis vector:  
$A=\sum_{{i}=1}^m \sum_{{j}=1}^n a_{{i}{j}}\outprod{f_{i}}{e_{j}}$.
The adjoint of $\hat{A}$ is the linear operator $\hat{A}^\dagger: \mathcal{W} \to \mathcal{V}$ represented by the complex conjugate transpose of the matrix $A$, denoted by $A^\dagger$. 

An operator $\hat{T}:\mathcal{V} \to \mathcal{V}$, represented by the matrix $T$, is called \textit{\h{Hermitian}} (or self-adjoint) if $T^\dagger=T$,  \textit{unitary} if $T^\dagger=T^{-1}$, 
and \textit{normal} if $T^\dagger T= T T^\dagger$. These operators have nice properties that are often exploited in quantum mechanics. For example\h{,} all the eigenvalues of a \h{Hermitian} operator are real. Unitary operators preserve inner products between vectors \h{and} the norm of the vectors. Both \h{Hermitian} and unitary operators are normal. Moreover, every normal operator acting on a finite-dimensional Hilbert space $\mathcal{V}$ is diagonalizable, and the set of its eigenvectors \h{forms} an orthonormal basis of $\mathcal{V}$ (spectral theorem). \h{When discussing} the Hamiltonian of a system, observables, and measurements\h{, these properties should be kept in mind.} For more in-depth coverage of the basic concepts of linear algebra on Hilbert spaces, the reader is referred to \citepA{kaye2007introduction,nielsen2002quantum}.

Another operation often used in quantum mechanics is the \textit{tensor product}, which is crucial to describe quantum states of multiparticle systems. The tensor product of $\mathcal{V}$ and $\mathcal{W}$, denoted by $\mathcal{V} \otimes \mathcal{W}$, is a $n m$-dimensional Hilbert space for which the set $\{\ket{e_1} \otimes \ket{f_1},\, \ket{e_1} \otimes \ket{f_2},\, \dots, \ket{e_n} \otimes \ket{f_m} \}$ is an orthonormal basis, where   $\{\ket{e_j}\}_{j=1, \dots,n}$ and $\{\ket{f_i}\}_{i=1,\dots,m}$ are given orthonormal basis of $\mathcal{V}$ and $\mathcal{W}$. The tensor product of two vectors $\ket{v}=\sum_{j=1}^n v_{j}\ket{e_{j}}$ and $\ket{w}=\sum_{{i}=1}^m w_{i}\ket{f_{i}}$ 
 is equal to 
\begin{equation}
    \ket{v} \otimes \ket{w}=\sum_{j=1}^n \sum_{i=1}^m v_{j} w_{i} \, \ket{e_{j}} \otimes \ket{f_{i}}
\end{equation}
that is the coordinate of the vector $\ket{v} \otimes \ket{w}$ can be calculated explicitly via the left Kronecker product of the coordinates of the vectors $\ket{v}$ with the coordinates of $\ket{w}$.
Note that the abbreviated notations $\ket{v}\ket{w}$, $\ket{v,w}$ or even $\ket{vw}$ are often used to denote the tensor product $ \ket{v} \otimes \ket{w}$. If $\hat{A}:\mathcal{V}\to \mathcal{V}$ and $\hat{B}:\mathcal{W}\to \mathcal{W}$ are linear operators represented by the matrices $A$ and $B$, respectively, then $\hat{A}\otimes \hat{B}$ is the \h{linear} operator on $\mathcal{V} \otimes \mathcal{W}$ defined by 
$
\hat{A}\otimes \hat{B} ( \ket{v} \otimes \ket{w})= A \ket{v} \otimes B \ket{w}
$, whose matrix is given by the left Kronecker product of the matrix A with the matrix B. Note that $(A \otimes B)^T=A^T \otimes B^T$, $(A \otimes B)^\dagger=A^\dagger \otimes B^\dagger$, and  thus $(\ket{v} \otimes \ket{w})^\dagger=\ket{v}^\dagger \otimes \ket{w}^\dagger=\bra{v} \otimes \bra{w}$. 

\subsection{Postulates of Quantum Mechanics}\label{app:postulates_quantum_mechanics}
The four postulates of quantum mechanics describe the behavior of an \textit{isolated} system, which is an ideal physical system that does not interact with its environment. In particular, they describe the state space of an isolated system, its evolution with time, how information is extracted from the isolated system by the interaction with an external system, and the state of a composite system in terms of its parts.

\subsubsection{State Space}
\begin{post} 
At each instant, the \emph{state} of any isolated physical system is completely described by a unit vector of a Hilbert space, also known as the \emph{state space}.
\end{post}
Quantum mechanics does not describe which is the state space of a particular physical system but rather assure us that there exists a Hilbert space whose vectors can be used to describe each state of the system. 
Interestingly, as the space of the states comes equipped with an inner product, we have a way to associate a complex number to any two states, and thus, as we see later, a way to extract information from an isolated system.
In theory, the state space may be infinite-dimensional. However, realistic models of quantum computation usually use states described by vectors in a finite-dimensional Hilbert space. In~\autoref{sec:qubit}, we introduce the concept of a qubit, \h{ the basic unit of quantum information. Here,} we observe that any physical system whose state space can be described by $\mathbb{C}^2$ can serve as a physical realization of a qubit. We referred to these kinds of systems as quantum \textit{two-level} systems, as their state can be described by a vector in 2-dimensional Hilbert space. \h{A further discussion on the physical realization of the qubit is provided in \autoref{sec:physical_realization}.}

\subsubsection{\h{Time} Evolution}
Since a physical system \h{changes} in time, a state vector $\psi$ of the system can be viewed as a function of time, \ie, $\psi(t)$. The second postulate of quantum mechanics asserts that the evolution of a state vector of an isolated system is linear, and it can be described \h{using} unitary operators:
\begin{post} 
The time evolution of an isolated system is described by a unitary operator. That is, the state $\ket{\psi(t_1)}$ of the system at time $t_1$ is related to the state $\ket{\psi(t_2)}$ of the system at time $t_2$ by an unitary operator $\hat{U}$ which depends on the times $t_1$ and $t_2$:
$$
\ket{\psi(t_2)}= \hat{U} \ket{\psi(t_1)}
$$
\end{post}

Note that the postulate does not specify how to compute $\hat{U}$. Instead, it asserts that for any discrete-time evolution of the isolated system, there exists such a unitary operator that describes the dynamics of the system state.
\h{Some models, like the Adiabatic Quantum Computing (discussed in \autoref{subsec:adiabatic_model}), }
allow for continuous-time evolution of an isolated system described by the \h{\textit{Schr{\"o}dinger equation} 
\begin{equation}
    {i \hbar\frac{d \ket{\psi(t)}}{dt}=\Ham \ket{\psi(t)},}
\end{equation}
 where $\Ham$ is the \textit{Hamiltonian} of the system, which is a Hermitian operator representing the total energy function for the system, and $\hbar$ is the Planck's constant.}
 Considering a \h{time-independent} Hamiltonian, we have $\ket{\psi(t)}=\hat{U}(t)\ket{\psi(0)}$ where $\hat{U}(t)= \exp{(-i{t}\Ham/{\h{\hbar}} )}$, and $\ket{\psi(0)}$ is the state at $t=0$. Thus, for those cases, the Evolution Postulate follows from the  Schr{\"o}dinger equation.
%
Also, note that for the case of a two-level system, the Evolution postulate assures us that a specific discrete-time evolution of the system is described by a unitary operator on $\mathbb{C}^2$ acting on a qubit, referred to as \textit{one-qubit gate}. 

\subsubsection{Measurements}\label{app:measurements}
When we perform any measurement/observation on an isolated system, we interact with the system itself, which is no longer isolated, and thus the Evolution Postulate is no longer appropriate for describing its evolution. 
The following postulate explains the effects of measurements of quantum systems:   
\begin{post}[Quantum measurement]
A \emph{measurement} with outcome set $S$ is a collection of $|S|$ \emph{measurement operators} $\{\hat{M}_k\}_{k\in S}$, each acting on the state space of the system being measured, that satisfy the \emph{completeness equation}
\begin{equation}
  \sum_{k\in S} \hat{M}_k^\dagger \hat{M}_k= \hat{I},  
\end{equation}
where $\hat{I}$ is the identity operator and each index $k$ refers to an outcome that may occur in the measurement process.

If the quantum system is in the state $\ket{\psi}$ immediately before the measurement, then the \textit{probability} of outcome $k\in S$ is given by 
\begin{equation}
p (k)= \bra{\psi} \hat{M}_k^\dagger \hat{M}_k\ket{\psi}=\| \hat{M}_k\ket{\psi} \|^2
\end{equation}
and the corresponding \textit{post-measurement state} $\ket{\psi_k}$ is 
\begin{equation}
 \ket{\psi_k}= 
 \dfrac{\hat{M}_k \ket{\psi}}{\| \hat{M}_k\ket{\psi} \|}  
\end{equation}
\end{post}

Note that the completeness equation expresses that, for any state $\ket{\psi}$, the probabilities of all the outcomes sum to 1: $\sum_{k\in S} p(s)= \sum_{k\in S}\bra{\psi} \hat{M}_k^\dagger \hat{M}_k\ket{\psi}= 1$. Moreover, two states that differ only by a global phase, \eg, $\ket{\psi}$ and $e^{i\varphi} \ket{\psi}$ are equivalent as the  statistics of any measurement we could perform on the state  $e^{i\varphi} \ket{\psi}$ is the same as the one for the state  $\ket{\psi}$ since $\| \hat{M}_k e^{i\varphi}\ket{\psi} \|^2=\| \hat{M}_k\ket{\psi} \|^2$ for every $k$. \h{The operators $\hat{M}_k^\dagger \hat{M}_k$ are called \textit{Positive-Operator-Valued Measure (POVM)}. }

An important class of measurements \h{consists of} \textit{projective measurements} based on orthogonal projections. We recall that a projection is an operator $\hat{P}$ such that $\hat{P}^2=\hat{P}$, and $\hat{P}^\dagger=\hat{P}$.  A projective measurement is a collection of orthogonal projections $\{\hat{P}_{k}\}_{k}$ that decompose the Identity operator as $\hat{I}=\sum_k \hat{P}_k$. Such a measurement outputs $k$ with probability $p(k)=\bra{\psi} \hat{P}_k\ket{\psi}$  and leave the system in the normalized state $\dfrac{\hat{P}_k \ket{\psi}}{\| \hat{P}_k\ket{\psi} \|}$. Projective measurements are often described in terms of an \textit{observable}, which is a Hermitian operator on the state space of the system being observed.  From the Spectral Theorem, we know that any observable $\hat{M}$ has a spectral decomposition $\hat{M}=\sum_k \lambda_k \hat{P}_k$, where $\hat{P}_k$ is the orthogonal projection onto the eigenspace of $\hat{M}$ with real eigenvalue $\lambda_k$. The possible outcomes of the measurements are the eigenvalues of $\hat{M}$, where the probabilities of getting result \h{$\lambda_k$} is given by $p(\h{\lambda_k})=\bra{\psi} \hat{P}_k\ket{\psi}\h{=\|\hat{P}_k\ket{\psi}\|^2}$. 
The expected \h{value} of a projective measurement of a state $\ket{\psi}$, \h{i.e., the expected value of the observable $\hat{M}$}, can be easily calculated as 
\begin{equation}
    \h{\langle\hat{M}\rangle}=\sum_k\lambda_k p(\h{\lambda_k})= \bra{\psi}\hat{M}\ket{\psi}.
\end{equation}
\h{Equivalently, for a system in a pure state represented by the density  $\rho=\outprod{\psi}{\psi}$  the expected value of the observable $\hat{M}$ can be expressed as $\langle\hat{M}\rangle=\mathrm{Tr}\{\rho \hat{M}\}$, where $\mathrm{Tr}$ is the trace operator.}

We also mention that the term \textit{orthogonal measurement} is used to refer to a  projective measurement whose  operators are of the type 
\begin{equation}
    \hat{P}_k=\outprod{e_k}{e_k}
\end{equation}
where $\{\ket{e_1}, \dots, \ket{e_n}\}_i$ is \h{an} orthonormal basis of the state space. When measuring the state $\ket{\psi}$, the probability of outcome $k\in \{1, \dots, n\}$ is 
\begin{equation}
    p(k)=|\inprod{\psi}{e_k}|^2
\end{equation}
and the corresponding post-measurement state is $\ket{\psi_k}=\ket{e_k}$.
For example, if the state $\ket{\psi}= \sum_k \alpha_k \ket{e_k}$ is provided as input to the orthogonal measurement, it will output label $k$ with probability $|\alpha_k|^2$ and leave the system in state $\frac{\alpha_k}{|\alpha_k|}\ket{e_k}$, which is equivalent to the state $\ket{e_k}$ since $\frac{\alpha_k}{|\alpha_k|}$ has modulus one.

Note that the space state basis used in the measurement can play a crucial role in \h{gaining measurement information}. For example, the Hadamard basis states $\ket{+}=\frac{1}{\sqrt{2}}(\ket{0}+\ket{1})$ and $\ket{-}=\frac{1}{\sqrt{2}}(\ket{0}-\ket{1})$ cannot be distinguished by an orthogonal measurement based on the computation basis $\{\ket{0}, \ket{1}\}$ since both outcomes \quotes{$0$} and \quotes{$1$} occur with probability $1/2$ (\ie, $ |\inprod{+}{0}|^2=|\inprod{-}{0}|^2=1/2$). However\h{,} if we use the states themselves as the measurement basis, then when measuring  $\ket{+}$ we always get the outcome \quotes{$+$} and when measuring $\ket{-}$ we always get the outcome \quotes{$-$} (\ie, $p(+)=|\inprod{+}{+}|^2=1$ and $p(-)=|\inprod{-}{-}|^2=1$). In general, it is always possible to consider a measurement that \h{distinguishes} orthonormal states, but it can be proved that there is no quantum measurement capable of distinguishing \h{non}-orthogonal states \citepA[pp. 87]{nielsen2002quantum}.

\subsubsection{Composition of Systems}
So far, we have considered the case of a single system only. However, we may be interested in a composite system of two or more \h{different} quantum systems. The fourth postulate of quantum mechanics describes the state space of composite systems in terms of the states of \h{their} component systems.

\begin{post}
The state-space of a composite physical system is the tensor product of the state spaces of the individual component physical systems. Moreover, if $n$ physical system are treated as one combined system and $\ket{\psi_i}$ is the state of the $i$-th system, then the state of the composite system is 
\begin{equation}
\ket{\psi_1} \otimes \dots \otimes \ket{\psi_n}    
\end{equation}
\end{post}
The joint state is often compactly denoted as $ \ket{\psi_1, \dots, \psi_n}$ where the symbol $\otimes$ is omitted. The postulate asserts that given the states of the component systems, we can compute the state of the combined system by using the tensor product. However, not all states of a combined system can be separated into the individual components' tensor product: 
A combined system is called \textit{entangled} if it cannot be expressed as the tensor product of states of the component systems.

\subsection{Physical Realization of Qubits}\label{sec:physical_realization}

\h{Since the qubit represents the fundamental information carrier for a quantum computer, its physical design and realization represented a milestone concerning the outspread of such technologies. Albeit the topic is fascinating, a comprehensive discussion about all the available technologies and implementations of physical qubits is out of scope for this survey. Nonetheless, we refer the curious reader to several specialized reviews available in the literature:~\citepA{wendin2017quantum,gu2017microwave,gambetta2017building,krantz2019quantum,kjaergaard2020superconducting}. 

However, in what follows, we briefly report about three different qubit designs, namely: \textit{trapped ions}~\citepA{cirac1995quantum}, \textit{transmon}~\citepA{peterer2015coherence}, and \textit{flux}~\citepA{harris2010demonstration}. We made such a selection based on the following reasons. Trapped ions offer very high fidelity for both single- and two-qubit entangling gates~\citepA{harty2014high,gale2020optimized,ballance2016high,levine2018high,benhelm2008towards,gyongyosi2019survey}, while superconducting-based qubits are typically available in free-access quantum computing platforms\footnote{\url{https://quantum-computing.ibm.com/}}$^{,}$\footnote{\url{https://cloud.dwavesys.com/leap}} for researchers. 

Trapped Ions (TI) based quantum computation dates back to the '90~\citepA{cirac1995quantum} when qubits were realized as single ions confined in radiofrequency (RF) traps~\citepA{neuhauser1980localized}. 
In TI, ions' internal electronic states are exploited to generate the two-level system required to represent qubits. Depending on the electronic states used, trapped ions-based qubits are typically divided into four categories, namely: optical, fine structure, Zeeman, and hyperfine. 

As mentioned before, flux~\citepA{mcgeoch2020theory} and transmon~\citepA{smith2018automated,barr2021characterization} qubits-based QPUs are the most famous in terms of free availability to end-users to develop custom applications. These two technologies account for a different theoretical formulation and physical implementation of the two-state system. Indeed, from a theoretical perspective, in the first case, we talk about the spin interpretation while, in the second case, we find the oscillator interpretation of a quantum bit.

Compared to flux and transmon qubits, TIs offer much longer coherence times ranging from $\sim1$ up to $\sim10$ minutes depending on the type of qubit and whether or not dynamical coupling techniques are used~\citepA{wang2017single,bollinger1991303,harty2014high}. Interestingly, TI-based qubits benefit from the fact that ions of a given isotope are always equal. Thus, all qubits show the very same properties. Differently, flux and transmon qubits have properties that might change from one another due to fabrication defects, thus typically requiring a more significant number of calibration steps. Another difference among these technologies stems from the number of qubits that can be implemented on a single QPU. Indeed, while for superconducting-based technologies, it has been possible to construct quantum processors with up to several thousands of qubits, in the case of trapped ions, such a number decreases down to 20~\citepA{friis2018observation}. However, in~\citepA{pagano2018cryogenic}, the authors designed a cryogenic ion trapping system for large-scale quantum simulation of spin models, capable of trapping over 100 ${}^{171}$Yb${}^{+}$ ions in a linear lattice. 

However, it is fundamental to state that the number of qubits itself is not a proper metric to compare the computational capability for a given quantum processor. There are available in the literature different metrics aiming at such a goal. For example, IBM proposed the so-called quantum volume~\citepA{cross2019validating,jurcevic2021demonstration}, a quantity that can be used to quantify the largest random circuit of equal width and depth that the computer successfully implements~\citepA{cross2019validating}.

}

\subsection{The Bloch Sphere representation} \label{sec:bloch}
\h{A \textit{qubit state} is a unit vector in $\mathbb{C}^{2}$ that, using the Dirac notation, can be expressed as $\ket{\psi}= \alpha\, \ket{0} + \beta\, \ket{1}$, where the \textit{amplitudes} $\alpha$ and $\beta$ are complex numbers such that $|\alpha|^2+|\beta|^2=1$. Since a qubit has more states available than simply two levels, often it is useful to visualize it as a point of a unit sphere in a Euclidean space.
}
As shown in the following, quantum states can be put in correspondence with the points of the so-called \textit{Bloch sphere}, which is a unit sphere in a three-dimensional Euclidean space with the north and south pole corresponding to the computational basis states $\ket{0}$ and $\ket{1}$, respectively.

\begin{figure}[tbp] 
\centering
{\includegraphics[trim= 35mm 35mm 35mm 35mm,clip,width=0.35\linewidth]{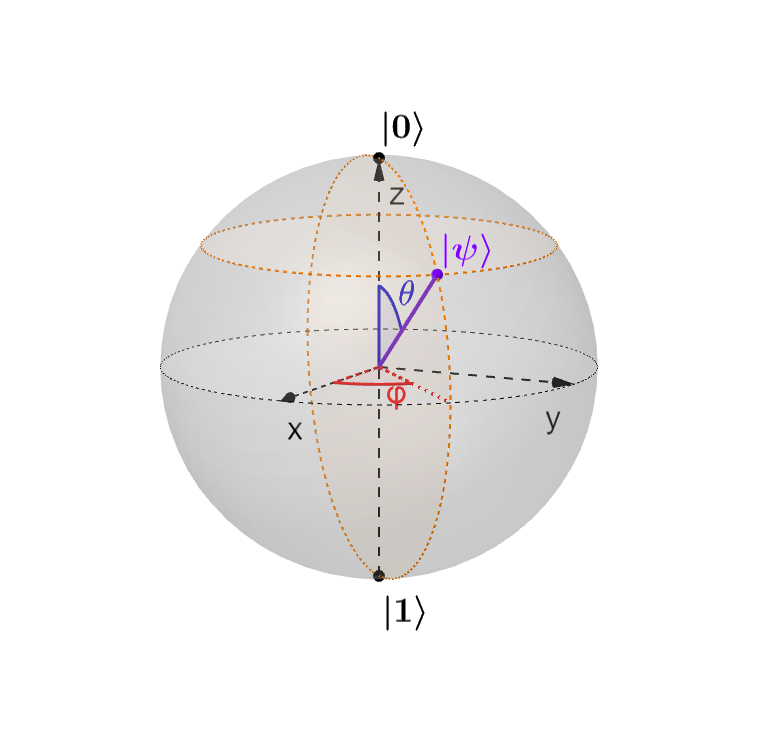}} 
\caption{
Bloch sphere representation of a qubit: 
a quantum state  $\ket{\psi}= \alpha\, \ket{0} + \beta\, \ket{1}$, where $\alpha=|\alpha| e^{i\varphi_\alpha}$ and $\beta=|\beta| e^{i\varphi_\beta}$, is represented as the point of the Bloch sphere with spherical coordinates given by the angles $\theta$ and $\varphi$, where $\theta=\arccos{(|\alpha|^2-|\beta|^2)}\in [0,\pi]$ and $\varphi=\varphi_\beta-\varphi_\alpha \in [0, 2\pi)$. Vice versa, a point of the unit sphere with spherical coordinates
$x=\sin \theta \cos \varphi$, $y=\sin \theta \sin \varphi$, and  $z= \cos \theta$ corresponds to the quantum state $\ket{\psi}\equiv\cos\frac{\theta}{2}\, \ket{0}+e^{i\varphi}\sin\frac{\theta}{2}\, \ket{1}$.   
}\label{fig:bloch}
\end{figure}

An immediate geometrical interpretation of a state as a vector of a unit sphere can be obtained by the map 
$$[ \alpha ,  \beta]\in \mathbb{C}^2 \to [\text{Re}(\alpha) ,\text{Im}(\alpha) , \text{Re}(\beta)  ,\text{Im}(\beta) ] \in \mathbb{R}^4$$ but this interpretation is not \h{helpful} for visualization purposes as it relies on a four-dimensional Euclidean space. However, since we are interested only in unitary 2-dimensional complex vectors, we can exploit the condition $|\alpha|^2+|\beta|^2=1$ to reduce the parameters needed to describe the quantum state, obtaining a three-dimensional representation that can be easily visualized. Specifically, we could use the following transformation 
\begin{equation}
\begin{cases}
x= 2\, \text{Re}(\alpha^*\beta) \\
y= 2\, \text{Im}(\alpha^*\beta)\\ 
z= |\alpha|^2-|\beta|^2\\ 
\end{cases}
\end{equation} 
It can be easily proved that for $|\alpha|^2+|\beta|^2=1$ then $x^2+y^2+z^2=1$. Thus, the above transformation maps a state $\ket{\psi}= \alpha\, \ket{0} + \beta\, \ket{1}$ onto a point of a \h{unit} sphere in $\mathbb{R}^{3}$, which is referred to as the {Bloch sphere} (\autoref{fig:bloch}). 
Note that any point of a \h{unit} sphere in $\mathbb{R}^{3}$ can be described using the spherical coordinate system
$$
\begin{cases}
x= \sin \theta \cos \varphi \\
y= \sin \theta \sin \varphi \\
z= \cos \theta
\end{cases} \qquad \theta \in [0,\pi], \quad \varphi \in [0, 2\pi)
 $$
 which further reduces the number of parameters needed to describe the quantum state. In fact, given the quantum state $\ket{\psi}= \alpha\, \ket{0} + \beta\, \ket{1}$, where $\alpha=|\alpha| e^{i\varphi_\alpha}$, $\beta= |\beta| e^{i\varphi_\beta}$, and $|\alpha|^2+|\beta|^2=1$, the spherical coordinates $(\theta, \varphi)$ can be computed by solving the system  
  \begin{equation}\label{eq:sist}
  \begin{cases}
 \sin \theta \cos \varphi = 2\, \text{Re}\left(\alpha^*\beta\right)\\
 \sin \theta \sin \varphi = 2\, \text{Im}\left(\alpha^*\beta \right)\\
 \cos \theta=|\alpha|^2-|\beta|^2
\end{cases}
 \end{equation}
 Since $|\alpha|^2-|\beta|^2=2|\alpha|^2-1=1-2|\beta|^2$, from the last condition in \autoref{eq:sist} we obtain that
\begin{align}\label{eq:aplphabeta}
|\alpha| = \sqrt{\frac{1}{2}+ \frac{1}{2}\cos \theta }= \cos\frac{\theta}{2}, \qquad
|\beta| = \sqrt{\frac{1}{2}- \frac{1}{2}\cos \theta}= \sin\frac{\theta}{2}
\end{align} 
Moreover, by combining \autoref{eq:aplphabeta} and the first two conditions in \autoref{eq:sist} we obtain $\varphi=\varphi_\beta-\varphi_\alpha$ because
\begin{equation}
 2\alpha^*\beta=2 |\alpha||\beta|e^{i(\varphi_\beta-\varphi_\alpha )}= 2\cos\frac{\theta}{2}\sin\frac{\theta}{2}e^{i(\varphi_\beta-\varphi_\alpha)}=\sin\theta e^{i(\varphi_\beta-\varphi_\alpha)}.  
\end{equation} 
Therefore, the spherical coordinates associated \h{with} the quantum state $\ket{\psi}= \alpha\, \ket{0} + \beta\, \ket{1}$ are 
\begin{align}
\theta=\arccos{(|\alpha|^2-|\beta|^2)}, \qquad \varphi=\varphi_\beta-\varphi_\alpha
\end{align}

Vice versa, any point of the Bloch sphere with spherical coordinates
$(\theta, \varphi)$ can be mapped to the quantum state $\cos\frac{\theta}{2}\, \ket{0}+e^{i\varphi}\sin\frac{\theta}{2}\, \ket{1}$.
At first thought, since $\cos\frac{\theta}{2}$ is a real number, it might seem that such a mapping from the Bloch sphere to the quantum state space is not surjective because the amplitudes $\alpha$  and $\beta$ of a generic quantum state $\ket{\psi}=\alpha\, \ket{0} + \beta\, \ket{1}$ are both complex numbers.
However, it should be noted that since the magnitudes $|\alpha|$ and $|\beta|$ are \textit{positive} real \h{numbers} constrained to lie on the \h{unit} circle then they can be specified by a single angle, \ie, there \h{exists} a $\gamma \in [0,\frac{\pi}{2}]$ such that $|\alpha|=\cos{\gamma}$ and $|\beta|=\sin{\gamma}$, or equivalently, $|\alpha|=\cos\frac{\theta}{2}$ and $|\beta|=\sin\frac{\theta}{2}$ for $\theta=2\gamma \in [0,\pi]$. 
Thus, we have $\alpha=e^{i\varphi_\alpha}\cos\frac{\theta}{2}$ and $\beta=e^{i\varphi_\beta}\sin\frac{\theta}{2}$ for some $\varphi_\alpha,\varphi_\beta \in [0,2\pi)$. It follows that any state $\ket{\psi}$ can be rewritten as
\begin{align}
 \ket{\psi}=e^{i\varphi_\alpha}\left(\cos\frac{\theta}{2}\, \ket{0}+e^{i(\varphi_\beta-\varphi_\alpha)}\sin\frac{\theta}{2}\, \ket{1}\right) \qquad \text{with} \quad \varphi_\alpha,\varphi_\beta \in [0,2\pi), \, \theta \in [0,\pi]
\end{align}
where $e^{i\varphi_\alpha}$ is called \textit{global phase factor} and $e^{i(\varphi_\beta-\varphi_\alpha)}$ is the \textit{relative phase factor}. As the global phase has no observable effect (\ie, quantum states $\ket{\psi}$ and $e^{-i\varphi_\alpha}\ket{\psi}$ are indistinguishable when performing a measurement\footnote{The quantum states $\ket{\psi}$ and $e^{i\gamma}\ket{\psi}$ are indistinguishable when performing a measurement because they have the same probabilities of observing the computational basis states, \ie $|e^{i\gamma}\alpha|^2=|\alpha|^2$ and $|e^{i\gamma}\beta|^2=|\beta|^2$ for any $\gamma \in [0, 2\pi)$.}) it can be ignored. 
In other words, any quantum state $\ket{\psi}$ can be expressed as
\begin{equation}
     \ket{\psi} \equiv \cos\frac{\theta}{2}\, \ket{0}+e^{i\varphi}\sin\frac{\theta}{2}\, \ket{1}
\end{equation}
where $\theta \in [0,\pi]$ and $\varphi \in [0,2\pi)$ are equal, respectively, to the polar and azimuthal angles of the Bloch Sphere representation.

Note that antipodal points in the Bloch sphere correspond to orthogonal states. In fact, two states $\ket{\psi_1}=\cos\frac{\theta_1}{2}\,\ket{0}+e^{i\varphi_1}\sin\frac{\theta_1}{2}\, \ket{1}$ and $\ket{\psi_2}=\cos\frac{\theta_2}{2}\,\ket{0}+e^{i\varphi_2}\sin\frac{\theta_2}{2}\, \ket{1}$, with $\theta_1, \theta_2\in [0,\pi]$ and $\varphi_1, \varphi_2\in [0,2\pi)$, are orthogonal if and only if $\inprod{\psi_1}{\psi_2}=0$. It can \h{be} easily proved that the orthogonality condition holds if and only if either $(\theta_1,\theta_2)=(0,\pi)$,  $(\theta_1,\theta_2)=(\pi,0)$, or $\theta_1 +\theta_2=\varphi_2-\varphi_1=\pi$, which \h{precisely} identifies antipodal points in the sphere.

It is worth mentioning that the normalized sum of the computational basis vectors is called \textit{diagonal state}, which is indicated by $\ket{+}=\dfrac{1}{\sqrt{2}}\ket{0}+\dfrac{1}{\sqrt{2}}\ket{1}$. It corresponds to the intersection of the positive x-axis and the Bloch sphere. The state $\ket{-}=\dfrac{1}{\sqrt{2}}\ket{0}-\dfrac{1}{\sqrt{2}}\ket{1}$ corresponds to the intersection point of the negative x-axis and the sphere. The intersection points of the y-axis and the sphere are the states $\dfrac{1}{\sqrt{2}}\ket{0} \pm \ii \dfrac{1}{\sqrt{2}}\ket{1}$.  The basis formed by $\{\ket{+}, \ket{-}\}$ is referred to as \textit{Hadamard basis}.

\begin{figure}[tb] 
\centering
{\includegraphics[width=0.5\linewidth]{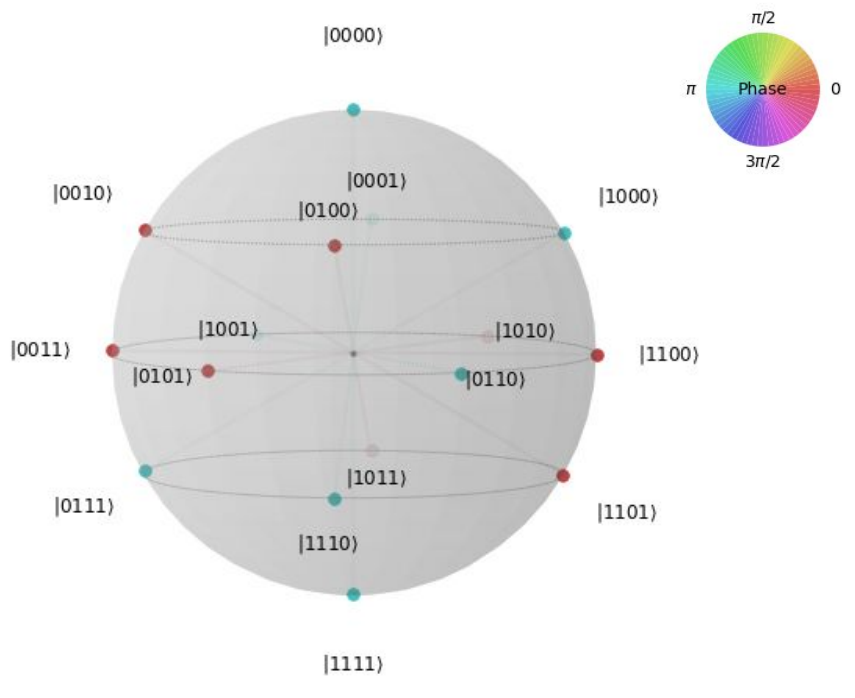}}
\caption{Q-sphere for a 4-qubit system. The represented state is the uniform superposition of all possible states of the 4-qubit system.}\label{fig:q_sphere}
\end{figure}
The Bloch sphere can be generalized to $n$-qubit systems and, in such a case, it is named \textit{Q-sphere}\footnote{\url{https://quantum-computing.ibm.com/composer/docs/iqx/visualizations}}. 
Concerning multi-qubit systems, the states 
$\ket{0}^{\otimes n }$ and $\ket{1}^{\otimes n}$
sit at the north and south pole, respectively, analogously to the single-qubit case. All the other states are placed along parallels with an increasing number of 1's from top to bottom.
For example, in~\autoref{fig:q_sphere}, we report the Q-sphere for a 4-qubit system in the uniform superposition of all the possible states.
Specifically, the size of the colored circles represents the probability amplitude for each state, which is the same for all of them, while the color represents the phase. Finally, we notice that the states $\ket{0000}$ and $\ket{1111}$ sit at the north and south pole while all the other states are distributed along parallels grouped by the same number of 1's (increasing from top to bottom), as mentioned above.
\newpage
\subsection{Frequently used single and multi-qubit gates}\label{sec-gates_tables}

\h{In~\autoref{tab:gates_examples} and ~\autoref{tab:gates_examples2}, we summarize some frequently used single- and multi-qubit gates, respectively. Please refer to \autoref{gate_model} for the definition of gates and Gate Computational Model.}

\renewcommand{\arraystretch}{1.2}
\begin{table}[tb]
	\caption{Basic one-qubit gates: Notations typically used in literature, the corresponding matrices with respect to the basis $\{\ket{0}, \ket{1}\}$, the action on a quantum state, and the corresponding action on the spherical coordinates in the Bloch sphere representation.}	\label{tab:gates_examples}
    \[\setcellgapes{3pt}\makegapedcells 
\footnotesize
	\begin{tabular}{p{0.05\columnwidth}cp{0.06\columnwidth}p{0.1\columnwidth}p{0.21\columnwidth}p{0.37\columnwidth}}
		\toprule
	\textbf{Name} & {\textbf{Symbols}} & {\textbf{Circuit}} & \textbf{Matrix}   &\textbf{Action on $\ket{\psi}=\alpha\ket{0}+ \beta \ket{1}$} & \textbf{Action on coordinates $(\theta,\varphi)$ of the Bloch sphere representation}  \\
		\midrule
       Identity&{$I$}   &
        {\Qcircuit @C=1em @R=.7em {& \gate{\,I\,} & \qw}}  &
             $ \begin{bmatrix}
           \,\, 1& 0 \,\,\\ 
          \,\, 0 & 1\,\,    
            \end{bmatrix}$ & $I\ket{\psi}= \ket{\psi}$
            & \makecell{$I(\theta,\varphi)=(\theta,\varphi)$}
            \\ 
            \hline 
         \makecell{Pauli-X\\(\textit{bit flip},\\ \textit{NOT}) }&  $X$, $\paul x$ &
        \makecell{{\Qcircuit @C=1em @R=.7em {& \gate{X} & \qw}} 
         \\or
   \\
         {\Qcircuit @C=1em @R=.7em {& \targ & \qw}}
         }
       &
             $ \begin{bmatrix}
          \,\, 0 & 1 \,\, \\ 
          \,\, 1 & 0  \,\,
            \end{bmatrix}$ & $X\ket{\psi}=\beta\ket{0}+ \alpha \ket{1}  $
            &
            \makecell{$X(\theta,\varphi)=(\pi-\theta,2\pi-\varphi)$ \\
            \textit{rotation around the $x$-axis by $\pi$ radians}}
            \\ 
            \hline 
           Pauli-Y & $Y$, $\paul y$ &
        {\Qcircuit @C=1em @R=.7em {& \gate{Y} & \qw}}  &
             $ \begin{bmatrix}
            0 &  -\ii \\ 
            \ii &   0 
            \end{bmatrix}$ & $Y\ket{\psi}= -\ii \beta\ket{0}+ \ii \alpha \ket{1}$ 
            &
             \makecell{$Y(\theta,\varphi)=(\pi-\theta,\pi-\varphi \,( \text{mod } (2\pi))$  \\
            \textit{rotation around the $y$-axis by $\pi$ radians}}
            \\ 
            \hline 
        \makecell{Pauli-Z \\ (\textit{phase flip})} &  $Z$,   $\paul z$ & 
        {\Qcircuit @C=1em @R=.7em {& \gate{Z} & \qw}} 
    &
             $ \begin{bmatrix}
           1 & 0 \\ 
           0 & -1   
            \end{bmatrix}$ & $Z\ket{\psi}=\alpha\ket{0}- \beta \ket{1} $
            &
              \makecell{ $Z(\theta,\varphi)=(\theta,\varphi+\pi \,( \text{mod } (2\pi))$  \\
            \textit{rotation around the $z$-axis by $\pi$ radians}}
            \\ 
            \hline 
            Phase &{$S$} & {\Qcircuit @C=1em @R=.7em {& \gate{S} & \qw}}  &
            $ \begin{bmatrix}
           1 & 0 \\ 
           0 & \ii  
            \end{bmatrix}$
            & 
            $S\ket{\psi}=\alpha\ket{0}+ \ii\beta\ket{1} $
            &  
            \makecell{ $S(\theta,\varphi)=(\theta,\varphi+\pi/2 \,( \text{mod } (2\pi))$  \\
            \textit{rotation around the $z$-axis by $\pi/2$ radians}}
            
            \\ \hline
             $\pi/8$ & {$T$} & {\Qcircuit @C=1em @R=.7em {& \gate{T} & \qw}}  &
        $ \begin{bmatrix}
           1 & 0 \\ 
           0 & e^{\ii\pi/4}  
            \end{bmatrix}$
             & 
             $T\ket{\psi}=\alpha \ket{0}+ e^{\ii\pi/4}\beta\ket{1} $
             &
            \makecell{  $T(\theta,\varphi)=(\theta,\varphi+\pi/4 \,( \text{mod } (2\pi))$  \\
            \textit{rotation around the $z$-axis by $\pi/4$ radians}}
             \\ \hline
             Hadamard & {$H$} & {\Qcircuit @C=1em @R=.7em {& \gate{H} & \qw}}  & $ \dfrac{1}{\sqrt{2}}\begin{bmatrix}
           1 & 1 \\ 
           1 & -1   
            \end{bmatrix}$
             & 
             $H\ket{\psi}=\dfrac{\alpha+\beta}{\sqrt{2}}\ket{0}+ \dfrac{\alpha-\beta}{\sqrt{2}}\ket{1} $
             &  
           $
           H(\theta,\varphi)= 
            \begin{cases}
             (\varphi,\frac{3}{2}\pi)  \quad \text{if} \,\, \theta= \frac{\pi}{2}, \, 0\leq \varphi \leq \pi\\
              (2\pi-\varphi,\frac{\pi}{2})  \quad \text{if} \,\, \theta= \frac{\pi}{2}, \, \pi< \varphi < 2\pi\\
             \left(\cos^{-1}(\sin\theta\cos\varphi ), \cos^{-1}\left(\frac{\cos \theta}{\sin(\cos^{-1}(\sin\theta\cos\varphi ))}\right)\right)\\ \qquad \qquad \qquad \qquad \qquad \qquad\quad\text{otherwise}
             \end{cases}
             $
             
             
            \textit{rotation around the y-axis by $\pi/2$ radians} 
           \textit{followed by a rotation about the $x$-axis by}
           \textit{ $\pi$ radians}
           %
             \\ 
         
        
        
        \bottomrule
	\end{tabular}
	\]

\end{table}

\begin{table}[htb]
    	\caption{Frequently used multi-qubit gates.}	\label{tab:gates_examples2}
    \[\setcellgapes{3pt}\makegapedcells 
	\centering 
\footnotesize
	\begin{tabular}{p{0.12\columnwidth}p{0.06\columnwidth}p{0.13\columnwidth}p{0.25\columnwidth} p{0.30\columnwidth}}
		\toprule
\textbf{Name} &	\textbf{Symbol} & \textbf{Circuit} & \textbf{Matrix representation} &\textbf{Action} \\
		\midrule
        Controlled-NOT & {$CNOT$} & 
       $\quad\qquad {\Qcircuit @C=1em @R=.9em {\lstick{\ket{\psi}} & \ctrl{1} & \qw \\
        \lstick{\ket{\gamma}} & \targ & \qw }} $
        &
        $\begin{bmatrix}
             1 & 0 & 0 & 0 \\
             0 & 1 & 0 & 0 \\
             0 & 0 & 0 & 1 \\
             0 & 0 & 1 & 0 \\
        \end{bmatrix}$
        & 
        $CNOT\left(\ket{\psi} \otimes \ket{\gamma}\right)=\ket{\psi} \otimes \ket{\psi\, \text{\texttt{XOR}}\,\gamma}$\\ \hline
        
        SWAP &  & 
        $\quad\qquad {\Qcircuit @C=1em @R=1.5em {
        \lstick{\ket{\psi}} & \qw & \link{1}{-1} & \rstick{\ket{\gamma}} \qw\\
        \lstick{\ket{\gamma}} &\qw & \link{-1}{-1} & \rstick{\ket{\psi}} \qw}} $
        &
        $\begin{bmatrix}
             1 & 0 & 0 & 0 \\
             0 & 0 & 1 & 0 \\
             0 & 1 & 0 & 0 \\
             0 & 0 & 0 & 1 \\
        \end{bmatrix}$
        & 
        It swaps two qubits \\ \hline
        
        Toffoli & $CCNOT$ & 
         $\quad\qquad
        {\Qcircuit @C=1em @R=.9em {\lstick{\ket{\psi}} & \ctrl{1} & \qw \\
        \lstick{\ket{\delta}} & \ctrl{1} & \qw \\
        \lstick{\ket{\gamma}} & \targ & \qw \\ }} $ 
        &
        $\begin{bmatrix}
             1 & 0 & 0 & 0 & 0 & 0 & 0 & 0 \\
             0 & 1 & 0 & 0 & 0 & 0 & 0 & 0 \\
             0 & 0 & 1 & 0 & 0 & 0 & 0 & 0 \\
             0 & 0 & 0 & 1 & 0 & 0 & 0 & 0 \\
             0 & 0 & 0 & 0 & 1 & 0 & 0 & 0 \\
             0 & 0 & 0 & 0 & 0 & 1 & 0 & 0 \\
             0 & 0 & 0 & 0 & 0 & 0 & 0 & 1 \\
             0 & 0 & 0 & 0 & 0 & 0 & 1 & 0 \\
        \end{bmatrix}$
        &  
        $
        \begin{matrix} 
           CCNOT\left(\ket{\psi} \otimes \ket{\gamma} \otimes \ket{\delta}\right) \\ 
            \qquad\qquad= \ket{\psi} \otimes \ket{\gamma} \otimes \ket{ (\psi \, \text{\texttt{AND}}\, \gamma) \, \text{\texttt{XOR}}\, \delta} 
           \end{matrix}
       $

        \\
        \bottomrule
	\end{tabular} 
	\]
\end{table}

\newpage

    \subsection{Variational Principle} \label{app:variational_principle}

The \textbf{Variational Principle}~\citepA{messiah1962quantum,binney2013physics,griffiths2018introduction} allows to set an upper bound to the ground state's energy of a quantum system for which we do not know the exact solution to the time-independent Schr{\"o}dinger equation:
\begin{equation}\label{eq:tise}
    \hat{H}\ket{\psi} = \xi\ket{\psi}
\end{equation}
where $\ket{\psi}$ is an eigenstate, with eigenvalue $\xi$, of the Hamiltonian operator $\hat{H}$. The eigenvalues of the Hamiltonian correspond to different energy levels of the system\h{. Thus,} they can be ordered using an index $i\geq 0$ so that $\xi_i > \xi_{i+1}$. We indicate with $\ket{\psi_i}$ the eigenstate with egenvalue $\xi_i$. We know that the ground state has the lowest eigenvalue, $\xi_0$. Thus the Variational Principle asserts that, for any given normalized $\ket{\psi}$, one has that: 
\begin{equation}\label{eq:vp}
    \xi_0 \leq \bra{\psi}\hat{H}\ket{\psi} = \langle \Ham \rangle
\end{equation}

\begin{proof}
We start the proof by considering the complete orthonormal set of eigenstates of $\hat{H}$, $\{\ket{\psi_n}\}$. We can express a generic quantum state as:
\begin{equation}
    \ket{\psi} = \sum_i c_i \ket{\psi_i}
\end{equation}
where any $\ket{\psi_i}$ \h{satisfies }~\autoref{eq:tise}. Based on the normalization property of wave functions and the orthonormality of the eigenfunctions, we have:
\begin{equation}\label{eq:norm}
    1 = \inprod{\psi}{\psi} 
    = \bigg\langle \sum_i c_i \psi_i  \,\bigg| \sum_j c_j \psi_j \bigg\rangle 
    = \sum_{i,j} c_i^* c_j \inprod{\psi_i}{\psi_j}
    = \sum_i |c_i|^2
\end{equation}

From~\autoref{eq:vp} and~\autoref{eq:norm}, the average value for the energy can \h{be} written as:
\begin{equation}\label{eq:meanh}
    \langle \hat{H} \rangle = \bigg\langle \sum_i c_i \psi_i  \,\bigg|\, \hat{H}\, \bigg|\, \sum_j c_j \psi_j \bigg\rangle
    = \sum_{i,j} c_i^* \xi_j c_j \inprod{\psi_i}{\psi_j}
    = \sum_i \xi_i |c_i|^2
\end{equation}

The energy of the ground state is encoded, by definition, in the smallest eigenvalue of $\hat{H}$. Thus,
\begin{equation}\label{eq:vp2}
    \langle \hat{H} \rangle \geq  \xi_0 \sum_i |c_i|^2 = \xi_0
\end{equation}
which concludes the proof.

\end{proof}

Translating into words, what the Variational Principle tells us is the following. Suppose we want to know the energy of the ground state of a given Hamiltonian operator $\hat{H}$ for which we do not know the eigenstates. We can pick a generic trial state, although with some constraints related to the continuity, characterized by a certain number of adjustable parameters and use it to evaluate~\autoref{eq:meanh}. Then, we can tweak the parameters to obtain the lowest possible value that resembles the best upper limit on the unknown $\xi_0$. Unfortunately, there is not an easy way to evaluate how far an upper bound is from $\xi_0$~\citepA{drake2002ground,korobov2002nonrelativistic}. We are only guaranteed that what we find is an upper bound.
    \subsection{Adiabatic Theorem}\label{appendix_adiabatic_theorem}

Generally speaking, if the external conditions of a system of interest are gradually changed, such a process can be considered adiabatic. Hence, such an approximation is typically named \textit{adiabatic approximation}. However, one must pay attention to the definition of what \quotes{gradually} means, and the \textit{Adiabatic Theorem}~\citepA{messiah1962quantum,binney2013physics,griffiths2018introduction} tells us exactly how to interpret such a term. 

\begin{adth*}
    Given a particle in the generic $i$-th eigenstate of the initial Hamiltonian, $\Ham_I = \Ham(t=0)$, it will follow the instantaneous $i$-th eigenstate of $\Ham(t)$ provided that the rate of change of $\Ham(t)$ is \h{low} enough.
\end{adth*}

Thus, the theorem deals with Hamiltonians that \h{vary} over time, i.e., $\Ham(t)$. 
Given such an operator, we have that eigenstates and eigenvalues change too. However, the orthonormality of the instantaneous eigenfunctions, $\ket{\psi_i(t)}$, is preserved. Thus, any solution to the Schr{\"o}dinger equation can still be expressed as a linear superposition of them. 

\begin{proof}
Given a time-dependent Hamiltonian, we can write the Schr{\"o}dinger equation as:
\begin{equation}\label{eq:tdse}
    i\hbar \frac{\partial}{\partial t} \ket{\psi(t)} = \Ham(t) \ket{\psi(t)}
\end{equation}
where
\begin{equation}\label{eq:tdse2}
    \ket{\psi(t)} = \sum_i c_i(t) \ket{\psi_i(t)} e^{i\theta_i(t)}
    \qquad \mathrm{and} \qquad
    \theta_i(t) = - \frac{i}{\hbar} \int_0^t \xi_i(t') dt'
\end{equation}
Note that differently from~\autoref{eq:tise}, currently, the energy is not constant. Instead, it is a function of time.

By inserting~\autoref{eq:tdse2} into~\autoref{eq:tdse}, we obtain
\begin{equation}\label{eq:tdse3}
    i \hbar \sum_i \big[\dot{c}_i \ket{\psi_i} + c_i \ket{\dot{\psi_i}} + i c_i \ket{\psi_i} \dot{\theta_i}\big] e^{i\theta_i} = \sum_i c_i \Ham \ket{\psi_i} e^{i\theta_i}
\end{equation}
where we drop the dependence from the time to compact the notation. By looking at the definition of $\theta_i$ in~\autoref{eq:tdse2} and from the Schr{\"o}dinger equation, we see that the last term of the left hand side in~\autoref{eq:tdse3} cancels out with the term on the right hand side. Thus, after taking the inner product with $\bra{\psi_j}$, we obtain:
\begin{equation}\label{eq:tdse4}
    \sum_i \dot{c}_i \delta_{ij} e^{i\theta_i} 
    = - \sum_i c_i \inprod{\psi_j}{\dot{\psi_i}}e^{i\theta_i} 
    \implies
    \dot{c}_j(t) = - \sum_i c_i \inprod{\psi_j}{\dot{\psi_i}}e^{i(\theta_i-\theta_j)} 
\end{equation}

By differentiating the Schr{\"o}dinger equation we obtain:
\begin{equation}\label{eq:tdse5}
    \dot{\Ham}\ket{\psi_i} + \Ham \ket{\dot{\psi}_i} = \dot{\xi}_i\ket{\psi_i} + \xi_i \ket{\dot{\psi}_i}
\end{equation}
and by taking the inner product with $\bra{\psi_j}$, we have:

\begin{equation}\label{eq:tdse6}
    \langle \psi_j | \dot{\Ham} | \psi_i \rangle = (\xi_i - \xi_j)\inprod{\psi_j}{\dot{\psi}_i}
\end{equation}

By inserting~\autoref{eq:tdse6} into~\autoref{eq:tdse4}, we obtain:
\begin{equation}\label{eq:tdse7}
    \dot{c}_j(t) = - c_j \inprod{\psi_j}{\dot{\psi}_j} - \sum_{i\neq j} c_i \frac{\langle \psi_j| \dot{\Ham} | \psi_i \rangle}{\xi_i - \xi_j} e^{i(\theta_i - \theta_j)}
\end{equation}
The results in~\autoref{eq:tdse7} is exact. We can now turn on the adiabatic approximation and consider the meaning of ``gradually". In such a context, we mean that the ratio in the second on the right hand side of~\autoref{eq:tdse7} is negligible. If such a condition is satisfied, then 
 $  \dot{c}_j(t) = - c_j \inprod{\psi_j}{\dot{\psi}_j}$
for which the solution is given by:
\begin{equation}\label{eq:tdse9}
    c_j(t) = c_j(t=0) e^{i\gamma_j(t)} 
    \qquad \mathrm{where} \qquad 
    \gamma_j(t) = i \int_0^t \big\langle \psi_j(t') | \dot{\psi}_j(t')  \big\rangle dt'
\end{equation}

\end{proof}

From~\autoref{eq:tdse9} we can see that if the system begins in the $j$-th eigenstate, i.e., $c_j(t=0) = 1$ and $c_i(t=0) = 0$ for $i \neq j$, then it will remain in such a state throughout the entire adiabatic evolution
\begin{equation}\label{eq:tdse10}
    \ket{\psi_j(t)} = \ket{\psi_j(t)} e^{i(\theta_j(t)+\gamma_j(t))} 
\end{equation}
Thus, the system stays in the $\ket{\psi_j}$ eigenstate and only picks up a phase factor.
 \bibliographystyleA{ACM-Reference-Format}
 \bibliographyA{bibl_sorted_by_year_abbr}
    
\end{document}